%% file: ms.tex
\documentclass[aps,prd,twocolumn,superscriptaddress,nofootinbib]{revtex4}

\usepackage{amsmath}   
\usepackage{graphicx}  
\usepackage{xspace}
\usepackage{units}
\usepackage{placeins}

\usepackage{natbib}
\usepackage{hyperref}
\usepackage{breakurl}
\usepackage{ulem}

\newcommand{\minerva}{MINERvA}
\newcommand{\qelike}{quasielastic-like}
\newcommand{\tune}{\minerva~GENIE tune v1}

\newcommand{\enuqe}{$E_{\nu,QE}$}
\newcommand{\pt}{$p_{t}$}
\newcommand{\pz}{$p_{||}$}
\newcommand{\qqqe}{$Q^{2}_{QE}$}

\newcommand{\sizecheck}{0} 
\newcommand{\PRDsupp}{0}   
\ifnum\PRDsupp=0
  
\else
  
\fi

\begin{document}

\title{Measurement of Quasielastic-Like Neutrino Scattering at $\left< E_\nu \right> \sim 3.5$~ GeV on a Hydrocarbon Target}

\newcommand{\Rutgers}{Rutgers, The State University of New Jersey, Piscataway, New Jersey 08854, USA}
\newcommand{\Hampton}{Hampton University, Dept. of Physics, Hampton, VA 23668, USA}
\newcommand{\Dortmund}{Institute of Physics, Dortmund University, 44221, Germany }
\newcommand{\Otterbein}{Department of Physics, Otterbein University, 1 South Grove Street, Westerville, OH, 43081 USA}
\newcommand{\JMU}{James Madison University, Harrisonburg, Virginia 22807, USA}
\newcommand{\Florida}{University of Florida, Department of Physics, Gainesville, FL 32611}
\newcommand{\UCIrvine}{Department of Physics and Astronomy, University of California, Irvine, Irvine, California 92697-4575, USA}
\newcommand{\CBPF}{Centro Brasileiro de Pesquisas F\'{i}sicas, Rua Dr. Xavier Sigaud 150, Urca, Rio de Janeiro, Rio de Janeiro, 22290-180, Brazil}
\newcommand{\PUCP}{Secci\'{o}n F\'{i}sica, Departamento de Ciencias, Pontificia Universidad Cat\'{o}lica del Per\'{u}, Apartado 1761, Lima, Per\'{u}}
\newcommand{\INRM}{Institute for Nuclear Research of the Russian Academy of Sciences, 117312 Moscow, Russia}
\newcommand{\Jlab}{Jefferson Lab, 12000 Jefferson Avenue, Newport News, VA 23606, USA}
\newcommand{\Pittsburgh}{Department of Physics and Astronomy, University of Pittsburgh, Pittsburgh, Pennsylvania 15260, USA}
\newcommand{\Guanajuato}{Campus Le\'{o}n y Campus Guanajuato, Universidad de Guanajuato, Lascurain de Retana No. 5, Colonia Centro, Guanajuato 36000, Guanajuato M\'{e}xico.}
\newcommand{\Athens}{Department of Physics, University of Athens, GR-15771 Athens, Greece}
\newcommand{\Tufts}{Physics Department, Tufts University, Medford, Massachusetts 02155, USA}
\newcommand{\WM}{Department of Physics, College of William \& Mary, Williamsburg, Virginia 23187, USA}
\newcommand{\FNAL}{Fermi National Accelerator Laboratory, Batavia, Illinois 60510, USA}
\newcommand{\Purdue}{Department of Chemistry and Physics, Purdue University Calumet, Hammond, Indiana 46323, USA}
\newcommand{\MCLA}{Massachusetts College of Liberal Arts, 375 Church Street, North Adams, MA 01247}
\newcommand{\UMD}{Department of Physics, University of Minnesota -- Duluth, Duluth, Minnesota 55812, USA}
\newcommand{\Northwestern}{Northwestern University, Evanston, Illinois 60208}
\newcommand{\UNI}{Universidad Nacional de Ingenier\'{i}a, Apartado 31139, Lima, Per\'{u}}
\newcommand{\Rochester}{University of Rochester, Rochester, New York 14627 USA}
\newcommand{\Austin}{Department of Physics, University of Texas, 1 University Station, Austin, Texas 78712, USA}
\newcommand{\USM}{Departamento de F\'{i}sica, Universidad T\'{e}cnica Federico Santa Mar\'{i}a, Avenida Espa\~{n}a 1680 Casilla 110-V, Valpara\'{i}so, Chile}
\newcommand{\Geneva}{University of Geneva, 1211 Geneva 4, Switzerland}
\newcommand{\Chicago}{Enrico Fermi Institute, University of Chicago, Chicago, IL 60637 USA}
\newcommand{\hired}{}
\newcommand{\OregonState}{Department of Physics, Oregon State University, Corvallis, Oregon 97331, USA}
\newcommand{\oxford}{Oxford University, Department of Physics, Oxford, United Kingdom}
\newcommand{\umiss}{University of Mississippi, Oxford, Mississippi 38677, USA}
\newcommand{\upenn}{Department of Physics and Astronomy, University of Pennsylvania, Philadelphia, PA 19104}
\newcommand{\AMU}{AMU Campus, Aligarh, Uttar Pradesh 202001, India}
\newcommand{\wroclaw}{University of Wroclaw, plac Uniwersytecki 1, 50-137 Wrocław, Poland}
\newcommand{\Mohali}{IISER, Mohali, Knowledge city, Sector 81, Manauli PO 140306}
\newcommand{\joelmousseauThanks}{now at University of Michigan, Ann Arbor, MI 48109, USA}
\newcommand{\cpatrickThanks}{Now at University College London, London WC1E 6BT, UK}
\newcommand{\jwolcottThanks}{now at Tufts University, Medford, MA 02155, USA}

\author{D.~Ruterbories}                   \affiliation{\Rochester}
\author{K.~Hurtado}                       \affiliation{\CBPF}  \affiliation{\UNI}
\author{J.~Osta}                          \affiliation{\FNAL}
\author{F.~Akbar}                         \affiliation{\AMU}
\author{L.~Aliaga}                        \affiliation{\WM}  \affiliation{\PUCP}
\author{D.A.~Andrade}                     \affiliation{\Guanajuato}
\author{M.~V.~Ascencio}                   \affiliation{\PUCP}
\author{A.~Bashyal}                       \affiliation{\OregonState}
\author{A.~Bercellie}                     \affiliation{\Rochester}
\author{M.~Betancourt}                    \affiliation{\FNAL}
\author{A.~Bodek}                         \affiliation{\Rochester}
\author{H.~Budd}                          \affiliation{\Rochester}
\author{G.~Caceres}                       \affiliation{\CBPF}
\author{T.~Cai}                           \affiliation{\Rochester}
\author{M.F.~Carneiro}                    \affiliation{\OregonState}
\author{J.~Chaves}                        \affiliation{\upenn}
\author{D.~Coplowe}                       \affiliation{\oxford}
\author{H.~da~Motta}                      \affiliation{\CBPF}
\author{S.A.~Dytman}                      \affiliation{\Pittsburgh}
\author{G.A.~D\'{i}az~}                   \affiliation{\Rochester}  \affiliation{\PUCP}
\author{J.~Felix}                         \affiliation{\Guanajuato}
\author{L.~Fields}                        \affiliation{\FNAL}  \affiliation{\Northwestern}
\author{A.~Filkins}                       \affiliation{\WM}
\author{R.~Fine}                          \affiliation{\Rochester}
\author{A.M.~Gago}                        \affiliation{\PUCP}
\author{R.Galindo}                        \affiliation{\USM}
\author{A.~Ghosh}                         \affiliation{\USM}  \affiliation{\CBPF}
\author{R.~Gran}                          \affiliation{\UMD}
\author{J.Y.~Han}                         \affiliation{\Pittsburgh}
\author{D.A.~Harris}                      \affiliation{\FNAL}
\author{S.~Henry}                         \affiliation{\Rochester}
\author{S.~Jena}                          \affiliation{\Mohali}
\author{D.Jena}                           \affiliation{\FNAL}
\author{J.~Kleykamp}                      \affiliation{\Rochester}
\author{M.~Kordosky}                      \affiliation{\WM}
\author{D.~Last}                          \affiliation{\upenn}
\author{T.~Le}                            \affiliation{\Tufts}  \affiliation{\Rutgers}
\author{X.-G.~Lu}                         \affiliation{\oxford}
\author{E.~Maher}                         \affiliation{\MCLA}
\author{W.A.~Mann}                        \affiliation{\Tufts}
\author{C.~Mauger}                        \affiliation{\upenn}
\author{K.S.~McFarland}                   \affiliation{\Rochester}  \affiliation{\FNAL}
\author{A.M.~McGowan}                     \affiliation{\Rochester}
\author{B.~Messerly}                      \affiliation{\Pittsburgh}
\author{J.~Miller}                        \affiliation{\USM}
\author{J.G.~Morf\'{i}n}                  \affiliation{\FNAL}
\author{J.~Mousseau}\thanks{\joelmousseauThanks}  \affiliation{\Florida}
\author{D.~Naples}                        \affiliation{\Pittsburgh}
\author{J.K.~Nelson}                      \affiliation{\WM}
\author{C.~Nguyen~}                       \affiliation{\Florida}
\author{A.~Norrick}                       \affiliation{\FNAL}  \affiliation{\WM}
\author{Nuruzzaman}                       \affiliation{\Rutgers}  \affiliation{\USM}
\author{A.~Olivier}                       \affiliation{\Rochester}
\author{V.~Paolone}                       \affiliation{\Pittsburgh}
\author{C.E.~Patrick}\thanks{\cpatrickThanks}  \affiliation{\Northwestern}
\author{G.N.~Perdue}                      \affiliation{\FNAL}  \affiliation{\Rochester}
\author{M.A.~Ram\'{i}rez}                 \affiliation{\Guanajuato}
\author{R.D.~Ransome}                     \affiliation{\Rutgers}
\author{H.~Ray}                           \affiliation{\Florida}
\author{D.~Rimal}                         \affiliation{\Florida}
\author{P.A.~Rodrigues}                   \affiliation{\oxford} \affiliation{\umiss} \affiliation{\Rochester}
\author{H.~Schellman}                     \affiliation{\OregonState}  \affiliation{\Northwestern}
\author{J.T.~Sobczyk}                     \affiliation{\wroclaw}
\author{C.J.~Solano~Salinas}              \affiliation{\UNI}
\author{H.~Su}                            \affiliation{\Pittsburgh}
\author{M.~Sultana}                       \affiliation{\Rochester}
\author{V.S.~Syrotenko}                   \affiliation{\Tufts}
\author{E.~Valencia}                      \affiliation{\WM}  \affiliation{\Guanajuato}
\author{J.~Wolcott}\thanks{\jwolcottThanks}  \affiliation{\Rochester}
\author{M.Wospakrik}                      \affiliation{\Florida}
\author{C.~Wret}                          \affiliation{\Rochester}
\author{B.~Yaeggy}                        \affiliation{\USM}
\author{L.~Zazueta}                       \affiliation{\WM}

\collaboration{The MINER$\nu$A Collaboration}\ \noaffiliation

\date{\today}

\begin{abstract}
\minerva~presents a new analysis of neutrino induced \qelike~interactions in a hydrocarbon tracking target. We report a double-differential cross section using the muon transverse and longitudinal momentum. In addition, differential cross sections as a function of the square of the four-momentum transferred and the neutrino energy are calculated using a quasielastic hypothesis. Finally, an analysis of energy deposited near the interaction vertex is presented. These results are compared to modified GENIE predictions as well as a NuWro prediction. All results use a dataset produced by $3.34\times10^{20}$ protons on target creating a neutrino beam with a peak energy of approximately 3.5 GeV.
\end{abstract}
\ifnum\sizecheck=0  
\maketitle
\fi

\input{Introduction.tex}
\input{Experiment.tex}

\input{Simulation.tex}

\input{SignalDef.tex} 
\input{EventReconstruction.tex}

\input{Selection.tex}

\input{Extraction.tex}

\input{Systematics.tex}
\input{Results.tex}

\clearpage
\input{Conclusions.tex}

\input{Acknowledgments.tex}

\clearpage

\bibliography{MINERvA_QE}
\clearpage
\input{Appendix.tex}

\ifnum\PRDsupp=1
  \clearpage
  \input{Supplemental.tex}

  \input{SupplementalQE.tex}

\fi

\end{document}

%% file: Introduction.tex
\section{Introduction}
\label{sec:Intro}
Charged current quasielastic (CCQE) scattering is an important signal process for neutrino oscillation experiments\citep{Abe:2015awa}\citep{Adamson:2016tbq}\citep{Acciarri:2015uup}\citep{Abe:2011ts}. In this process an incoming neutrino exchanges a $W$ boson with a neutron to produce an outgoing charged lepton and a proton.
The initial state neutron is not independent but bound inside the nucleus with an initial Fermi momentum. Thus the kinematics of this apparently simple process are significantly modified compared to the predictions from a straightforward two body scattering event.
In addition, the presence of the nucleus can cause other processes, for example resonance production and decay followed by pion absorption, to appear quasielastic in a detector.  All of these modifications taken together affect not only the total cross section and outgoing hadronic system, but also the outgoing lepton kinematics.  

We present in this paper the two-dimensional cross sections for quasielastic-like scattering as a function of muon transverse and longitudinal momentum.  Momentum of the final state lepton can be measured precisely and without reference to an interaction model in such neutrino interactions. These muon observables are suitable for comparing to models of inclusive neutrino scattering that can separate events by final state. Using the reconstructed muon angle and momentum the quantities \enuqe~and \qqqe~ are reconstructed. These are the reconstructed neutrino energy (Eq. \ref{eq:enuqe}) and square of the four-momentum transferred (Eq. \ref{eq:qqqe}) for the limiting case where the initial state nucleon was at rest and there were no nuclear effects. The binding energy, $E_b$, used in this analysis is 34 MeV. 

\begin{equation}
\label{eq:enuqe}
E_{\nu,QE} = \frac{M_n^2-(M_p-E_b)^2-M_\mu^2+2(M_p-E_b)E_\mu}{2(M_p-E_b-E_\mu+P_\mu cos(\theta_\mu))}
\end{equation}

\begin{equation}
\label{eq:qqqe}
Q^{2}_{QE} = 2E_\nu(E_\mu-P_\mu cos(\theta_\mu)) - M_\mu^2
\end{equation}

A differential cross section in \qqqe~is reported which extends to higher and lower \qqqe~than previously reported in Ref.~\citep{Fiorentini:2013ezn}, which was based on a subset of the data presented in this work. Additional studies associated with the vertex energy are presented in this paper, which build on the results of Ref. \citep{Fiorentini:2013ezn}.

%% file: Experiment.tex
\section{Experiment}
The \minerva~experiment employs a fine-grained tracking detector for recording neutrino interactions produced by the NuMI beamline at Fermilab\citep{Adamson:2015dkw}. 
Neutrinos are produced by directing 120 GeV protons from the Main Injector onto a graphite target. 
The produced charged pions and kaons are focused by two magnetic horns. 
The charged particles travel along a 675~m long, helium-filled pipe and decay primarily into muon-flavored neutrinos and muons. 
The polarity of the horns can be switched to produce either a neutrino-dominated or antineutrino-dominated beam at the peak neutrino energy.  
Approximately 97$\%$ of the muon neutrinos that reach the \minerva\  detector are produced by pion decay; the remainder come from kaon decay\cite{Adamson:2015dkw,Aliaga:2016oaz}.  The neutrino beam is inclined at an angle of 58~mrad down relative to the horizontal ($z$) direction at the \minerva\ detector.

The \minerva\ detector\cite{Aliaga:2013uqz} consists of 120 hexagonal modules.
These modules comprise an active tracking volume and a suite of nuclear targets positioned upstream, which are not used for this analysis.  This analysis studies the interactions in the scintillator tracking volume which has a fiducial mass of 5.48 tons,  
and is surrounded by electromagnetic and hadronic calorimeters.   

Each tracking module consists of two planes of triangular nested  polystyrene scintillator strips with a 1.7 cm strip-to-strip pitch. 
There are three different plane configurations rotated 0$^{\circ}$ and $\pm$ 60$^{\circ}$ relative to the longitudinal axis of the detector. 
The modules are arrayed in a stereoscopic combination along the horizontal axis. 
This enables a three dimensional reconstruction of the neutrino interactions via their charged particles in the final state. 
The downstream and side electromagnetic calorimeter consists alternating layers of scintillator and 2~mm thick lead planes while alternating scintillaor and 2.54~cm thick steel planes constitute the downstream hadronic calorimeter.  

Each scintillator strip is read out by a wavelength-shifting fiber connected to a multi-anode photomultiplier tube. 
The 3.0 ns timing resolution of the readout electronics is sufficient for separating multiple interactions within a single NuMI beam spill. 

The MINOS near detector\cite{Michael:2008bc}, a scintillator-based magnetized iron spectrometer, is located 2~m downstream of the \minerva\  detector.  
It is used to reconstruct the momentum and charge of muons that originate inside \minerva\ and enter MINOS. 

This analysis uses data that comprise an integrated $3.34\times 10^{20}$ protons on target received between March 2010 and May 2012, when the focusing horns were set to produce a broad band neutrino beam peaked at 3.5 GeV with a purity of 95\% muon neutrinos at the peak.

%% file: Simulation.tex
\section{Simulation}
\label{sec:Simulation}

The neutrino beam is simulated using a GEANT4-based description of the NuMI beamline\citep{Adamson:2015dkw}.  Because GEANT4\citep{Agostinelli2003250} does not exactly reproduce NA49's hadron production measurements of 158~GeV protons on carbon\citep{Alt:2006fr}, or other relevant hadron production measurements, the simulation is reweighted as a function of the pion transverse and longitudinal momentum to agree with those data, as described in Ref.~\cite{Aliaga:2016oaz}.  In addition, the flux is constrained by the measurement, in the same beamline, of neutrino scattering off atomic electrons, as described in Ref.~\cite{Park:2015eqa}.  

Neutrino interactions are simulated using the GENIE 2.8.4 neutrino event generator~\cite{Andreopoulos:2009rq}.  The quasielastic interaction (1p1h) is simulated by the Llewellyn-Smith formalism\citep{LlewellynSmith:1971zm}, and the vector form factors are modeled using the BBBA05 model~\cite{Bradford:2006yz}.  The dipole form is used for the axial vector form factor, with an axial mass of $M_A=0.99$ GeV/c$^2$.  The Rein-Sehgal model~\cite{Rein:1980wg} is used to simulate resonance production, and the resonant axial mass used is $M_A^{RES}=1.12$ GeV/c$^2$.  The DIS cross sections used by GENIE are a leading order model that use the Bodek-Yang prescription\citep{Bodek:2004pc} for the low $Q^2$ modification.  

The nuclear environment is simulated in GENIE using the relativistic Fermi gas model~\cite{Smith:1972xh} in combination with the Bodek-Ritchie high momentum tail~\cite{Bodek:1981wr} on the nucleon momentum distribution.  The maximum momentum for Fermi motion is assumed to be $k_F=0.221$ GeV/c.  GENIE models intranuclear rescattering, or final state interactions (FSI), of the produced hadrons using the INTRANUKE-hA package~\cite{Dytman:2007zz}.

\minerva~makes three sets of corrections to this version of GENIE that are based on improved models or measurements of the underlying processes.  First, we reduce the cross section for quasielastic events as a function of energy ($q_0$) and three momentum transfer ($q_3$) ( or four-momentum transfer ($Q^2$) above 3 GeV$^2$) based on the RPA part of the Valencia model~\cite{Nieves:2004wx}\citep{Gran:2017psn} appropriate for a Fermi gas\citep{Martini:2016eec}\citep{Nieves:2017lij}.  Second, we add a prediction for multinucleon (2p2h\footnote{The notation "2p2h" is meant to distinguish hard scattering process with momentum and energy transferred to a pair of nucleons from the standard quasielastic process, "1p1h", where the momentum and energy is transferred to a single nucleon in the hard scattering process.}) scattering from this same Valencia model~\citep{Nieves:2011pp}\citep{Gran:2013kda}\citep{Schwehr:2016pvn}.  However, we know from studies of an inclusive sample of \minerva~data at low momentum transfer that this model significantly underpredicts the measured cross section in a specific region of $q_0$ and $q_3$ where 2p2h processes contribute up to one-third of the rate~\cite{Rodrigues:2015hik}. We therefore enhance the 2p2h prediction from the Nieves model in this region.  Integrated over all phase space the rate of 2p2h is increased by 53\%. Third, we decrease the non-resonant pion production by 43\% to better agree with a fit to measurements on deuterium of that process~\cite{Rodrigues:2016xjj}, and take the uncertainties on that process from the same reference rather than use the GENIE-recommended uncertainties.  This modified version of the simulation is referred to later in this paper as \tune~and MnvGENIE-v1 in figures.  

The response of the \minerva~detector is simulated using a GEATNT4-based model which uses version 4.9.4p6 GEANT4~\cite{Agostinelli2003250}.  The optical and electronics performance is also simulated, and the absolute energy scale of minimum ionizing energy depositions is set by requiring the average and RMS of the energy deposits by through-going muons to match between the data and simulation as a function of time, as described in Ref. \cite{Aliaga:2013uqz}.  The absolute energy response of the detector to charged hadrons was determined by placing a scaled-down version of the MINERvA detector in a charged particle test beam, as described in Ref.~\cite{Aliaga:2015aqe}. In order to simulate the effects of accidental activity, such as deadtime and reconstruction failures, each simulated neutrino interaction is embedded in an event from the data in both the MINERvA and MINOS detector.  These spills from the data are taken from the same periods (typically determined by a configuration change) the Monte Carlo simulates.

%% file: SignalDef.tex
\section{Signal Defintion}
\label{sec:signaldef}

Quasielastic interactions are those where there are one or more nucleons in the final state, but no mesons or excited or heavy baryons. Charged current neutrino interactions where a pion is produced in the primary process but is absorbed in final state interactions appear as quasielastic in the \minerva~detector and are therefore also considered signal events in this analysis.  

This analysis uses a \qelike~signal, defined as those events with the following final state particles: 
\begin{itemize}
\item One negatively charged muon of angle $<20^{\circ}$ with respect to the neutrino beam
\item Any number of protons or neutrons
\item No mesons 
\item No heavy or excited baryons 
\item Any number of photons with energy $\le$ 10 MeV
\end{itemize}   
Events containing low energy ($\le$ 10 MeV) photons are accepted because they can arise from nuclear de-excitation in atoms and might not be part of the primary neutrino interaction products. 
Because of the geometric acceptance (MINOS-\minerva~overlap) and the need for muon charge and momentum identification, a restriction on the muon angle of $<20^{\circ}$ is required. 

%% file: EventReconstruction.tex
\section{Event Reconstruction}
\label{sec:eventreco}
In order for the quasielastic-like cross sections to be reported as a function of transverse and longitudinal muon momentum, event reconstruction must, at a minimum, measure the momentum, charge, and direction of the muon.  If another track consistent with a proton is found, then the proton's kinetic energy and direction can also be measured.  Recoil energy is defined as the event activity that is not coming from either the muon or any tracked protons. The absence of pions can be ensured by a cut on the recoil energy, and by requiring no evidence for a Michel electron from the $\pi\to\mu\to e$ decay chain.   Because low energy protons can deposit energy close to the vertex and not be reconstructed as tracks, the recoil energy does not include energy that is less than 150~mm from the vertex.  

Every event is required to have a reconstructed primary interaction vertex
inside the fiducial volume of the scintillator region of the \minerva~detector. 
The fiducial volume is defined as the region between scintillator modules 27 and 79 of the detector with an apothem of 850 mm giving a 5.48 metric ton fiducial mass. The tracker composition is described in Sec. \ref{sec:normalization}. The remaining aspects of event reconstruction are described in the following sections.
\subsection{Muon Track Reconstruction}
\label{sec:long_tracking}

This analysis requires events to contain one long track that originates in the fiducial volume defined above, traverses the remainder of the \minerva\ detector, and extrapolates to a track in MINOS whose momentum and charge are measured.  
The momentum of the muon as it enters the upstream face of the MINOS detector is measured on the basis of its range or track curvature information, as described in Ref.\cite{Michael:2008bc}. The total muon momentum must also include the momentum required to traverse the distance from the event vertex to the downstream end of the \minerva\ detector.  This momentum is determined by  computing the minimum ionization loss for the material traversed.

Long track signatures in \minerva\ are categorized as those that possess a contiguous set of clusters for at least 8 planes of the detector. 
Most of these tracks are made from muons, although  
sometimes they are made from energetic protons or pions. 

The \minerva\ detector uses planes of scintillator bars in three different orientations (vertical and $\pm$60$^{\circ}$ from vertical) to reconstruct tracks in three dimensions. 
A long track is found by looking for tracks in at least two of the three views, and for events with a long track found in all three views, they must all be geometrically consistent.  Given the geometry of the detector, a muon must traverse at least 9 (11) planes in order to create tracks in two (three) views.  

A track matching algorithm ensures that the reconstructed part of the muon track in \minerva\ matches with its corresponding momentum-analyzed counterpart in MINOS, by comparing their track timings and extrapolated positions at the back/front of the \minerva /MINOS detector.

Once a muon track has been identified in a beam spill, an event is formed with a specific (length-corrected) average time for the muon, a vertex, and a set of clusters which will no longer be considered in subsequent reconstruction steps that characterize other aspects of the event.  For the remainder of Section \ref{sec:eventreco}, only clusters that are between 20~ns before and 35~ns after the average muon time are considered.  In addition, the energy in each cluster is corrected for the passive material that is upstream of the cluster.

The corrected energy of clusters are further reconstructed into different objects:  short tracks, showers, vertex energy, residual energy, and recoil energy. The remainder of this section describes how these objects are formed.  

\subsection{Short Track Reconstruction}
\label{sec:short_tracking}
Low energy (less than 100 MeV) protons or other heavy baryons typically produce short tracks. These traverse just a few planes in the \minerva~detector. There are two tracking algorithms employed for reconstructing short tracks at the neutrino interaction vertex. 

Both the algorithms input clusters that are not yet associated with any reconstructed object in the event. The tracking threshold is 4 clusters, which means that particles must have deposited energy in at least 4 different planes to be considered for short tracks.  

The first algorithm starts by making track seeds out of sets of at least 4 contiguous planes with clusters near the interaction vertex. An iterative procedure merges these seeds into larger ones based on their relative proximity and angular distributions. A newly made short track is then anchored to the interaction vertex by adding the vertex cluster to the track’s clusters.  

The second algorithm performs a scan around the interaction vertex looking for clusters that have the same angular orientations. Most of the short tracks made here consist of contiguous clusters, but some (stub-like, present in just a few planes) can also contain clusters that are not contiguous. The anchoring of the newly made short track to the interaction vertex is based on the proximity of its nearest node to the vertex.

\subsection{Shower Reconstruction}
\label{sec:shower_objects}
Showers in MINERvA are defined as activity that is correlated in space but is not track-like. The shower reconstruction routine starts by using in time clusters that are not part of tracks. 
It combines cluster information from all three of the stereoscopic plane orientations and delivers one or many three dimensional shower objects. 
The routine first utilizes a peak finding algorithm to select the highest energy clusters in each of the three orientations separately, inside the tracker, the downstream electromagnetic calorimeter, and the hadron calorimeter.  
These are used for initiating the shower building and are referred to as shower seeds. 
A particular cluster is considered to be adjacent to the shower seed, if it is present in the next detector plane of the same orientation as the shower seed. 
The cluster is also required to overlap with the shower seed with respect to its transverse positions. 
If both the criteria are satisfied, the candidate cluster is added on to the growing shower seed.  
The shower reconstruction routine follows an iterative procedure, whereby newly formed shower seeds are also merged with each other, depending on their conditions of adjacency and overlapping positions. 

The seeds made inside each of the orientations are now matched across the different plane orientations, to transform them into three dimensional objects. 
This growing occurs inside each sub detector of the \minerva\  detector.  
Two seeds are said to be adjacent if they are present in the same or neighboring detector modules and overlap in their transverse dimensions.  
In this way a three dimensional shower seed is formed. 
An iterative combining procedure merges these shower seeds to form large, three dimensional shower-like objects. 

The energy of the shower is estimated by correcting each cluster in the shower for the passive material traversed just upstream of that cluster. 
\subsection{Vertex Energy Reconstruction}  
\label{sec:vtxBlobber}

The vertex energy in this analysis is defined as the untracked energy within a radius of 150~mm from the muon track vertex. The energy in this region is most likely coming from low energy protons which were not tracked, but could also come from other particles as well.  Because the current models of neutrino interactions do not model low energy protons accurately, we collect the energy of clusters in all three views in this region to study it, but we do not make any selection in this analysis based on the energy in this region.    

\subsection{Residual Energy Reconstruction} 
\label{sec:dispBlobber}
The residual energy is simply the scalar sum of the energy in all the unused clusters in all three views: those that are not already part of tracks, showers or within the primary vertex region. 
The sum can only be made after the track, shower and vertex energy reconstruction has occurred.

\subsection{Recoil Energy Reconstruction}
\label{sec:nonvertex_activity}
The ``recoil" for this analysis is defined as the activity not associated with the muon and proton candidate tracks, which is at least 150~mm away from the primary vertex. 
The total recoil energy is determined by taking the calorimetric sum of all the showers and the residual energy.

%% file: Selection.tex
\section{Event Selection}
\label{sec:eventselection}
In order to be sensitive to both low momentum transfer events where the outgoing nucleons are below tracking threshold, and to take advantage of the additional kinematic handles when a final state proton is identified, this analysis selects events with one negatively charged muon and any number of additional tracked particles originating from the muon vertex.   
Quasielastic-like interactions with Q$^{2} \lesssim $ 0.5 GeV/c$^{2}$,  often have soft nucleons in the final state which are below the tracking threshold.  
Final state protons from interactions with higher Q$^{2}$ are energetic enough to be tracked. 
Quasielastic-like candidate events for this analysis therefore fall into the categories of muon-only, and muon plus N$>$0 protons. The final cross sections reported in this paper combine these samples, but some reconstructed quantities like vertex energy and recoil energy show distinct differences which are of interest. 

In order to be considered in this analysis, an event  must have a muon track in \minerva\ that is matched into the MINOS detector, and whose charge has been identified in MINOS to be negative.  The origin of the muon track must be within the fiducial volume of the tracker volume of the \minerva\ detector.  
Once a muon track is found in a slice of time within the beam spill, the selection process removes activity coming from additional tracks which are not associated with the primary interaction.  These tracks occur because of the high intensity of the NuMI beam, and are uncorrelated with the muon track that defines the primary interaction.  The activity from  tracks whose time difference with respect to the muon time is outside the window of -5 to 10 ns are ignored. The recoil and vertex energy algorithms described above ignore this energy. If the track is within the timing window then particle identification cuts (Sec. \ref{subsec:pid}) are applied to the track.

\subsection{Charged Pion Rejection}
\label{subsec:pid}

All non-muon tracks passing the above timing cut have a proton particle identification algorithm applied to them. This algorithm compares the charge deposition spectrum to Bethe-Bloch expectations to estimate the momentum as well as provide a hypothesis of particle type. This analysis has a cut varying with \qqqe~which is relaxed as \qqqe~ increases due to the higher probability of secondary interactions between protons and the detector material.

\begin{figure}[tph!]
\centering
  \includegraphics[width=0.9\linewidth]{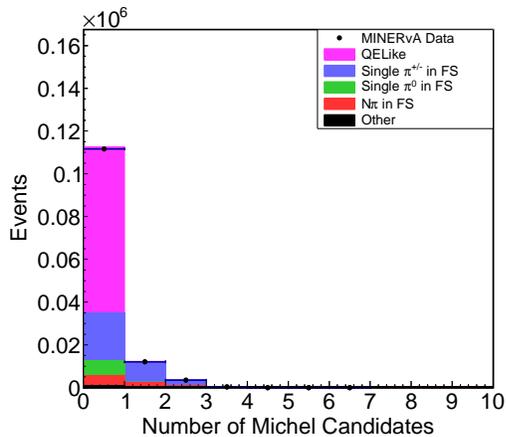} 
  \includegraphics[width=0.9\linewidth]{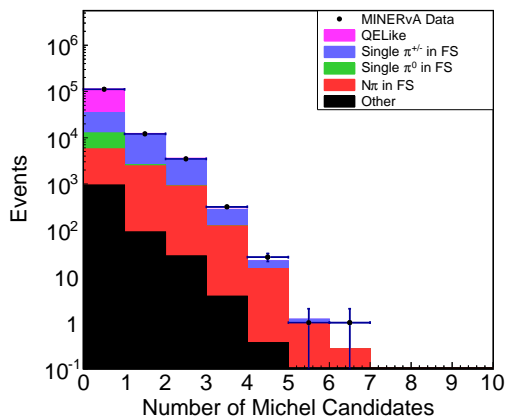} 
\caption{Number of reconstructed Michel candidates in the data and simulation for both a linear (top) and logarithmic (bottom) scale. The analysis signal events are labeled ``QELike". The background is broken down by the type of pion content in the final state (FS). Events with exactly one pion in the final state are labeled by the charge of the pion. If more than one pion is present it is labeled ``N$\pi$". The ``other" category is predominately from events with kaons in the final state.}
\label{fig:evsel_michelN}
\end{figure}

\begin{figure}[tp]
\centering
  \includegraphics[width=0.9\linewidth]{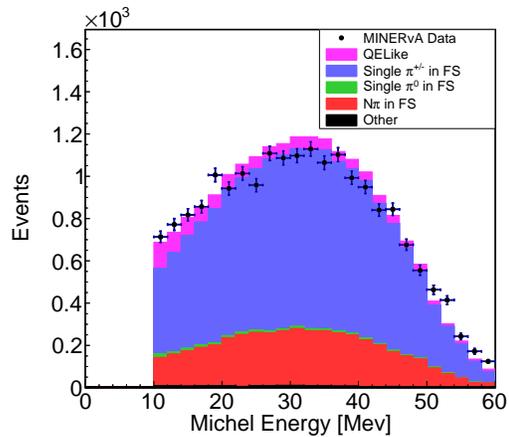} 
\caption{Reconstructed energy spectrum of Michel candidates in the data and simulation.}
\label{fig:evsel_michelEnergy}
\end{figure}

\begin{figure}[tp]
\centering
  \includegraphics[width=0.9\linewidth]{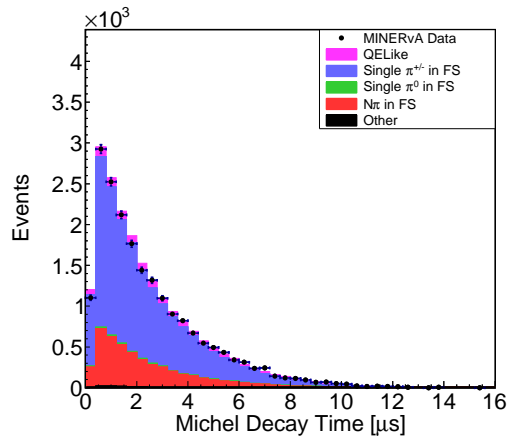} 
  \includegraphics[width=0.9\linewidth]{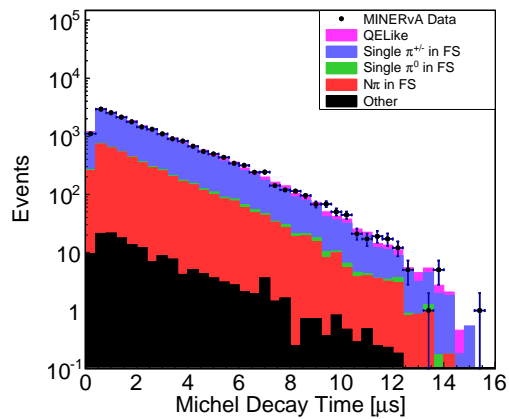} 
\caption{Decay time of the Michel candidate with respect to the vertex time in both a linear (left) and logarithmic (right) scale for the data and simulation.}
\label{fig:evsel_michelTime}
\end{figure}

Another way to reject interactions producing charged pions is to search for candidate Michel electrons that result from the pion to muon to electron decay chain.  These electrons are produced later in time than the primary neutrino interaction, and are defined as being either within 200~mm of the muon vertex because they originate from very low energy pions, or within 200~mm of the end points of any tracks passing the timing cut applied above.  
Michel candidates have at most  70~MeV as constrained by the muon mass, and occur at a later time.  Figures~\ref{fig:evsel_michelN} and \ref{fig:evsel_michelEnergy} show the multiplicity and energy of the Michel candidates respectively, and Fig.~\ref{fig:evsel_michelTime} shows the  time difference between the initial neutrino interaction and the Michel candidates from the data and simulation.  Events with one or more Michel candidates are rejected from the signal sample.

\subsection{Multi-pion and DIS Rejection}  

\begin{figure}[tp]
\centering
  \includegraphics[width=0.9\linewidth]{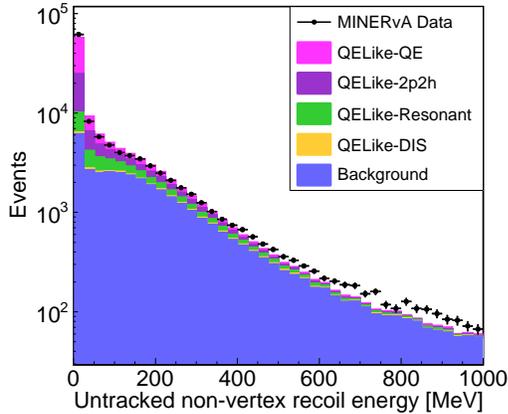} 
\caption{Recoil spectrum of candidate signal events after all cuts except recoil cut. The analysis signal is divided up into the initial state processes , as defined by GENIE, include true quasielastic events (QELike-QE) and multinucleon effects (QELike-2p2h). The contribution from pion production with subsequent pion absorption (QELike-Resonant and QELike-DIS) is shown. All backgrounds are categorized ``Background".}
\label{fig:evsel_recoil}
\end{figure}

Because of the broad energy range of the NuMI beamline, the \minerva~detector has access to events with large hadronic recoil mass (W). As a result, a large fraction of events can pass the event selection while having large amounts of recoil energy in the detector. An earlier analysis published in Ref.~\cite{Fiorentini:2013ezn} placed a tighter requirement on the recoil energy. This event selection choice introduced a model dependence on the recoil shape of the 2p2h events, and an efficiency dependence is introduced with this tighter selection which is not desirable. Instead, the selection criteria is modified to remove events with recoil energy greater than 500 MeV and the extra background accepted is controlled via control samples described in Sec. \ref{sec:extraction}. Figure \ref{fig:evsel_recoil} is the untracked recoil for events passing all selection cuts except the recoil cut.

The clustering algorithm to identify showers described in Sec.~\ref{sec:shower_objects} identifies isolated energy deposits that may come from photons as well as neutrons.  No more than one isolated shower is allowed in the event. Events with two or more showers are used in the two sideband samples described later in Section \ref{subsec:bkgconstraint}.

\begin{figure}[tp]
\centering
  \includegraphics[width=0.9\linewidth]{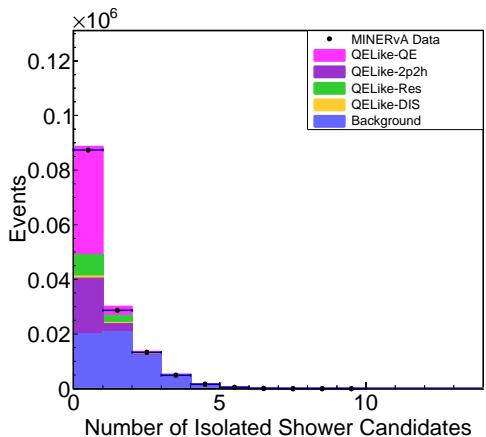} 
\caption{Number of reconstructed isolated showers.}
\label{fig:evsel_blobn}
\end{figure}

\begin{figure}[tp]
\centering
  \includegraphics[width=0.9\linewidth]{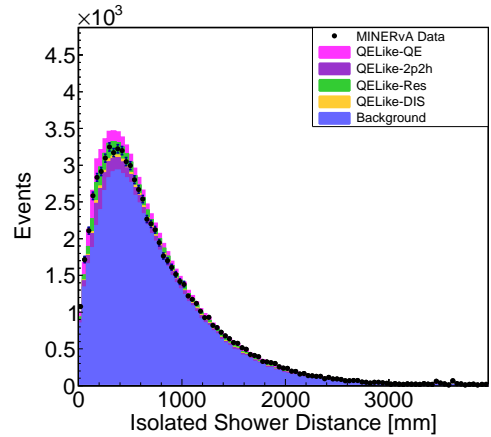} 
\caption{Distance of isolated showers from the vertex.}
\label{fig:evsel_blobdis}
\end{figure}

\begin{figure}[tp]
\centering
  \includegraphics[width=0.9\linewidth]{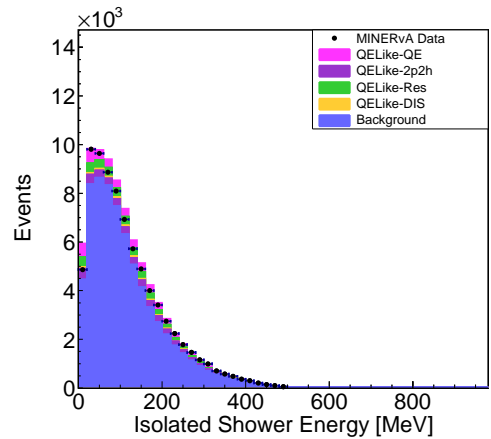} 
\caption{Energy spectrum of isolated showers.}
\label{fig:evsel_blobenergy}
\end{figure}

Figures \ref{fig:evsel_blobn}, \ref{fig:evsel_blobdis}, and \ref{fig:evsel_blobenergy} show the number of shower objects, distance from the interaction vertex and energy respectively.

\subsection{Final Event Sample}

\begin{figure*}[p]
\centering
  \includegraphics[width=\textwidth]{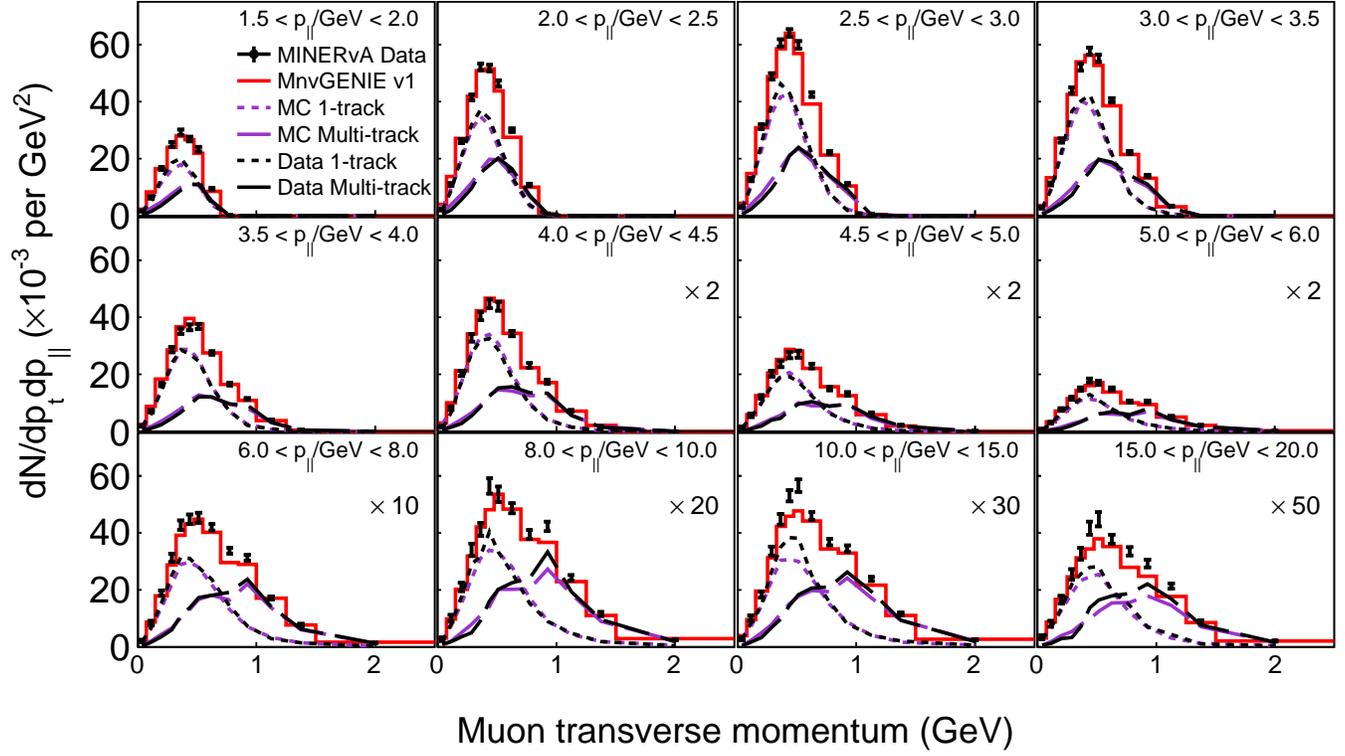} 
  \includegraphics[width=\textwidth]{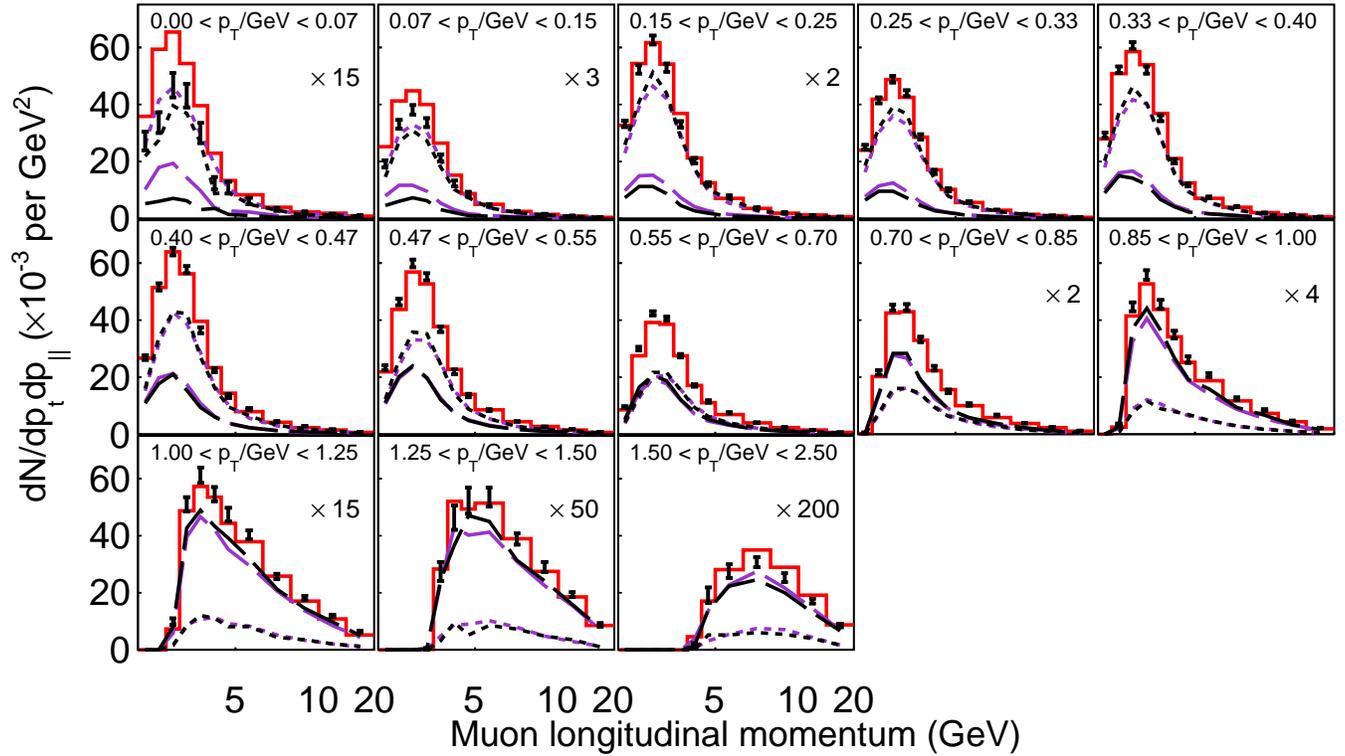} 
\caption{Selected event rates broken down by track multiplicity. The solid lines are contributions from events with more than one reconstructed track. The dotted lines are events where only the muon was reconstructed. The MC prediction for the solid/dotted lines are red while the measured rates are in black. A multiplier, which scales all content in the panel, has been applied to some panels to improve visibility.}
\label{fig:evsel_evtbymultptpz}
\end{figure*}

\begin{figure*}[p]
\centering
  \includegraphics[width=\textwidth]{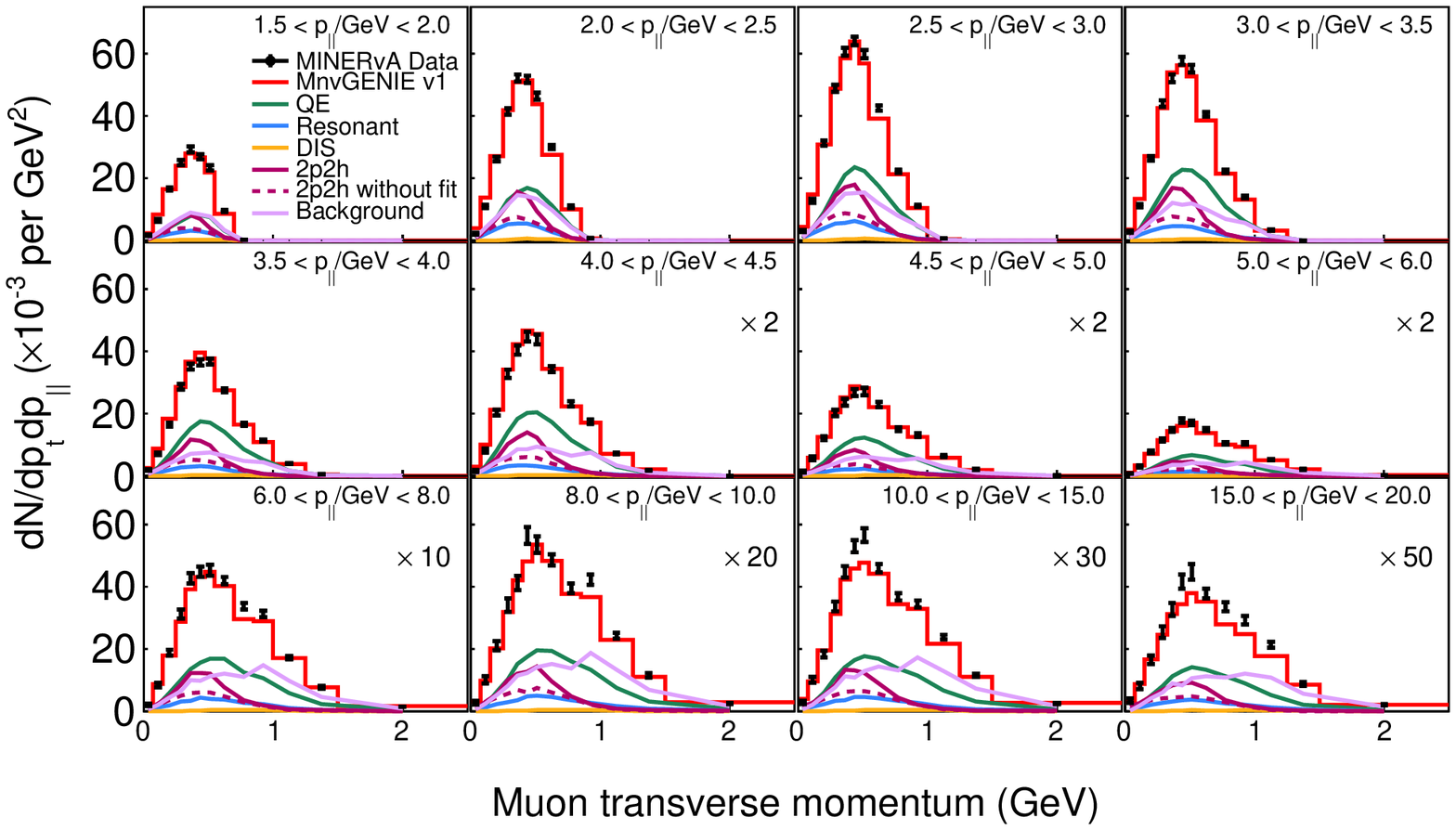} 
  \includegraphics[width=\textwidth]{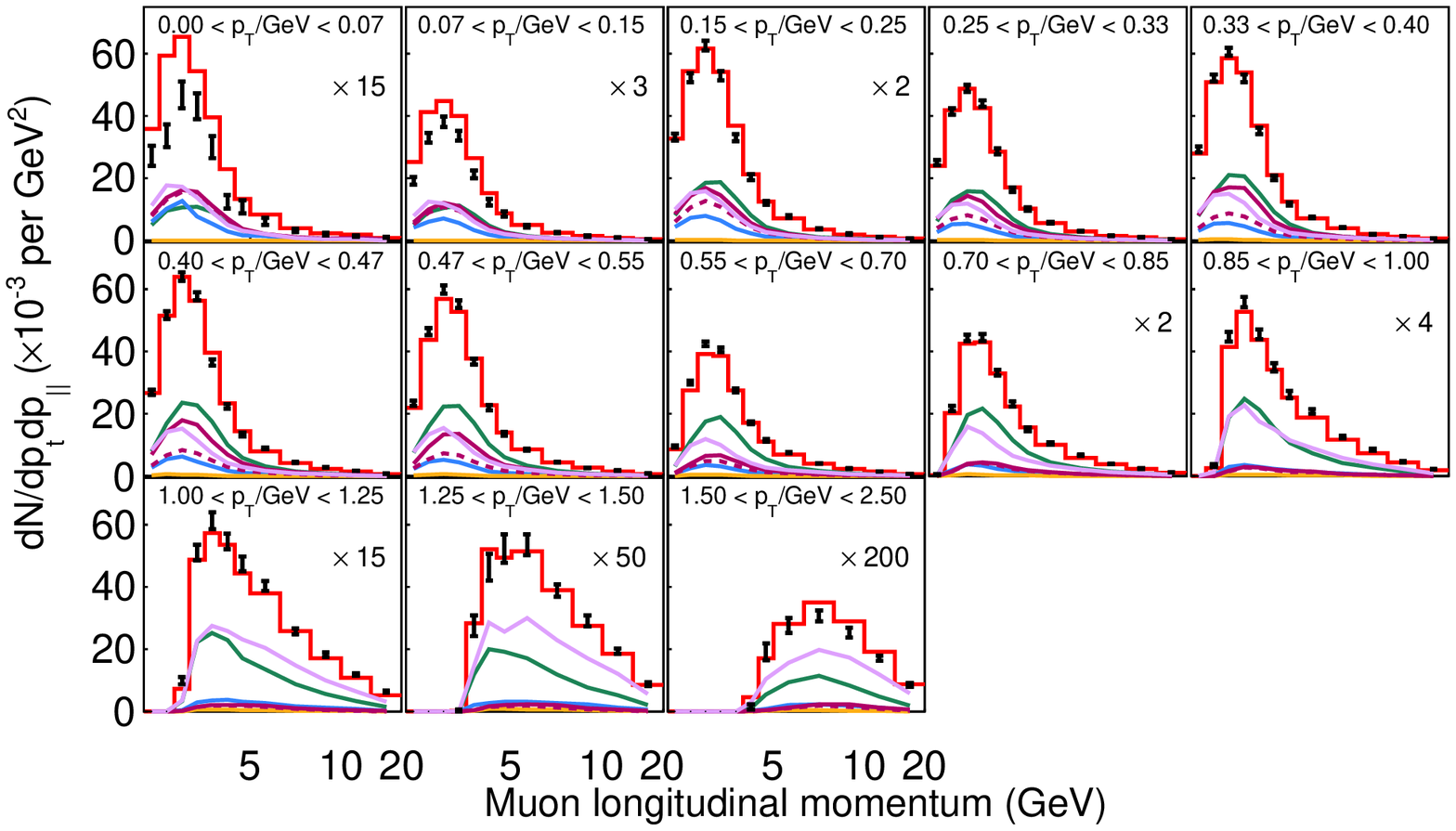} 
\caption{Selected event rates with constrained backgrounds. Model curves are \tune~predictions. Solid lines add up to \tune, while dotted lines are model component modifications. Signal events are classified as quasielastic (QE), resonant with the pion absorbed in FSI, 2p2h with an enhancement (solid line) and without (dotted), deep inelastic scattering (DIS) are events defined by GENIE as DIS.}
\label{fig:evsel_evtbymodelptpz}
\end{figure*}

After all event selections are applied the analysis has two populations of events: the muon only sample and the muon plus one or more proton track sample. The resulting event rates for each population as well as the combined sample as shown in Fig. \ref{fig:evsel_evtbymultptpz} for longitudinal momentum (\pz) and transverse momentum(\pt) versions of the two-dimensional results. In addition, a breakdown of the predicted event rate by type is presented in Fig. \ref{fig:evsel_evtbymodelptpz}.  The event samples after all cuts but before background subtraction are 62776 and 46499 events for the single  and multi- track sample, respectively.  The predicted purity, defined as the number of signal events per selected events, of these samples are 79.9\% and 56.0\%, respectively.

\begin{figure}[tp]
\centering
  \includegraphics[width=0.9\linewidth]{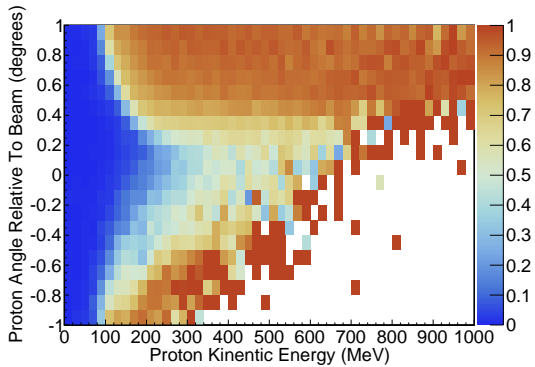} 
\caption{Reconstruction efficiency of protons as a function of proton angle with respect to the beam axis and the leading proton kinetic energy.}
\label{fig:evsel_evtprotonreco}
\end{figure}

The reconstructed proton efficiency as a function of the highest energy proton and angle with respect to the z-axis of the detector is shown in Fig. \ref{fig:evsel_evtprotonreco}.
This plot demonstrates how often \qelike~candidates end up in the multi-track events versus the single track event category.

%% file: Extraction.tex
\section{Cross section Extraction}
\label{sec:extraction}

To extract cross sections, the background must first be determined and subtracted.  In order to reduce model dependence the backgrounds are constrained using independent data samples.  After background subtraction the remaining event samples are unfolded to remove detector resolution effects.  After that the samples are corrected for detector efficiency.  Finally, the event samples are divided by the flux and number of targets to arrive at cross sections.  The following subsections describe each of those steps.  

\subsection{Background Constraint}
\label{subsec:bkgconstraint}
\begin{figure}[tp]
\centering
  \includegraphics[width=0.95\linewidth]{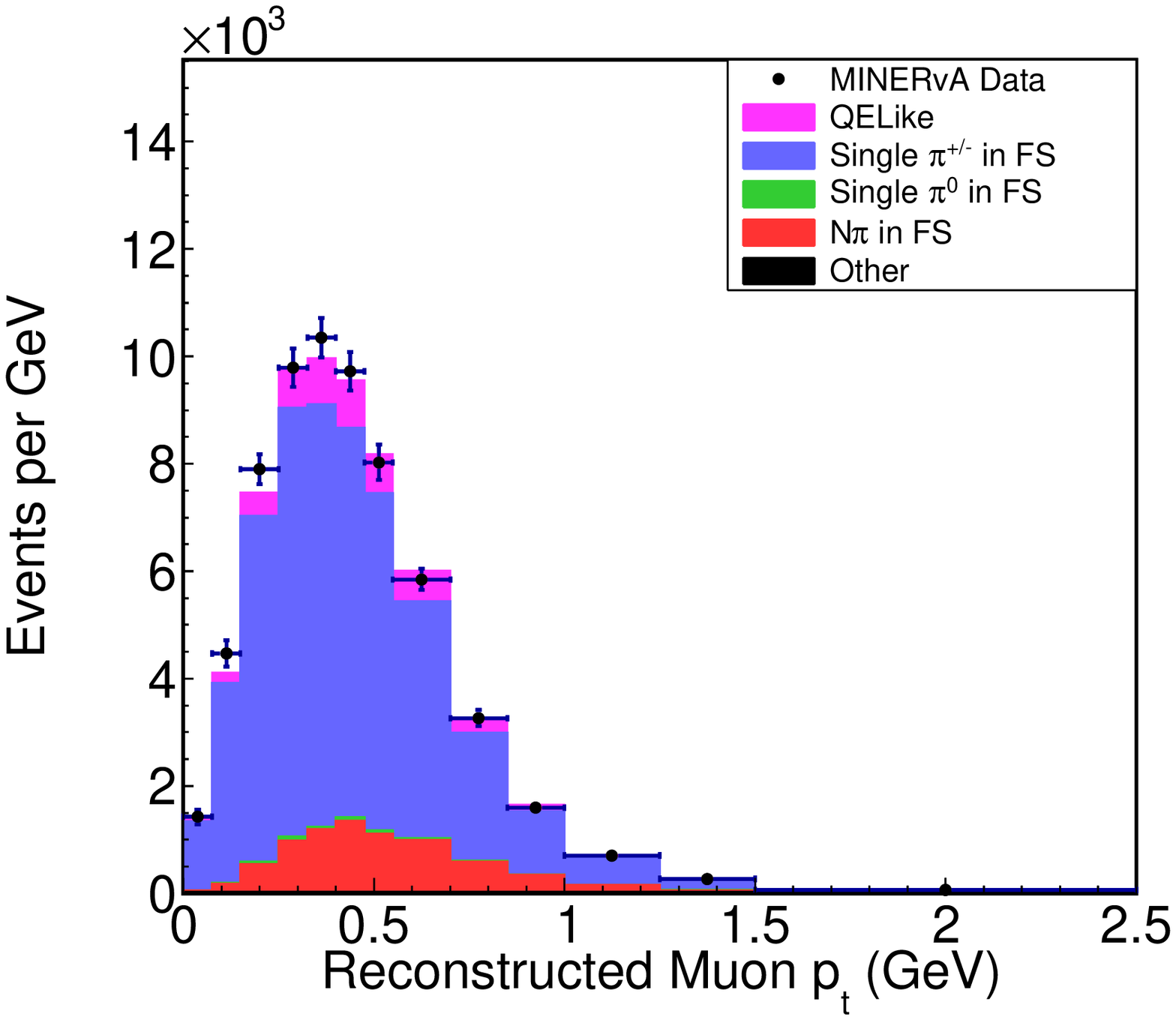} 
  \includegraphics[width=0.95\linewidth]{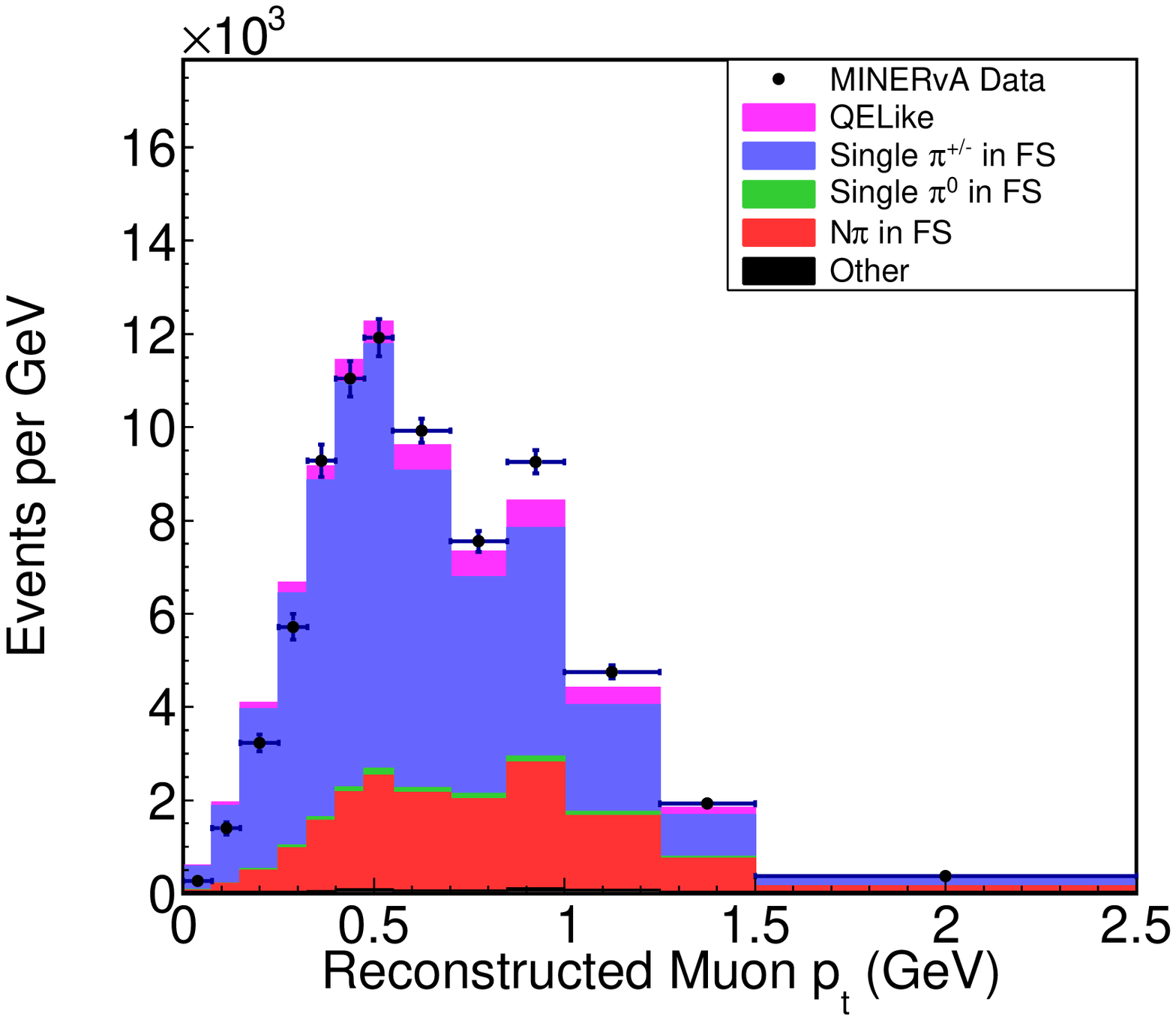} 
\caption{Control sample containing a reconstructed Michel electron with only the muon reconstructed (top), and with muon and proton candidates reconstructed (bottom).}
\label{fig:bkgMich}
\end{figure}

\begin{figure}[tp]
\centering
  \includegraphics[width=0.95\linewidth]{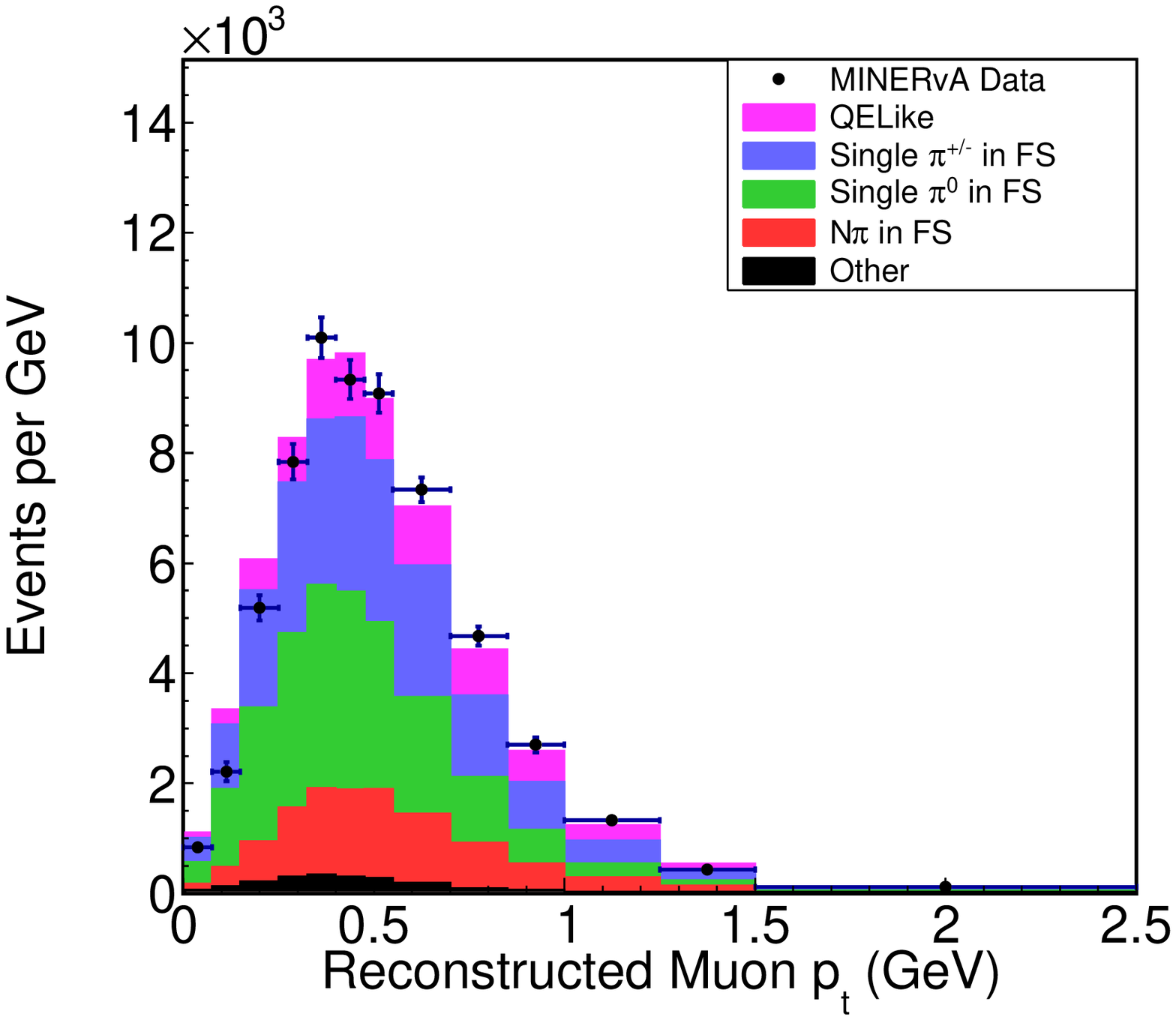} 
  \includegraphics[width=0.95\linewidth]{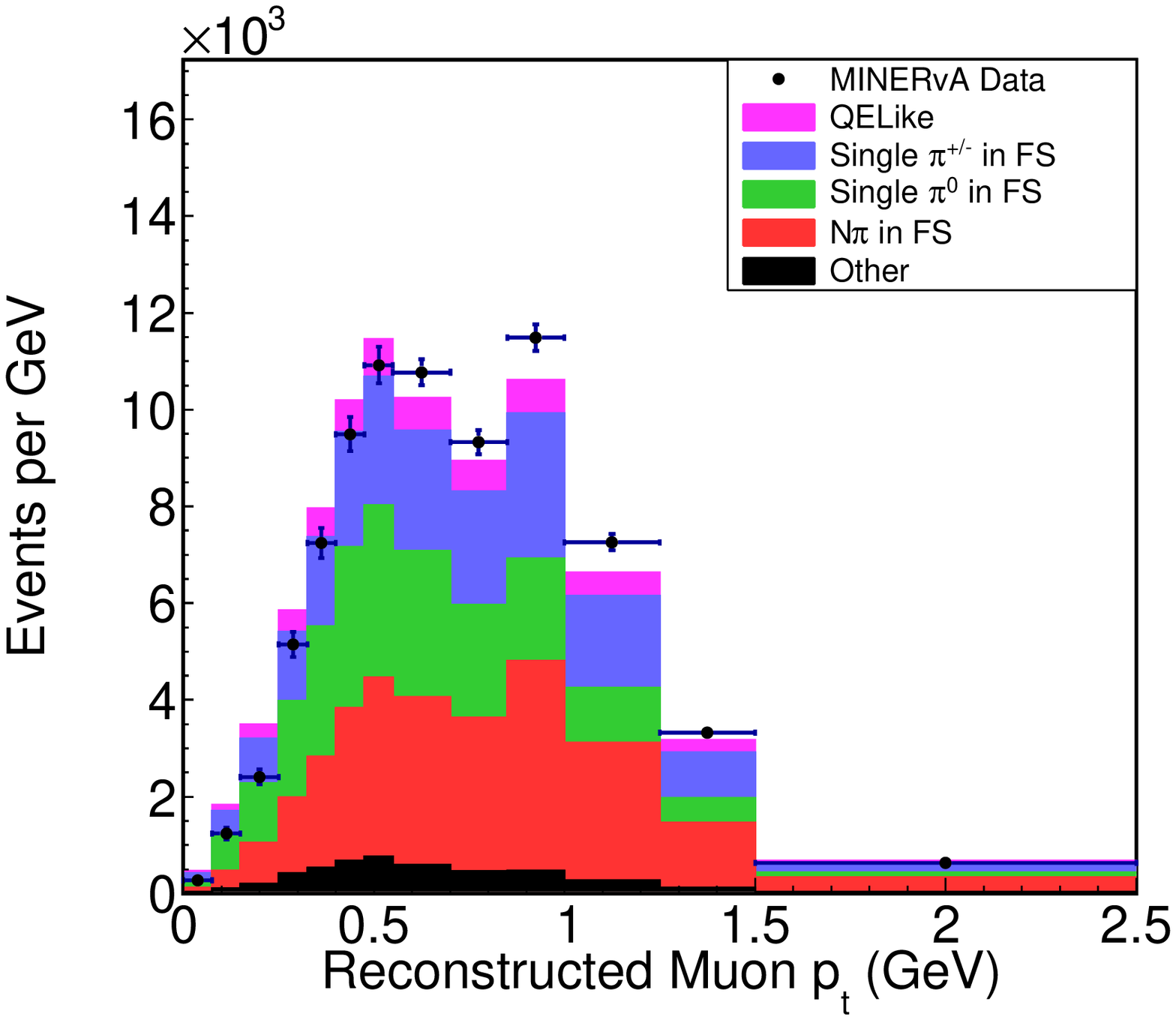} 
\caption{Control sample containing more than one shower candidate with only the muon reconstructed (top), and with muon and proton candidates reconstructed (bottom).}
\label{fig:bkgBlob}
\end{figure}

\begin{figure}[tp]
\centering
  \includegraphics[width=0.95\linewidth]{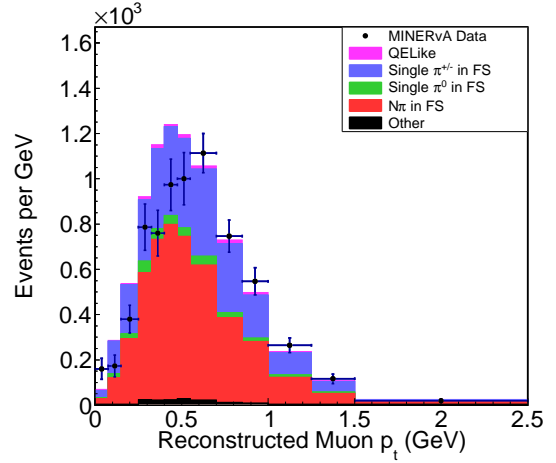} 
  \includegraphics[width=0.95\linewidth]{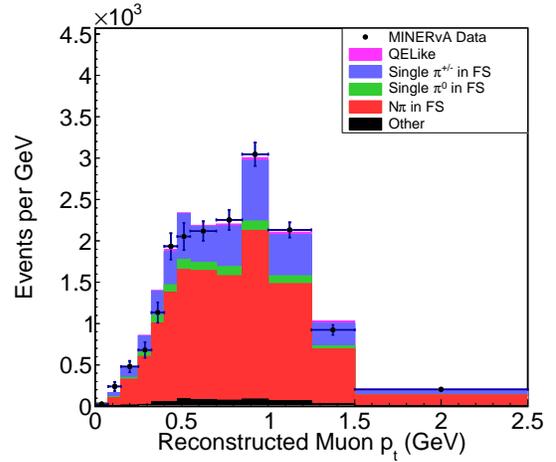} 
\caption{Control sample containing a reconstructed Michel electron and more than one shower candidate with only the muon reconstructed (top), and with muon and proton candidates reconstructed (bottom).}
\label{fig:bkgMichBlob}
\end{figure}

This analysis uses three independent control samples for the pion backgrounds: a Michel only sample, a shower only ($>$ 1 shower) sample, and a shower plus Michel sample. These are designed to characterize the single $\pi^+$ backgrounds (Michel), the neutral pion backgrounds (showers), and the multipion backgrounds (Michel plus showers). Figures \ref{fig:bkgMich}, \ref{fig:bkgBlob} and \ref{fig:bkgMichBlob} show the input distributions for the sideband constraint.

\begin{figure}[tp]
\centering
  \includegraphics[width=0.95\linewidth]{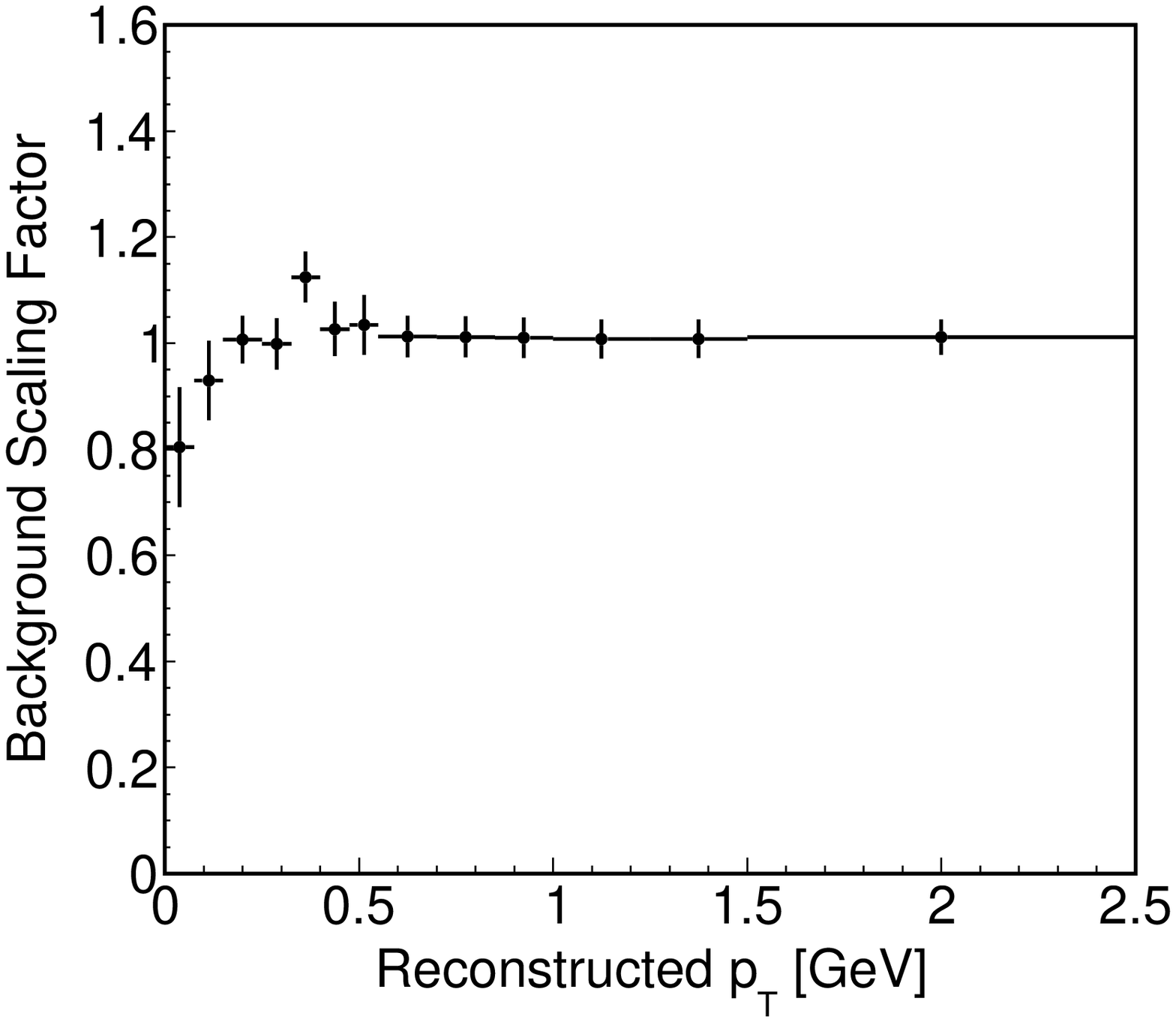} 
  \includegraphics[width=0.95\linewidth]{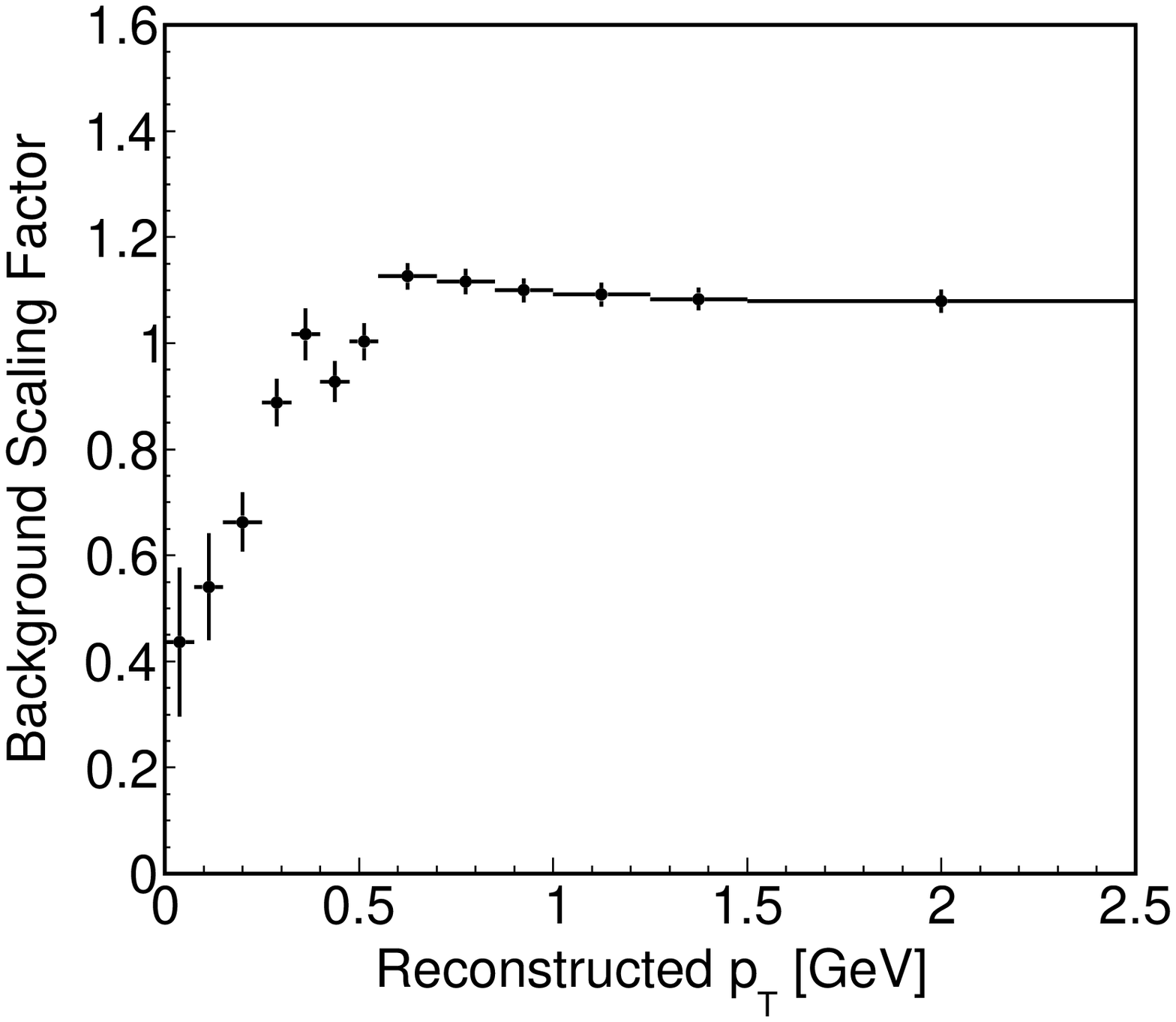} 
\caption{Combined scaling factor for $\pi^{\pm}$,$\pi^0$, and $N\pi$ backgrounds. Top is the sample with only the muon reconstructed. The bottom is the sample with the muon and proton candidates reconstructed.}
\label{fig:bkgscaling}
\end{figure}

All three samples are fit in bins of transverse muon momentum with binning compatible with the signal sample, with reduced numbers of bins in cases of small sample size. They are fit simultaneously with a standard $\chi^{2}$ minimization using ROOT’s TMinuit class\citep{BRUN199781}. Three scaling factors are extracted as a function of transverse momentum for the single track sample as well as the multi-track sample. These are then applied to the signal sample and the resulting overall scaling factor is derived and is shown in Fig. \ref{fig:bkgscaling}.

The sideband corrections reduce the pion content at moderate to low transverse momentum. This is consistent with previous \minerva~results which indicate a need for an overall reduction in the pion rate as well as preference for even lower rates at low \qqqe.

\subsection{Unfolding}
The D'Agostini unfolding method\citep{D'Agostini:1994zf}\citep{DAgostini:2010xxxxx}, via the implementation in RooUnfold\cite{Adye:2011gm} is used to remove detector resolution effects from the sample. The number of iterations is chosen based on pseudodata studies which take warped input pseudodata and unfold with a different sample's migration matrix. A wide variety of pseudodata samples were used to investigate the number of iterations necessary to unfold to the true pseudodata distribution. The study used model variations ranging from a  GENIE 2.8.4 prediction without 2p2h or other modifications to using \tune~with an additional modification of resonant pion RPA introduced via a MINOS empirical model\citep{Adamson:2014pgc}. This was done by a $\chi^2$ comparison between the unfolded pseudodata and the sample used to generate the pseudodata. In all cases the preferred number of iterations was no more than four, and if the pseudodata and migration model were similar even fewer iterations were required. This result uses four iterations.

During the study it was discovered that certain pseudodata samples would result in large $\chi^2$ at higher numbers of iterations. This was due to fluctuations of particular \pt-\pz~bins with low statistics. At four iterations the probability of the sample producing a large $\chi^2$ was 1 in 1000. We also analyzed the real data sample, using all model variations listed in Tab. \ref{tab:DDModelComp}, to more than a hundred iterations and did not see a $\chi^2$ increase. The combination of the very low probability and the performance with the real data gives us confidence the procedure is robust.

\begin{figure}[tp]
\centering
  \includegraphics[width=0.95\linewidth]{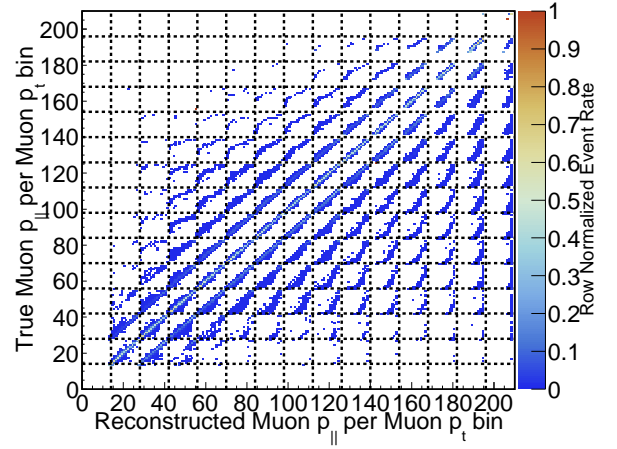} 
\caption{The mapped \pt-\pz~migration matrix. The large blocks correspond to increasing \pt. The subblocks are the \pz~bins.}
\label{fig:mig_full}
\end{figure}

\begin{figure}[tp]
\centering
  \includegraphics[width=0.95\linewidth]{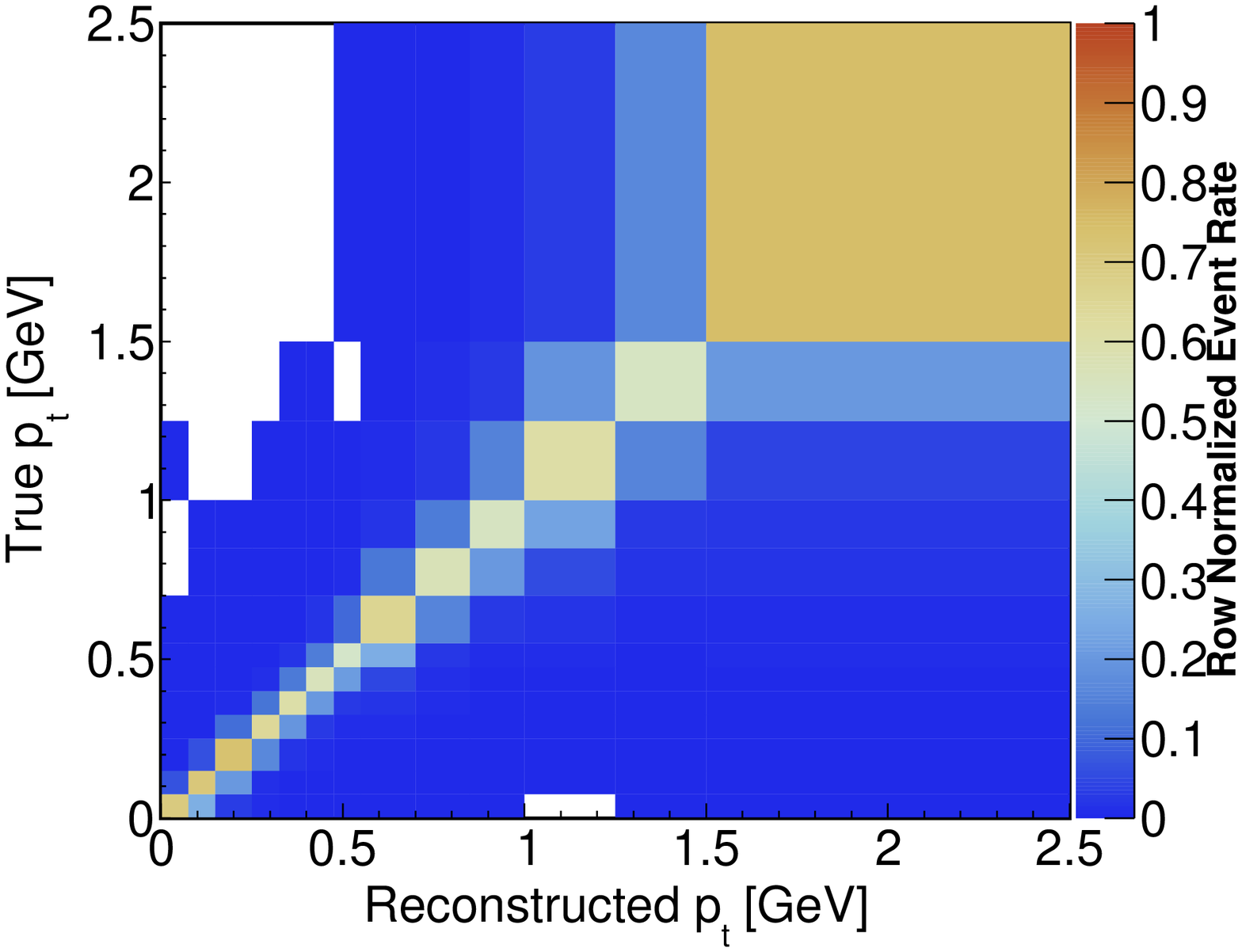} 
\caption{The projection of the mapped \pt-\pz~migration onto \pt.}
\label{fig:mig_pt}
\end{figure}

\begin{figure}[tp]
\centering
  \includegraphics[width=0.95\linewidth]{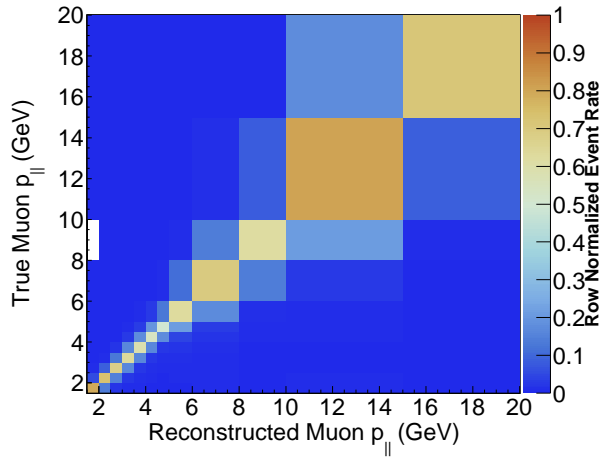} 
\caption{The projection of the mapped \pt-\pz~migration onto \pz.}
\label{fig:mig_pz}
\end{figure}

\begin{figure}[tp]
\centering
  \includegraphics[width=0.95\linewidth]{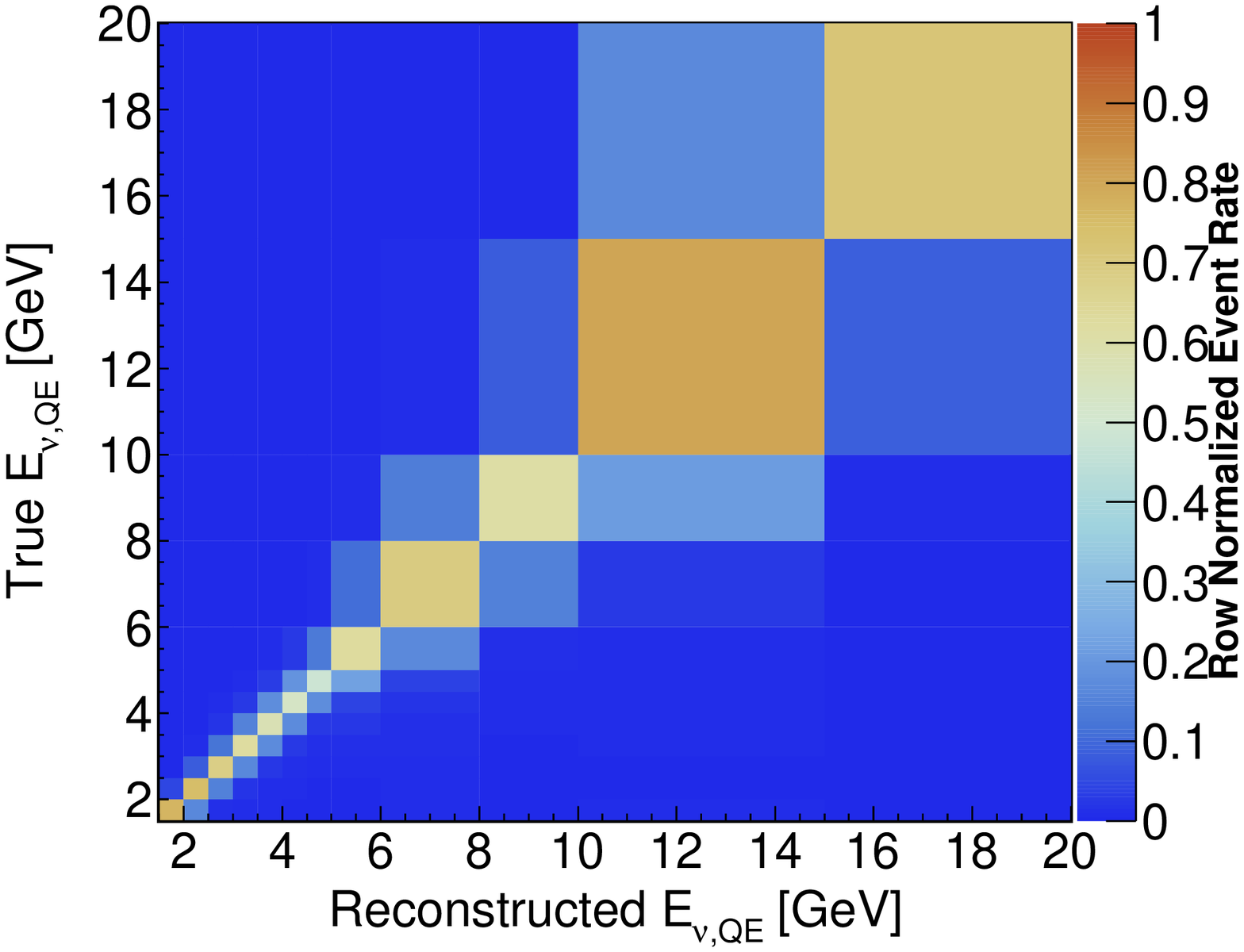} 
\caption{The migration matrix of \enuqe.}
\label{fig:mig_enuqe}
\end{figure}

\begin{figure}[tp]
\centering
  \includegraphics[width=0.95\linewidth]{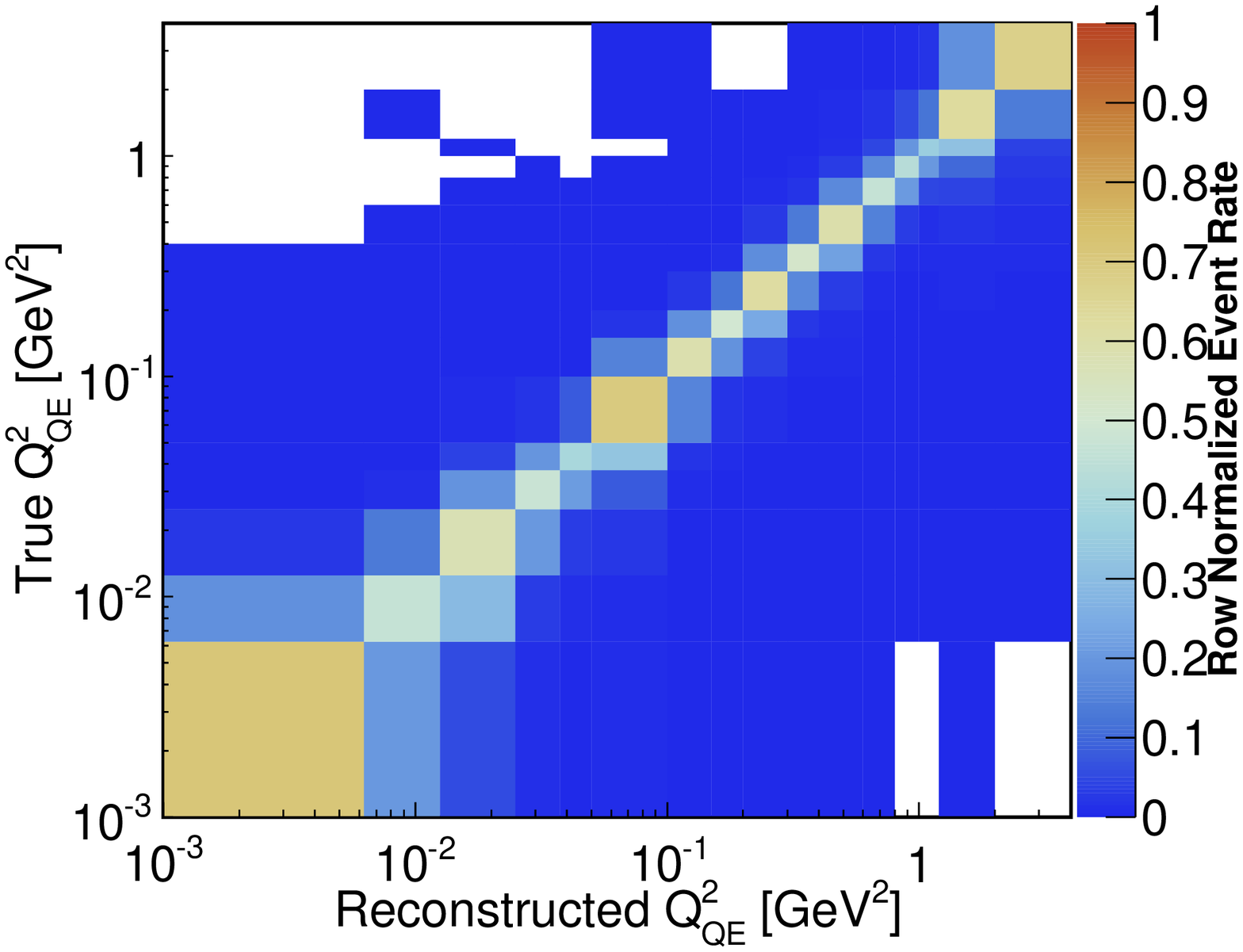} 
\caption{The migration matrix of \qqqe.}
\label{fig:mig_qqqe}
\end{figure}

The migration matrices for the \pt-\pz~result and their projections are in Figs. \ref{fig:mig_full}, \ref{fig:mig_pt}, \ref{fig:mig_pz}, \ref{fig:mig_enuqe}, and \ref{fig:mig_qqqe}. In general the populations are mostly diagonal.

\subsection{Efficiency Correction}

\begin{figure}[tp]
\centering
  \includegraphics[width=0.95\linewidth]{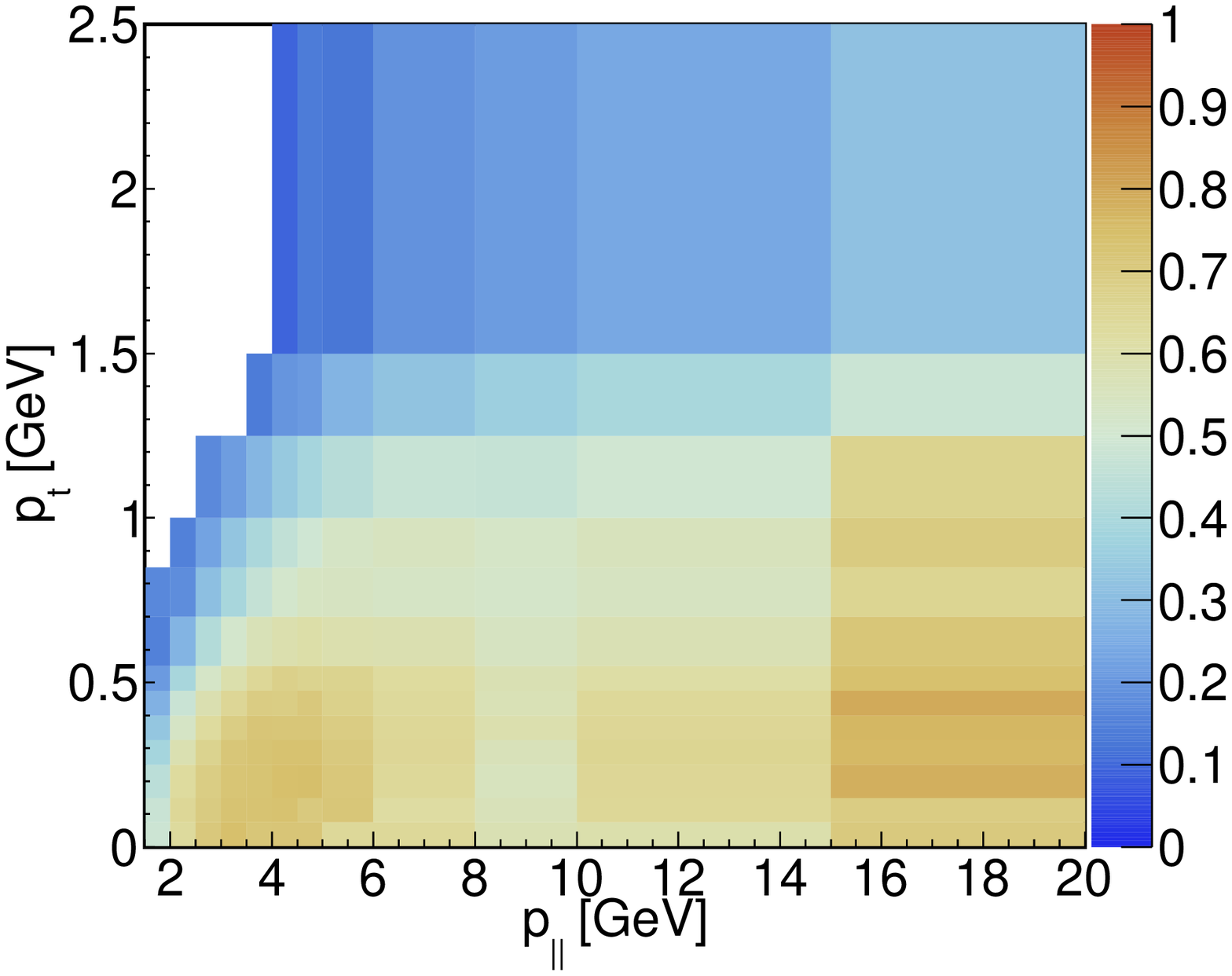} 
\caption{Selection efficiency for \qelike~events.}
\label{fig:eff_ptpz}
\end{figure}

The efficiency correction in the \pt-\pz~space is shown in Fig. \ref{fig:eff_ptpz}. The zero efficiency at lower \pz~for higher \pt~bins is the effect of the muon angle requirement in the signal definition. Bins where this requirement still accepts a very small component of signal events near bins edges are not reported in the final cross section result due to extremely small efficiencies and event rates. The efficiency in these boundary regions were kept as smooth as resolutions would allow. 

\begin{figure}[tp]
\centering
  \includegraphics[width=\columnwidth]{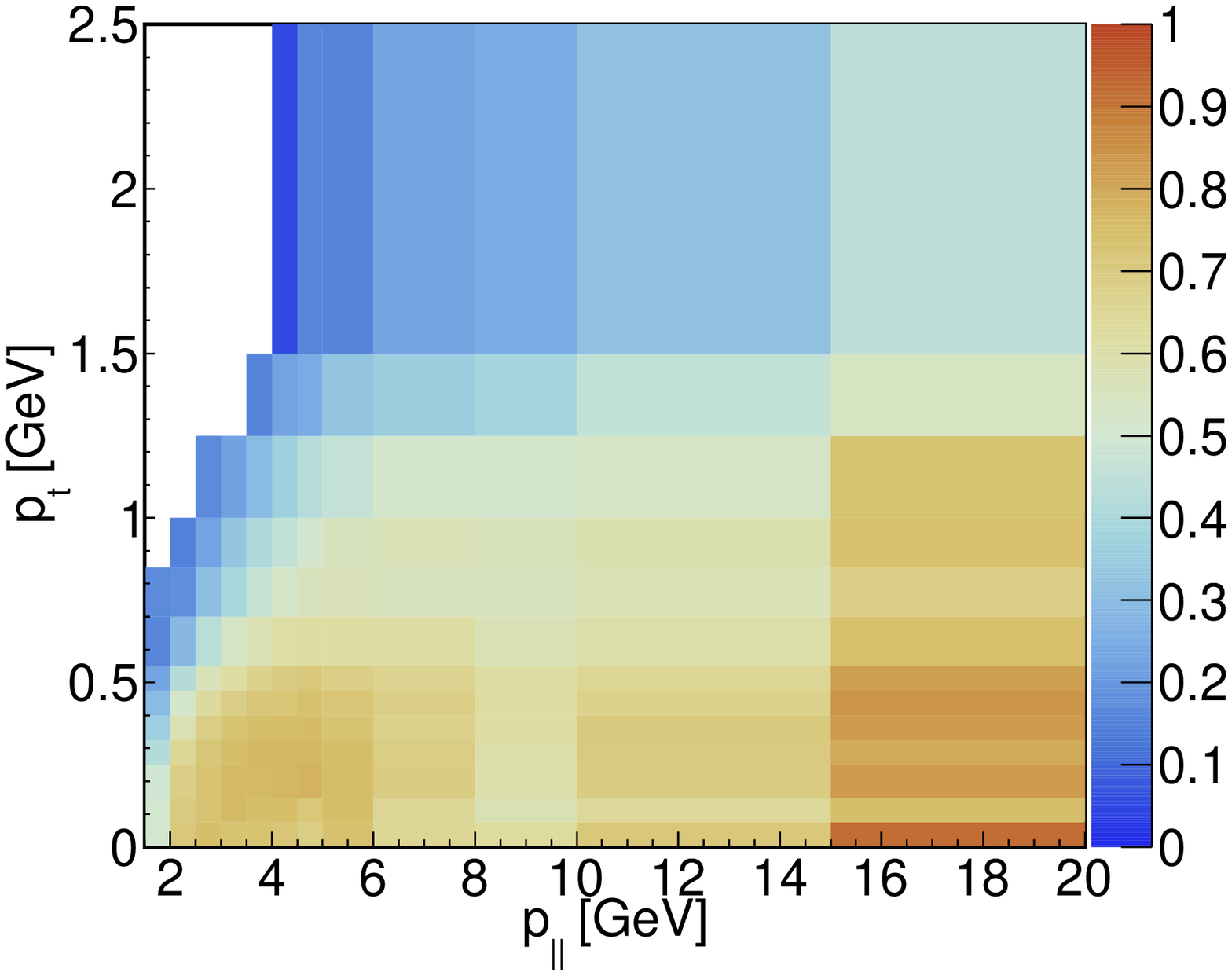} 
  \includegraphics[width=\columnwidth]{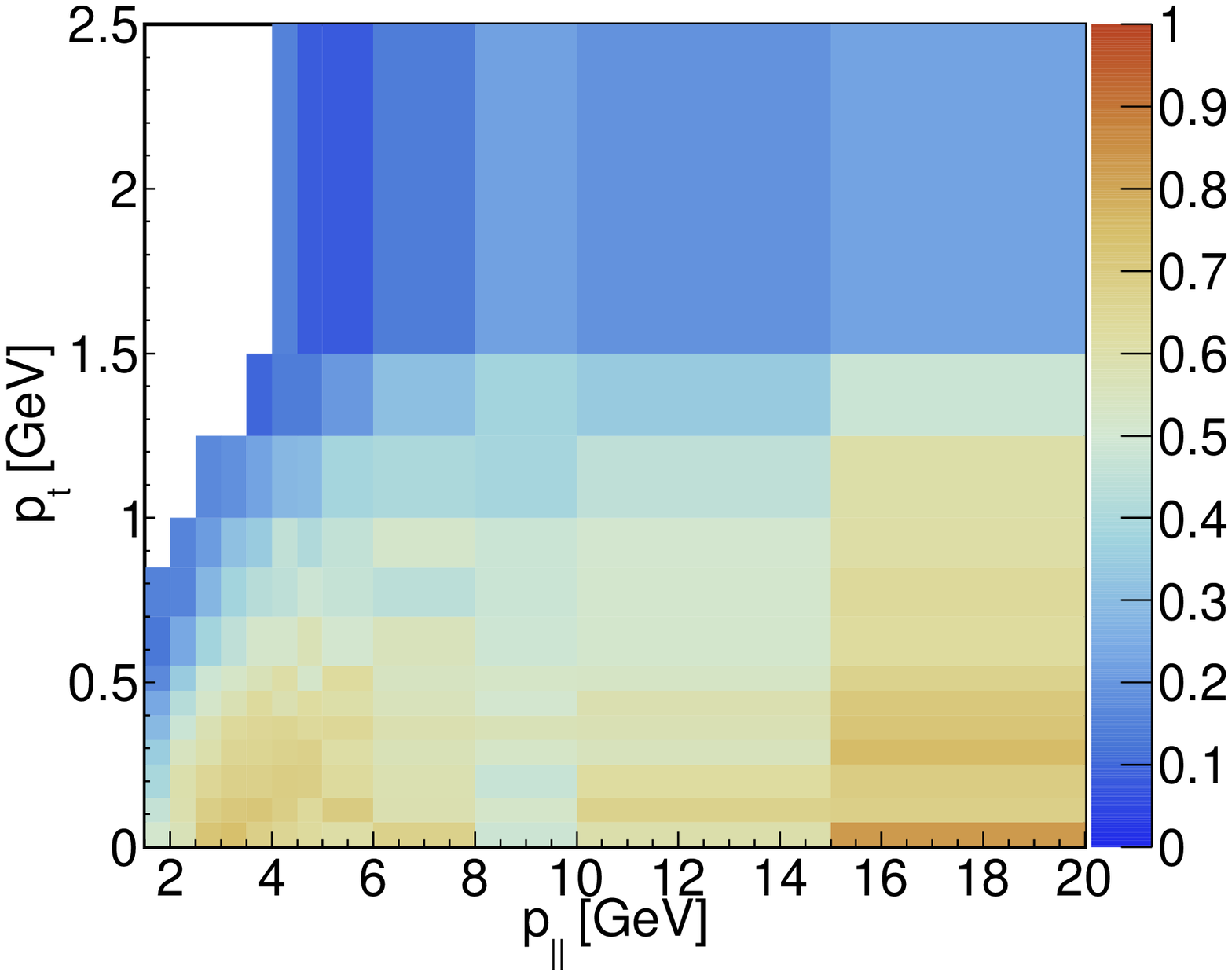} 
  \includegraphics[width=\columnwidth]{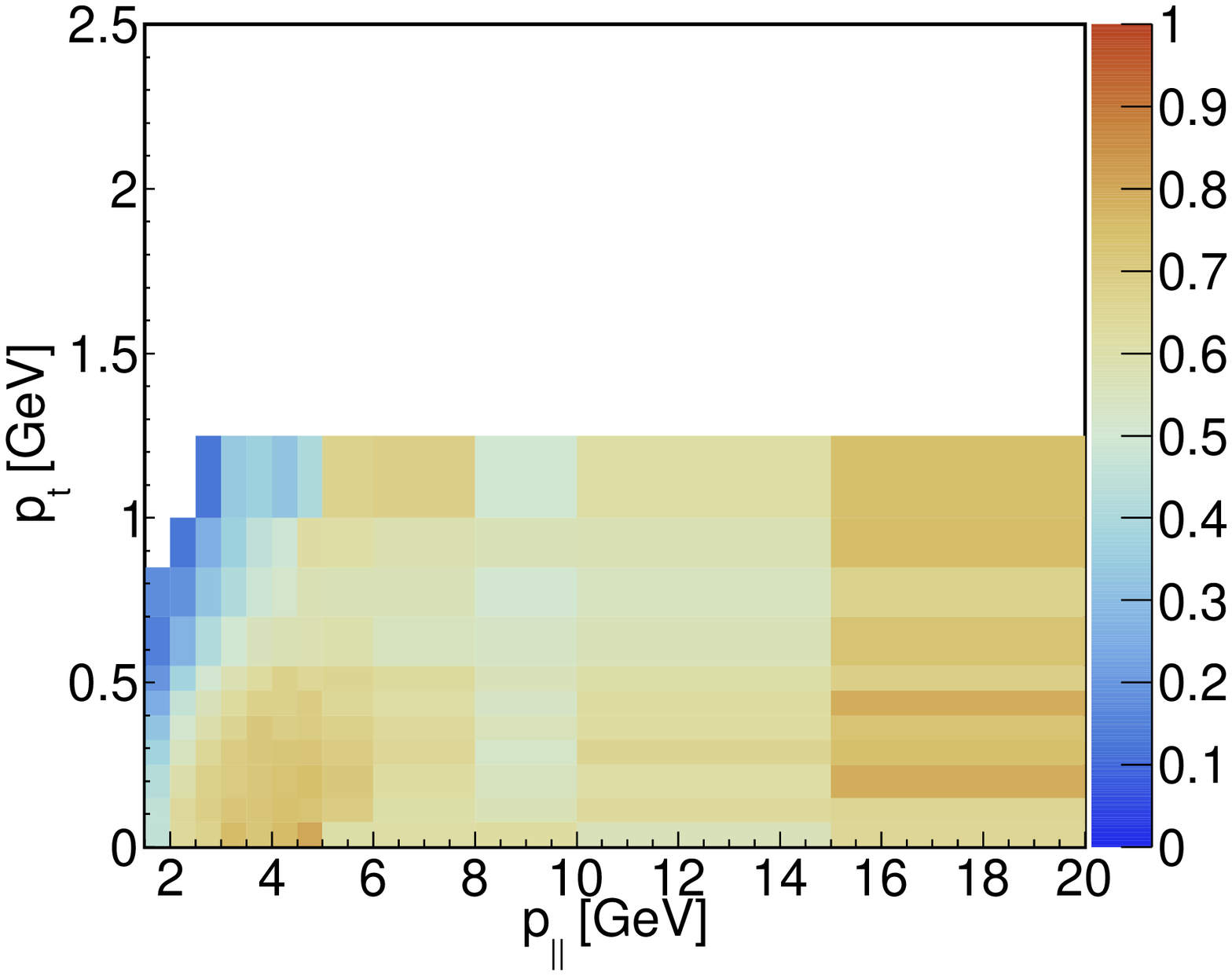} 
\caption{Selection efficiency for \qelike~events broken down by the GENIE component, QE(top), Resonant(middle), 2p2h(bottom).}
\label{fig:eff_qelike_comps}
\end{figure}

In addition to the overall efficiency of the signal, the generator level (QE, 2p2h, resonant) components of the signal efficiency are also presented in Fig. \ref{fig:eff_qelike_comps}, and show that the relative efficiencies for each predicted component are similar. The cutoff at \pt~of 1.2 GeV is due to the restriction of the 2p2h model having a cutoff at $q_{3}$ at 1.2 GeV.

\begin{figure}[tp]
\centering
  \includegraphics[width=0.95\linewidth]{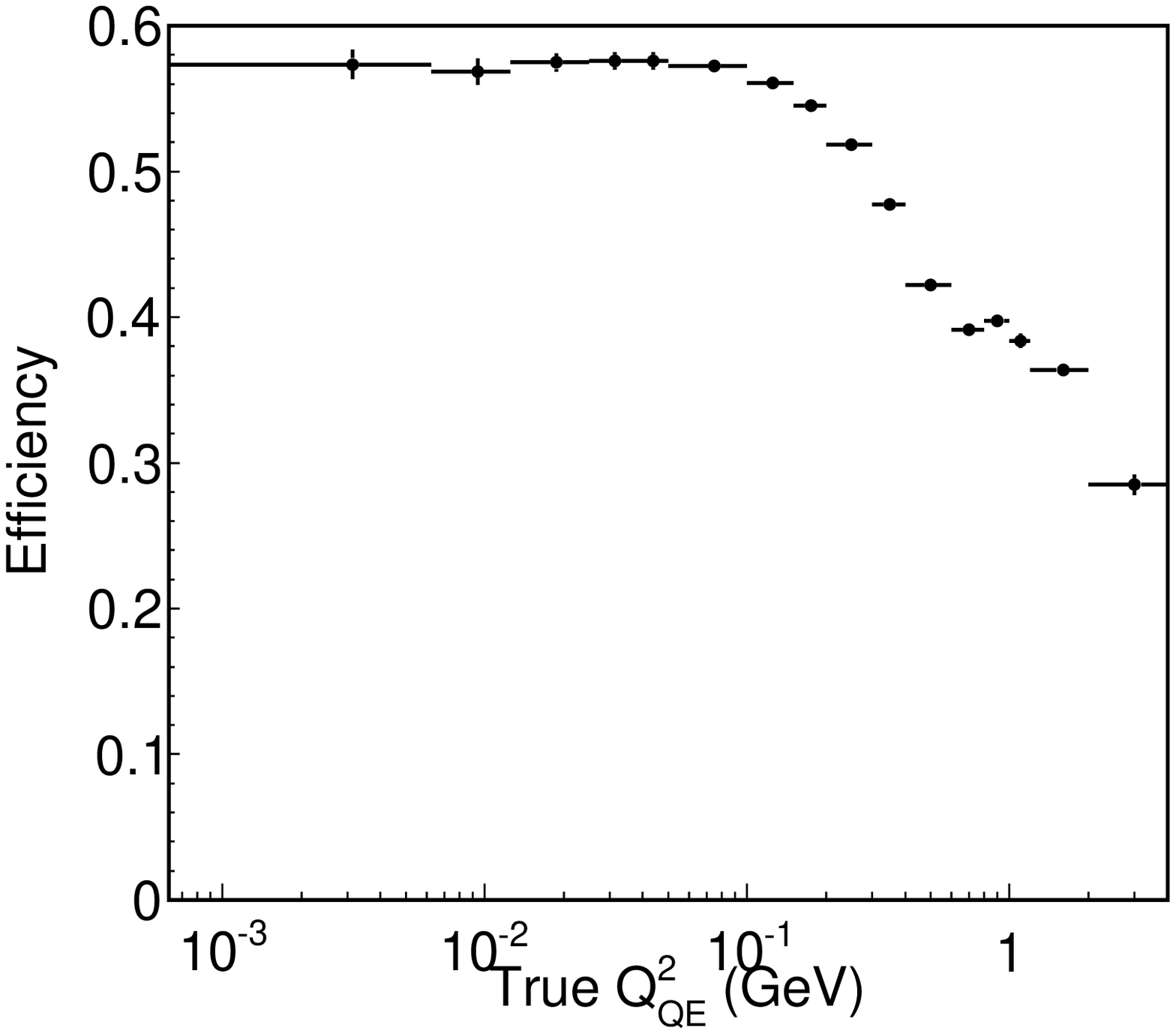} 
\caption{Selection efficiency for \qelike~events in \qqqe. The slight increase in efficiency at \qqqe~0.8 $GeV^2$ is due to the proton PID cut definition in Sec. \ref{subsec:pid}}
\label{fig:eff_q2}
\end{figure}

The \qqqe~efficiency is shown in Fig. \ref{fig:eff_q2}. The reduction at \qqqe~greater than 0.1 GeV$^{2}$ is due to the tracking thresholds and PID performance. The small increase of efficiency at 0.75 GeV$^{2}$ is due to the removal of the PID requirements, see Sec. \ref{subsec:pid}.

\begin{figure}[tp]
\centering
  \includegraphics[width=0.95\linewidth]{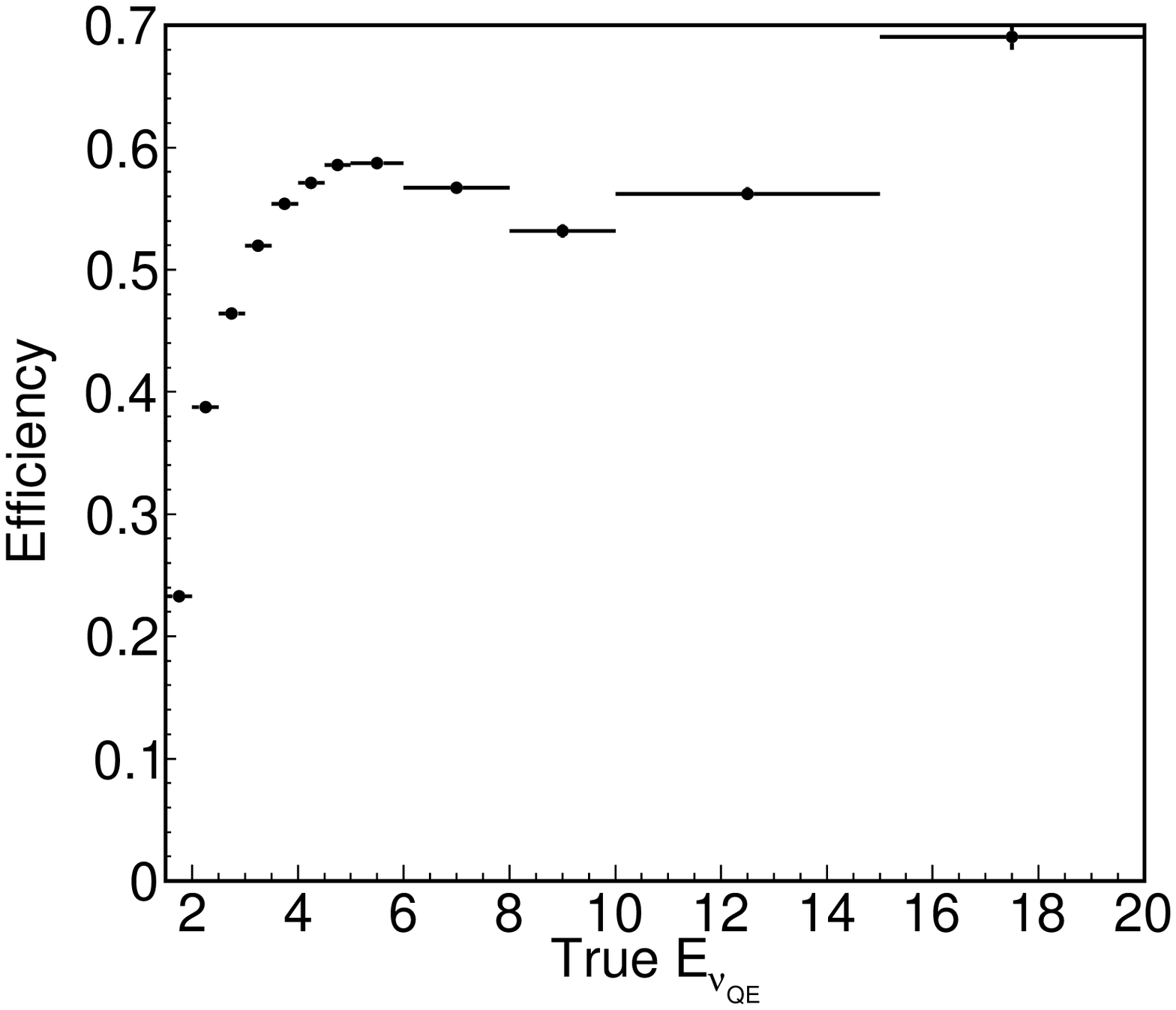} 
\caption{Selection efficiency for \qelike~events in \enuqe.}
\label{fig:eff_enuqe}
\end{figure}

The \enuqe~efficiency is shown in Fig. \ref{fig:eff_enuqe}. The lower efficiency at lower \enuqe~is due to the MINOS match requirement, but in general the sample is 50\% efficient or better. The point of inflection at 8 GeV occurs when the population of tracks move from the forward fine grained region of the MINOS detector to the less instrumented spectrometer downstream.

\subsection{Normalization}
\label{sec:normalization}
The analysis uses the flux described in Ref. \citep{Aliaga:2016oaz}. In addition, the each result is scaled by the number of nucleons in the fiducial volume, including hydrogen, with a mix of 88.51\% carbon, 8.18\% hydrogen, 2.5\% oxygen, 0.47\% titanium, 0.2\% chlorine, 0.07\% aluminum, and 0.07\% silicon. The number of neutrons in the region is $1.49\times10^{30}$. 

The \pt-\pz~and \qqqe~results are flux averaged results. The normalization factor is the result of the integral of the predicted flux from 0 to 120 GeV. This corresponds to $2.877\times10^{-8}$ cm$^{-2}$ per proton on target. For the \enuqe~result the normalization is done in the same manner as Ref. \citep{Patrick:2018gvi}. For each bin of \enuqe~the events are normalized by the integral, in true $E_\nu$, of the flux prediction with the same limits. This results in a cross section which is not a true total cross section, but is a well-defined quantity.

%% file: Systematics.tex
\section{Systematic Uncertainties}
\label{systematics}
\begin{figure*}[tp]
\centering
  \includegraphics[width=1\linewidth]{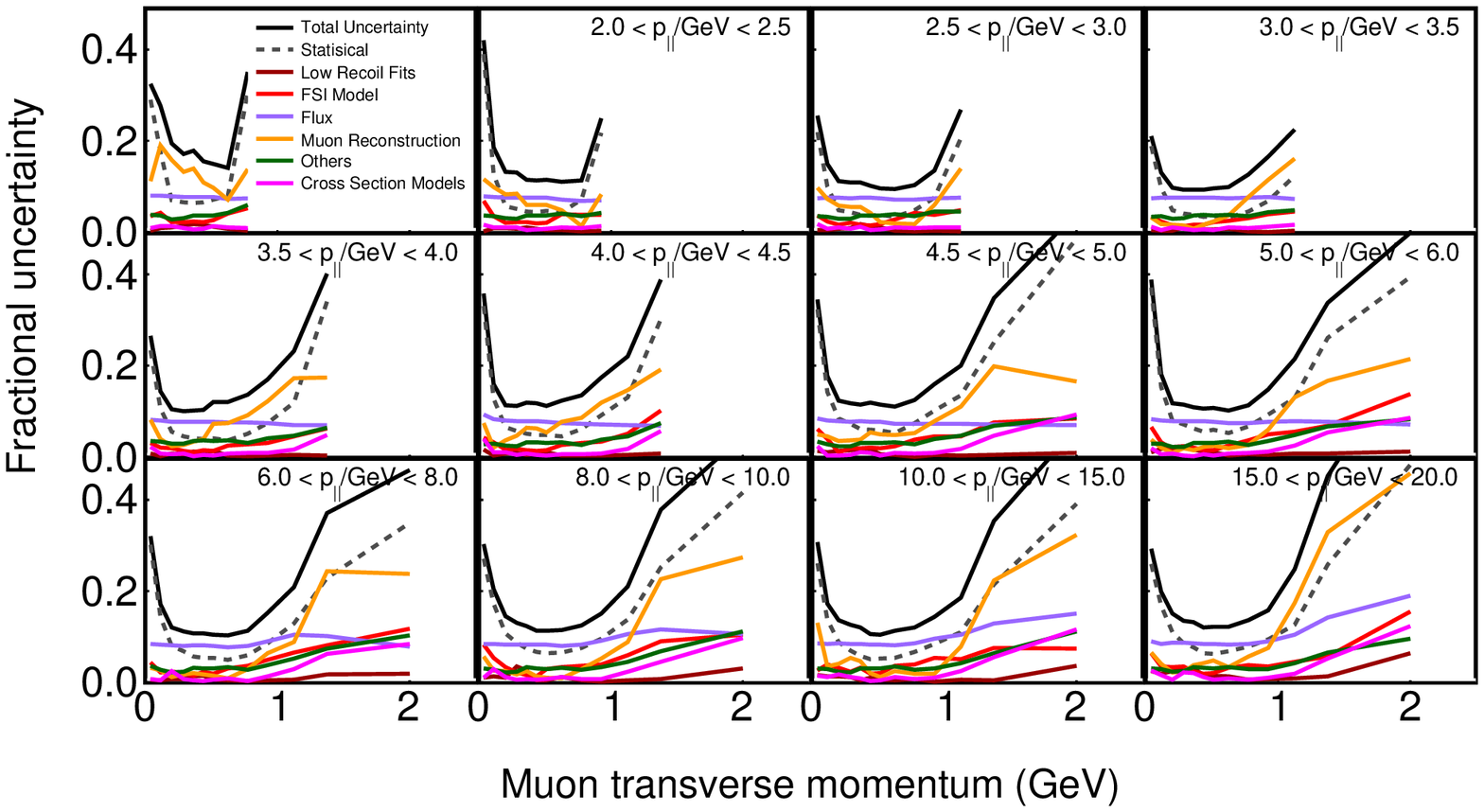} 
\caption{Systematic uncertainty breakdown for the \pt\pz~result.}
\label{fig:sys_ptpz}
\end{figure*}

\begin{figure}[tp]
\centering
  \includegraphics[width=1\linewidth]{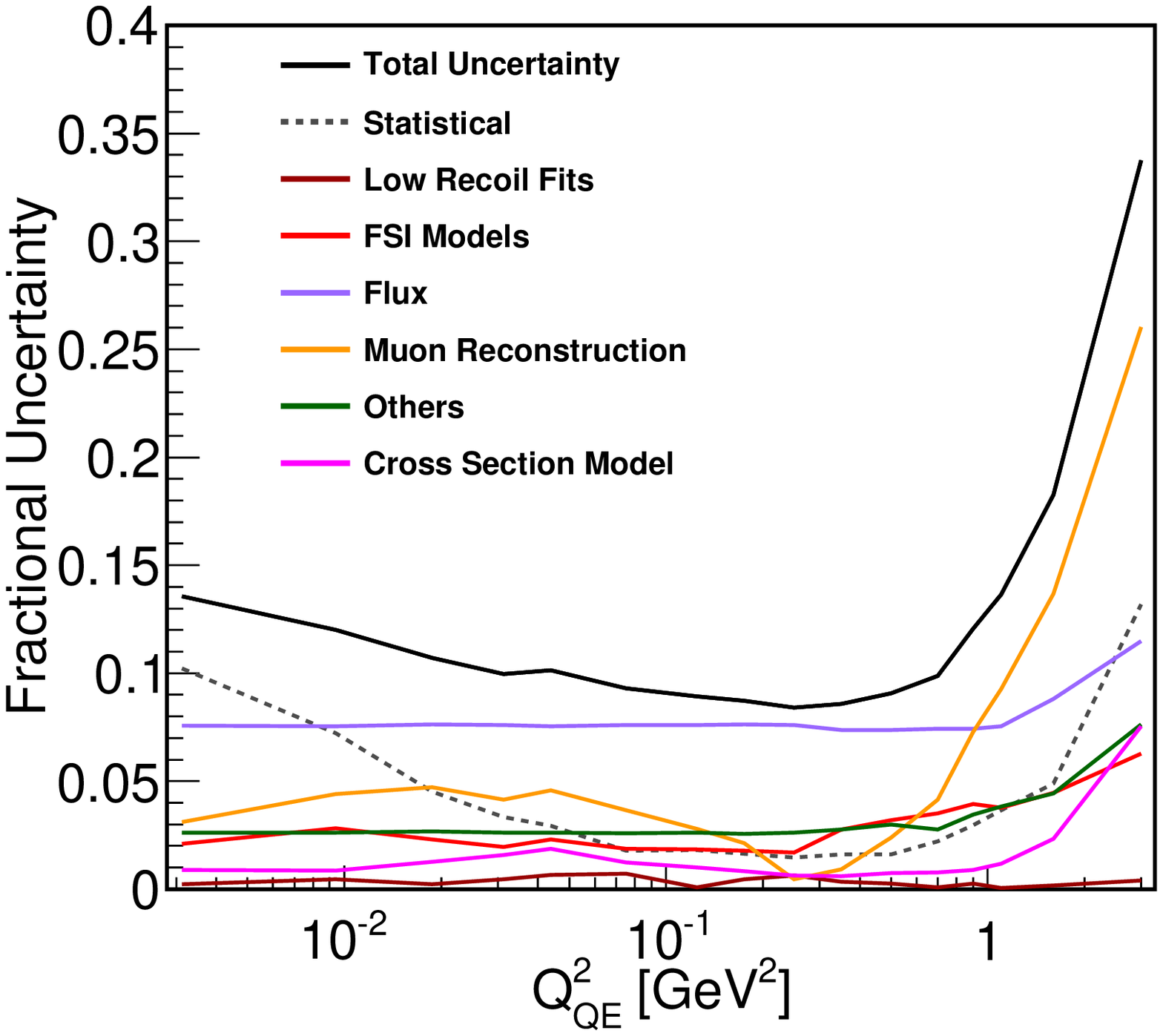} 
\caption{Systematic uncertainty breakdown for the $Q^{2}_{QE}$ result.}
\label{fig:sys_q2}
\end{figure}

\begin{figure}[tp]
\centering
  \includegraphics[width=1\linewidth]{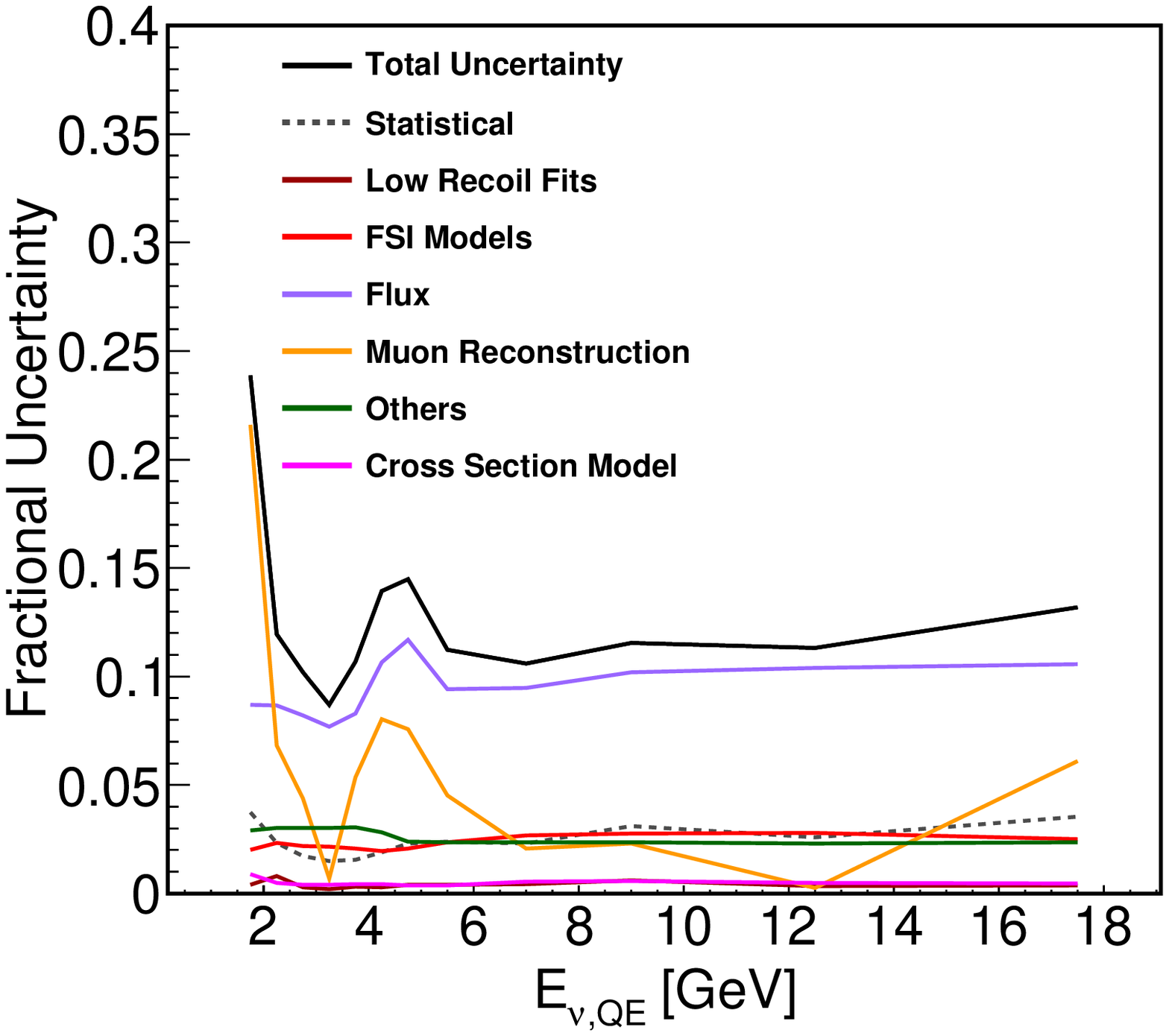} 
\caption{Systematic uncertainty breakdown for the \enuqe~result.}
\label{fig:sys_enuqe}
\end{figure}

The systematic uncertainties for the \pt-\pz, \qqqe, and \enuqe~are shown in Figs. \ref{fig:sys_ptpz}, \ref{fig:sys_q2}, and \ref{fig:sys_enuqe} respectively. 


Flux is typically the dominant uncertainty in most of the phase space, at the 8-9\% level for flux averaged results and up to 12\% in the falling edge of the beam focusing regions and higher neutrino energy regions for non-flux-integrated results. Another large systematic uncertainty is the muon reconstruction which dominates in particular regions of phase space due to the peaked nature of the muon momentum spectrum. 

The flux uncertainty is largest except at very large transverse momentum. The FSI model in particular plays a larger role in this region at the 8\% level. This is also a region where backgrounds are larger than signal and the sideband constraint is limited by statistics.

To evaluate the uncertainty of the enhancement to 2p2h described in Sec.~\ref{sec:Simulation}, the application of the 2p2h enhancement is varied in three ways: the enhancement is applied only to the np nucleon pairs in the 2p2h model, the enhancement is applied only to the nn nucleon pairs, and the enhancement is applied instead to the 1p1h model.  Each of these variations consistently describes the data in Ref.~\citep{Rodrigues:2015hik}, but attributes it to a different hard scattering process.  The measurement of \qelike~scattering is largely immune to these variations because the primary effect of these variations is to modify the recoil spectrum in the detector.  However, the selection only rejects events with greater than 500 MeV of visible recoil energy, much larger than is typical for 1p1h or 2p2h interaction. 

The FSI model uncertainties are evaluated using the standard GENIE reweighting infrastructure. Due to the correlated nature of the parameters we do not use the uncertainty associated with inelastic pion interactions, which is correlated with the pion absorption uncertainty. The uncertainty associated with the mean free path of pions and inelastic interactions of nucleons starts to contribute at the several per cent level beyond \pt$>$1 GeV.

The interaction model uncertainties are evaluated using the standard GENIE reweighting infrastructure. No source of uncertainty dominates for most of the phase space. The expection is non-resonant backgrounds where an interaction on a proton results in two pions, and two high-twist modifications in the Bodek-Yang model (A,B) which are individually greater than three percent at \pt$>$1.5 GeV.

The systematic uncertainty resulting from the application of RPA to the CCQE events is broken into two kinematic regions as described in the Ref. \citep{Gran:2017psn} analysis of the model from Ref. \citep{Nieves:2004wx}. Variations of uncertainties on the input parameters to the relativistic model are used as one source of uncertainty, though the non-relativistic version of the model is used technically to implement this.  Above 0.4 GeV$^2$, the uncertainty is 60\% of the difference between the two on the high side, and no modification on the low side.    For Q$^2 \leq 0.4$ GeV$^2$, the uncertainty is set by the comparison of the model with muon capture data on multiple nuclei \cite{Valverde:2006zn, Nieves:2017lij}.   This is set at 25\% change in the strength of suppression due to the poor description of muon capture on carbon.

The other category includes a large number of systematics including reconstruction efficiencies, calorimetric response to single particles constrained by Ref. \citep{Aliaga:2015aqe}, inelastic interaction cross sections for pions, protons, neutrons in the detector material, and detector modeling. A subset of these--proton reconstruction efficiency, particle identification by $\frac{dE}{dx}$ in the scintillator, and \minerva~-MINOS matching efficiencies--contribute significantly at large \pt.

%% file: Results.tex
\section{Results}
\label{sec:results}
The analysis provides five different cross section results: a double differential cross section in \pt-\pz~with comparisons to \tune~and additional modifications to the model as well as a comparison to NuWro\citep{PhysRevC.86.015505}, and four single differential results. In addition, the double differential cross section is projected into \qqqe~which is highly correlated with \pt~and \enuqe~which is correlated with \pz. An analysis measuring the reconstructed energy near the vertex is presented. Furthermore, an extraction of the double differential cross section with a modified signal definition can be found in Sec. \ref{sec:appccqe}.
   
\subsection{Double Differential Cross Section}
\begin{figure*}[p]
\centering
  \includegraphics[width=0.95\linewidth]{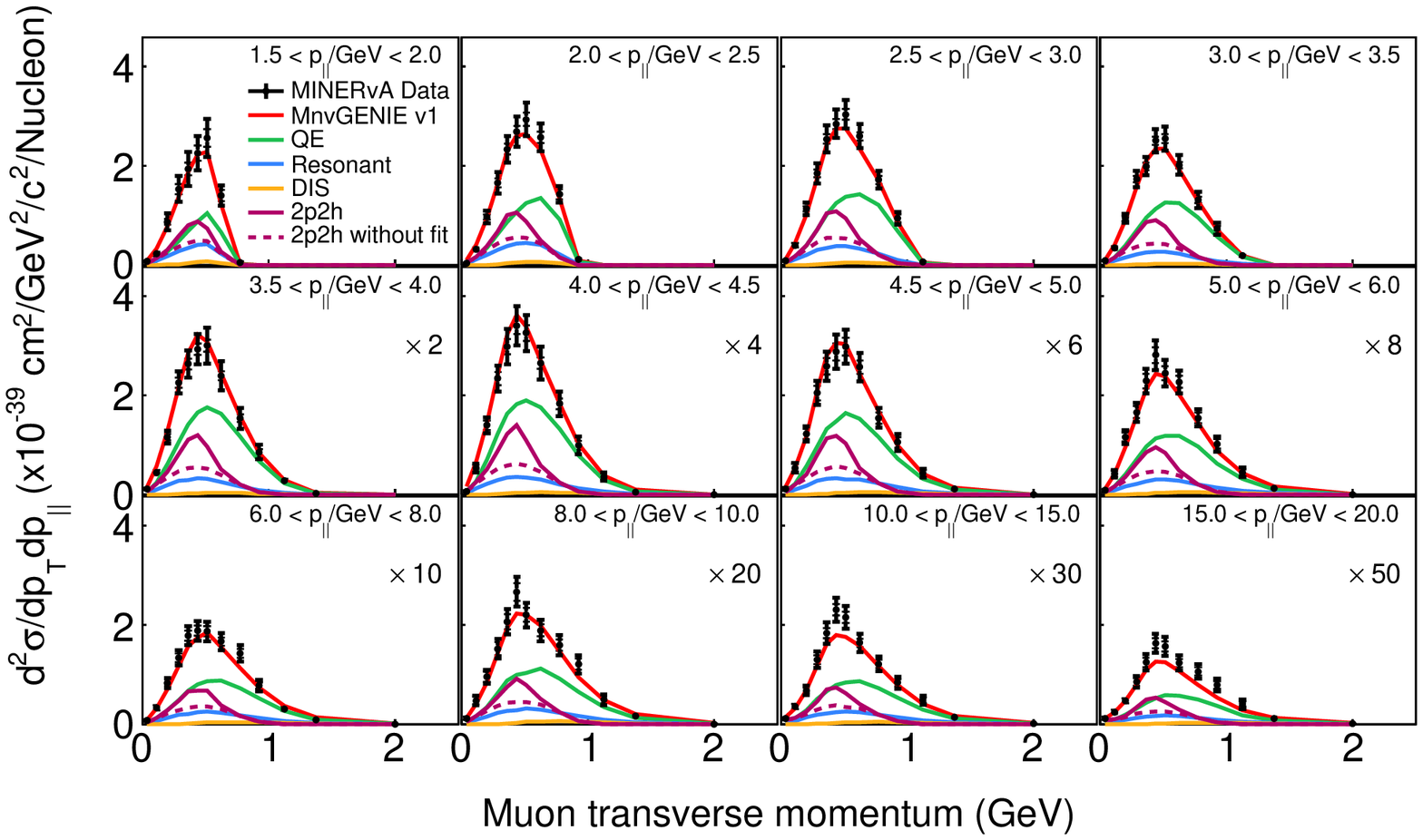} 
  \includegraphics[width=0.95\linewidth]{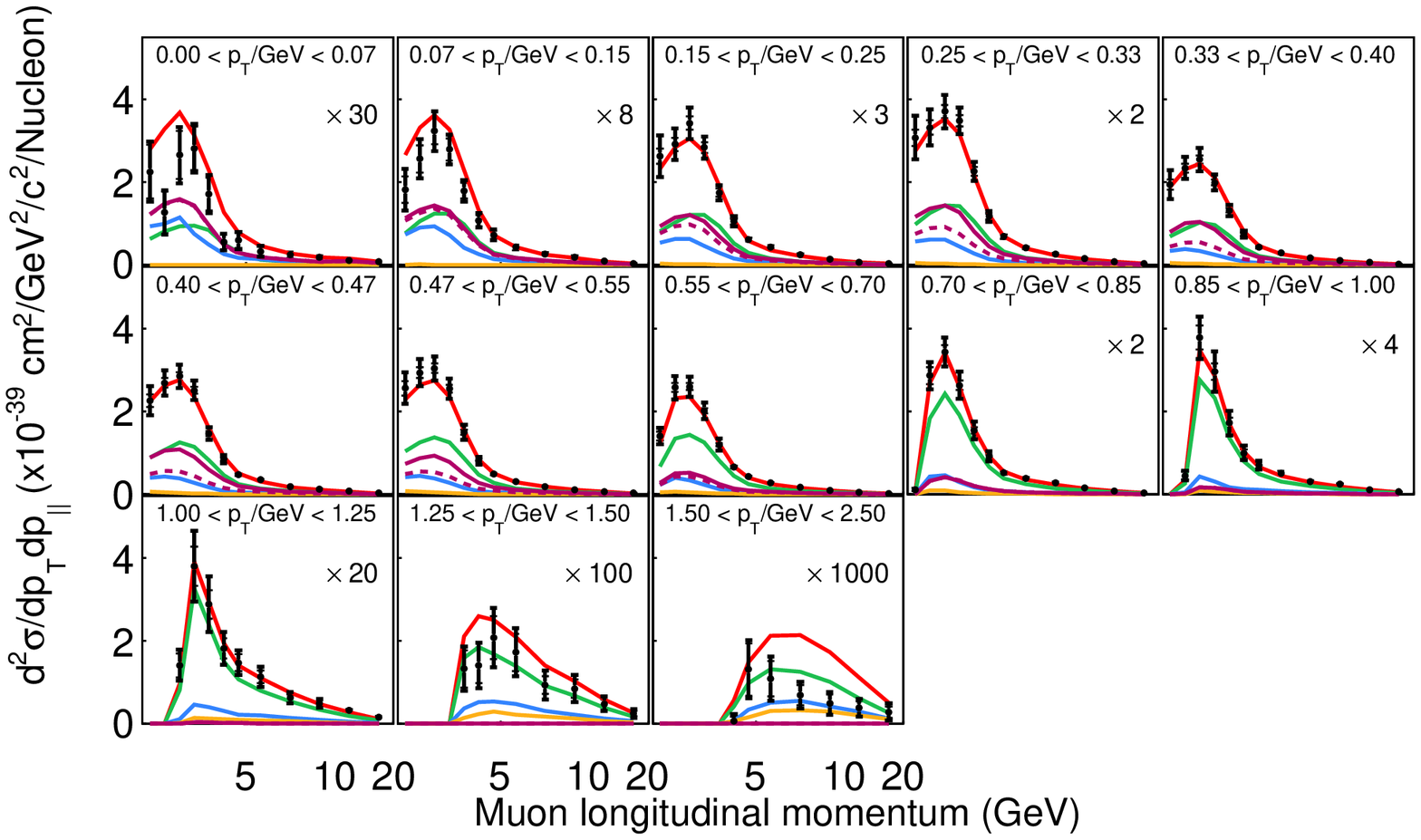} 
\caption{Full double differential results projected as a function of \pt~ and also a function of \pz. Unstacked curves show the various components of the \tune. The dotted line denotes original strength and shape of the 2p2h contribution.}
\label{fig:ptpzdoublediff}
\end{figure*}

The double differential cross section of \pt~versus \pz~is shown in Fig. \ref{fig:ptpzdoublediff}. The effect of the 2p2h enhancement and RPA reweighting are clearly visible and improve agreement with the data. The \tune~agrees with the data except in regions of low and high \pt~where the data cross section is smaller than predicted.

\begingroup
\squeezetable
\begin{table}[p]
\begin{tabular}{lcr}
\hline
Process Variant & Standard $\chi^{2}$ & Log-normal $\chi^{2}$ \\ 
\hline \hline
\tune & 253 & 207 \\ \hline
GENIE 2.8.4 & 282 & 320 \\ \hline
GENIE + RPA & 187 & 318 \\ \hline
GENIE + RPA + 2p2h & 284 & 294 \\ \hline
GENIE + 2p2h & 511 & 409 \\ \hline
\tune + \\ MINOS pion correction & 212 & 188 \\\hline
NuWro & 455 & 360 \\ 
\hline 
\end{tabular}
\caption{$\chi^{2}$ of various model variants compared to data using the standard and log-normal $\chi^{2}$ where there are 144 degrees of freedom.}
\label{tab:DDModelComp}
\end{table}
\endgroup
\begin{figure*}[p]
\centering
  \includegraphics[width=0.95\linewidth]{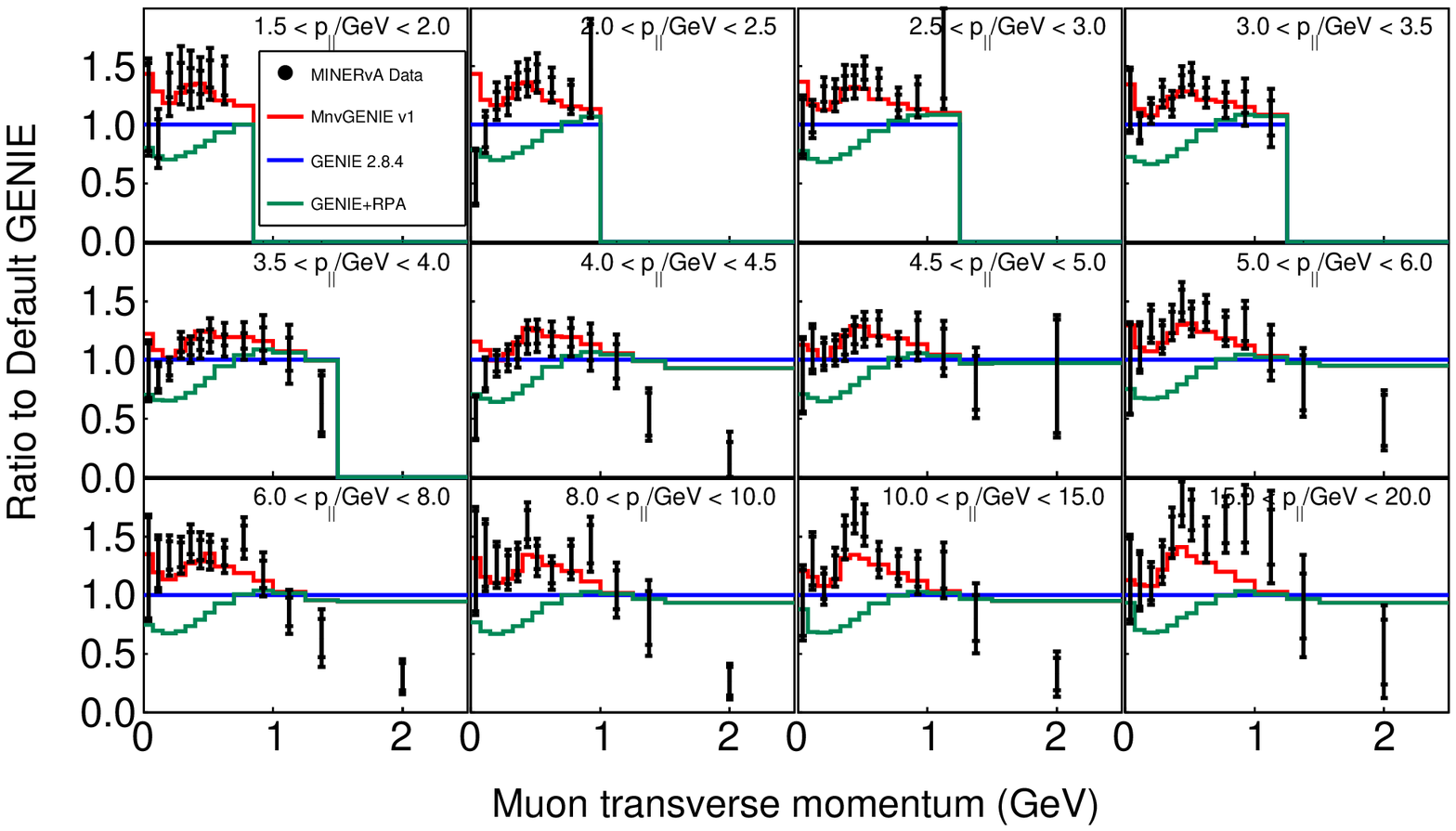}
  \includegraphics[width=0.95\linewidth]{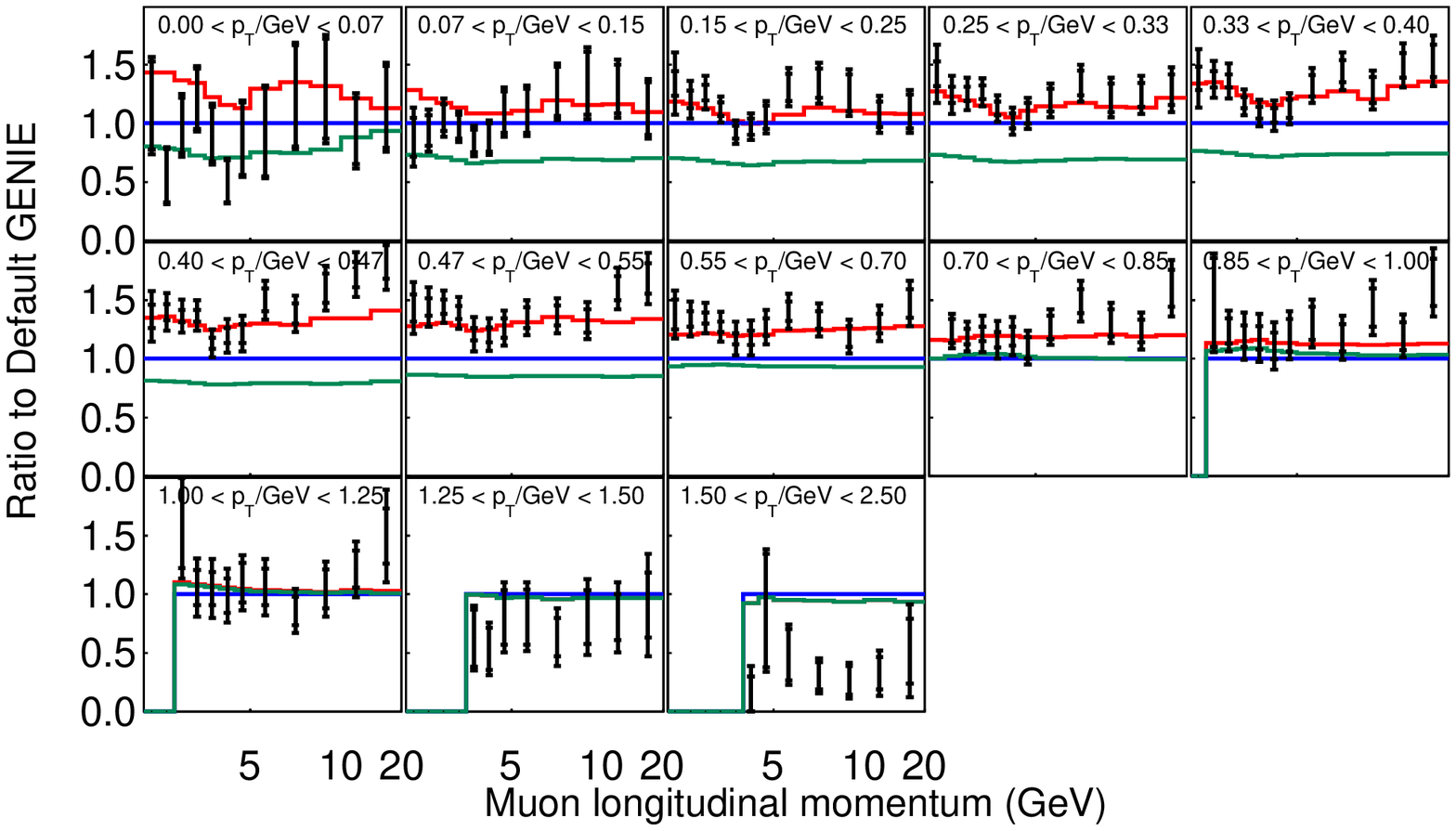}
\caption{Ratio of both the data and the prediction when only applying RPA to the CCQE channel in default GENIE 2.8.4 with respect to the \tune}
\label{fig:modcomp_rpa}
\end{figure*}

\begin{figure*}[p]
\centering
  \includegraphics[width=0.95\linewidth]{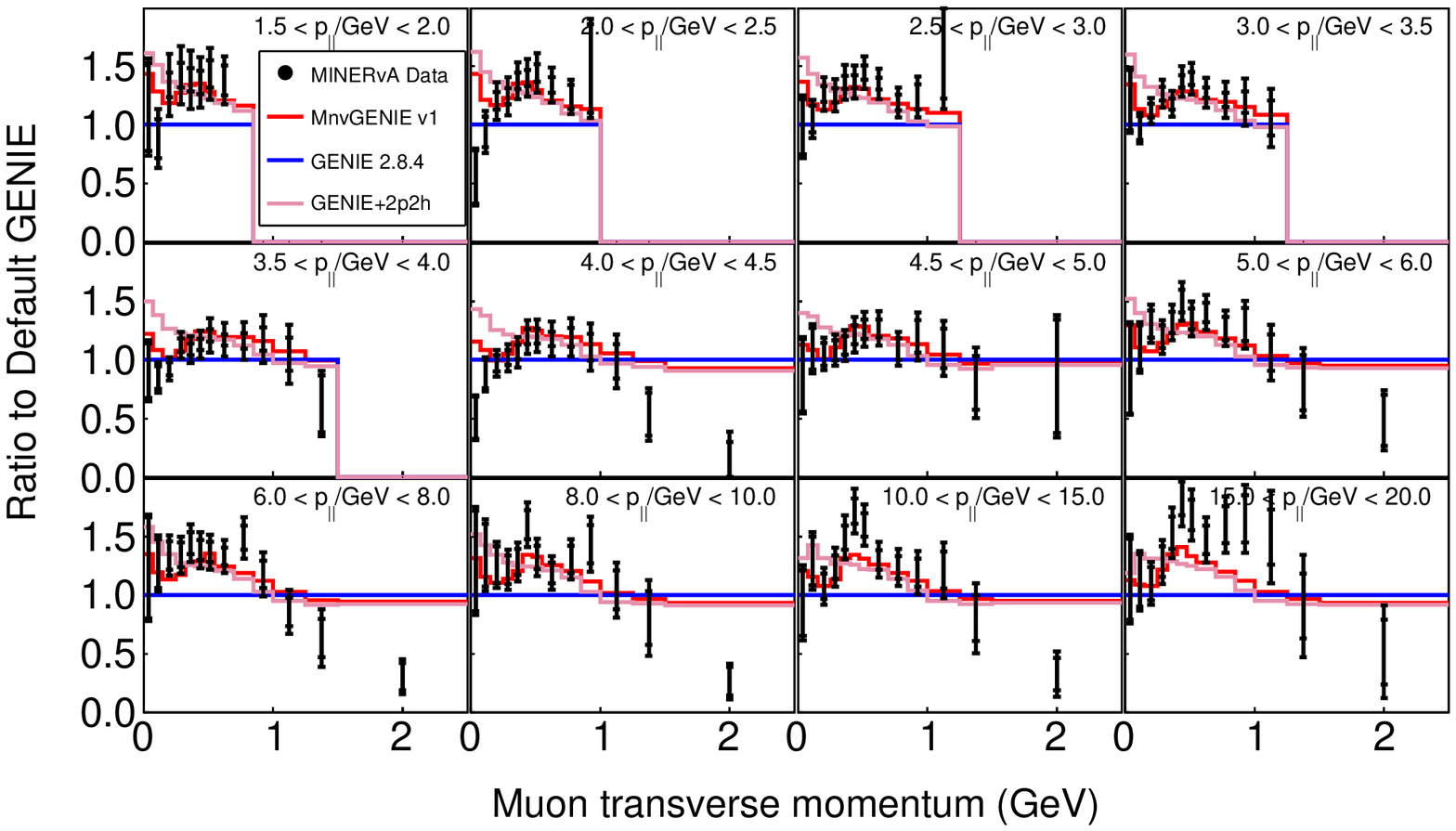}
  \includegraphics[width=0.95\linewidth]{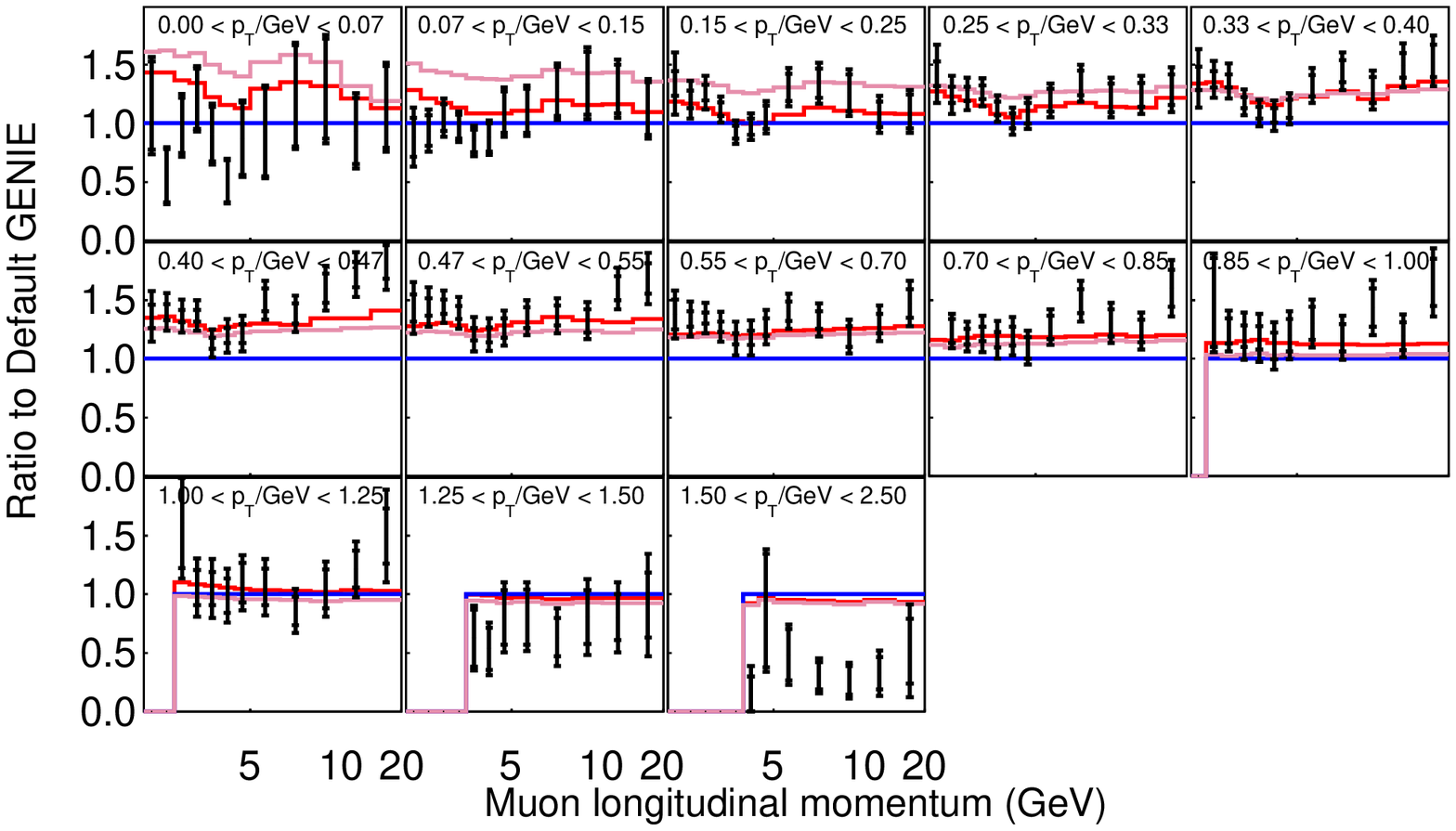}
\caption{Ratio of data, \tune, and the addition of 2p2h to default GENIE 2.8.4
Ratio of both the data and the prediction when 2p2h is added to default GENIE 2.8.4 with respect to the \tune}
\label{fig:modcomp_2p2h}
\end{figure*}
\begin{figure*}[p]
\centering
  \includegraphics[width=0.95\linewidth]{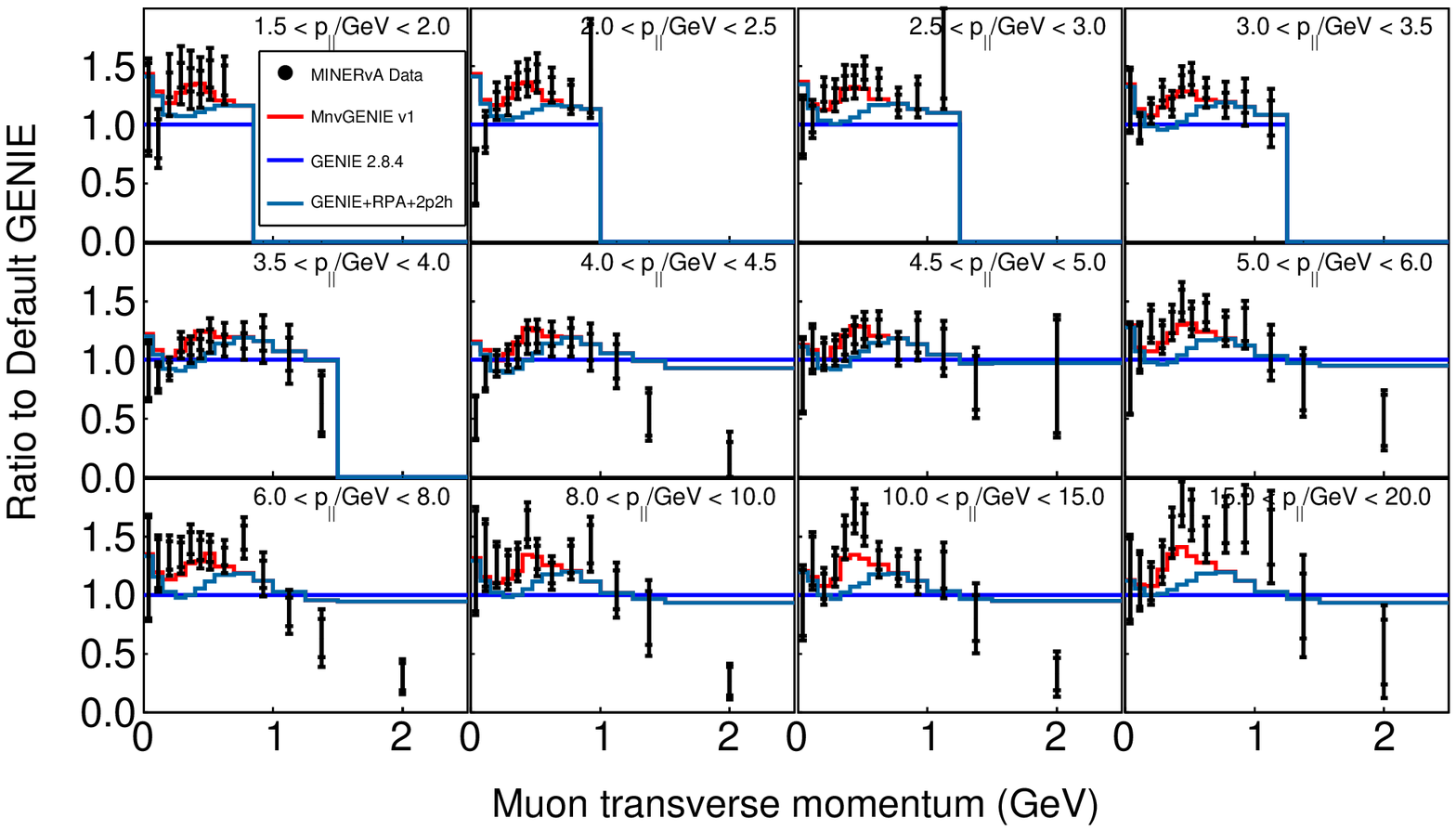}
  \includegraphics[width=0.95\linewidth]{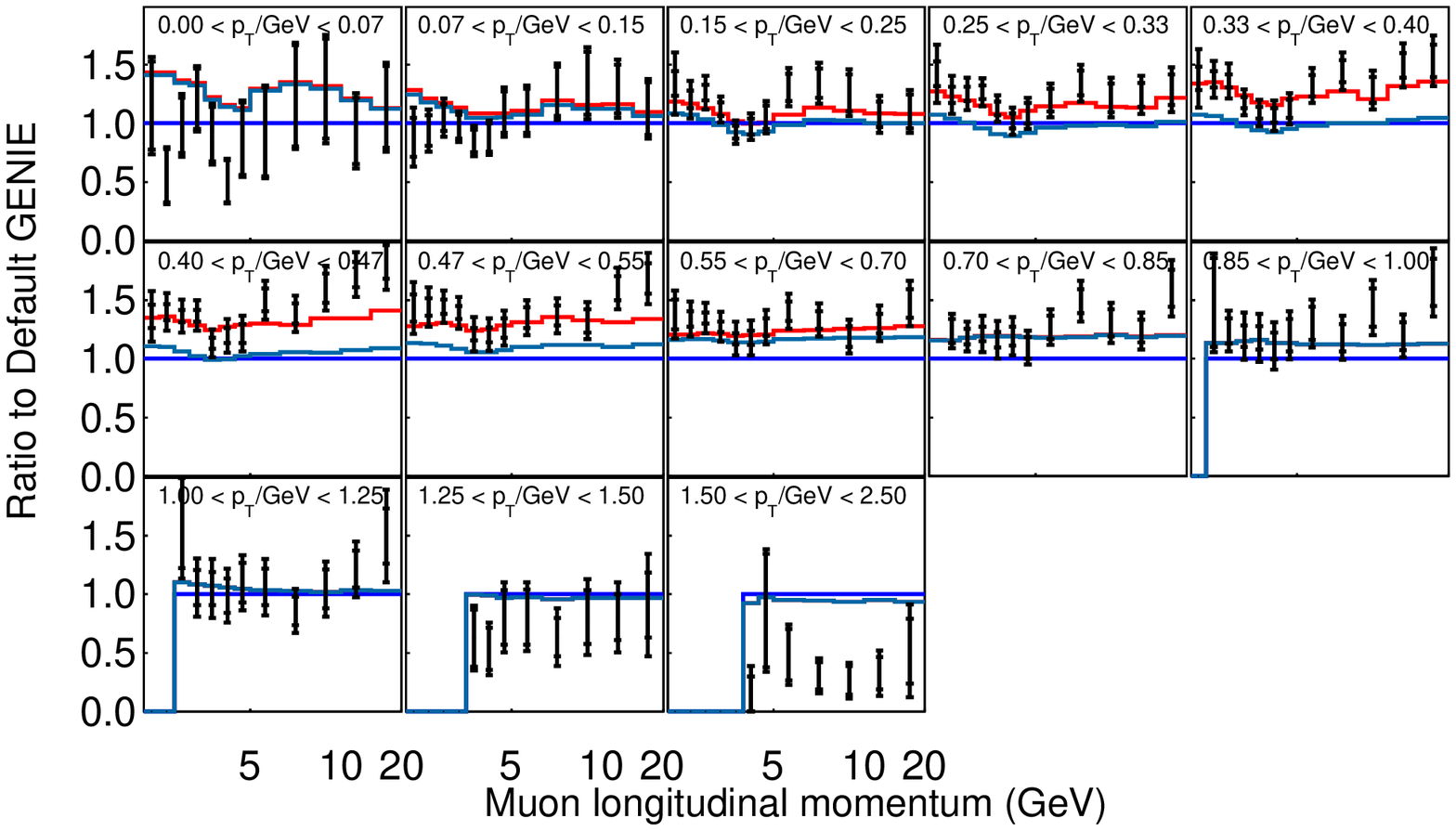}
\caption{Ratio of data, \tune, the addition of 2p2h and application of RPA to default GENIE 2.8.4
Ratio of both the data and the prediction when applying RPA to the CCQE channel and adding 2p2h to default GENIE 2.8.4 with respect to the \tune}
\label{fig:modcomp_2p2hrpa}
\end{figure*}

\begin{figure*}[p]
\centering
  \includegraphics[width=0.95\linewidth]{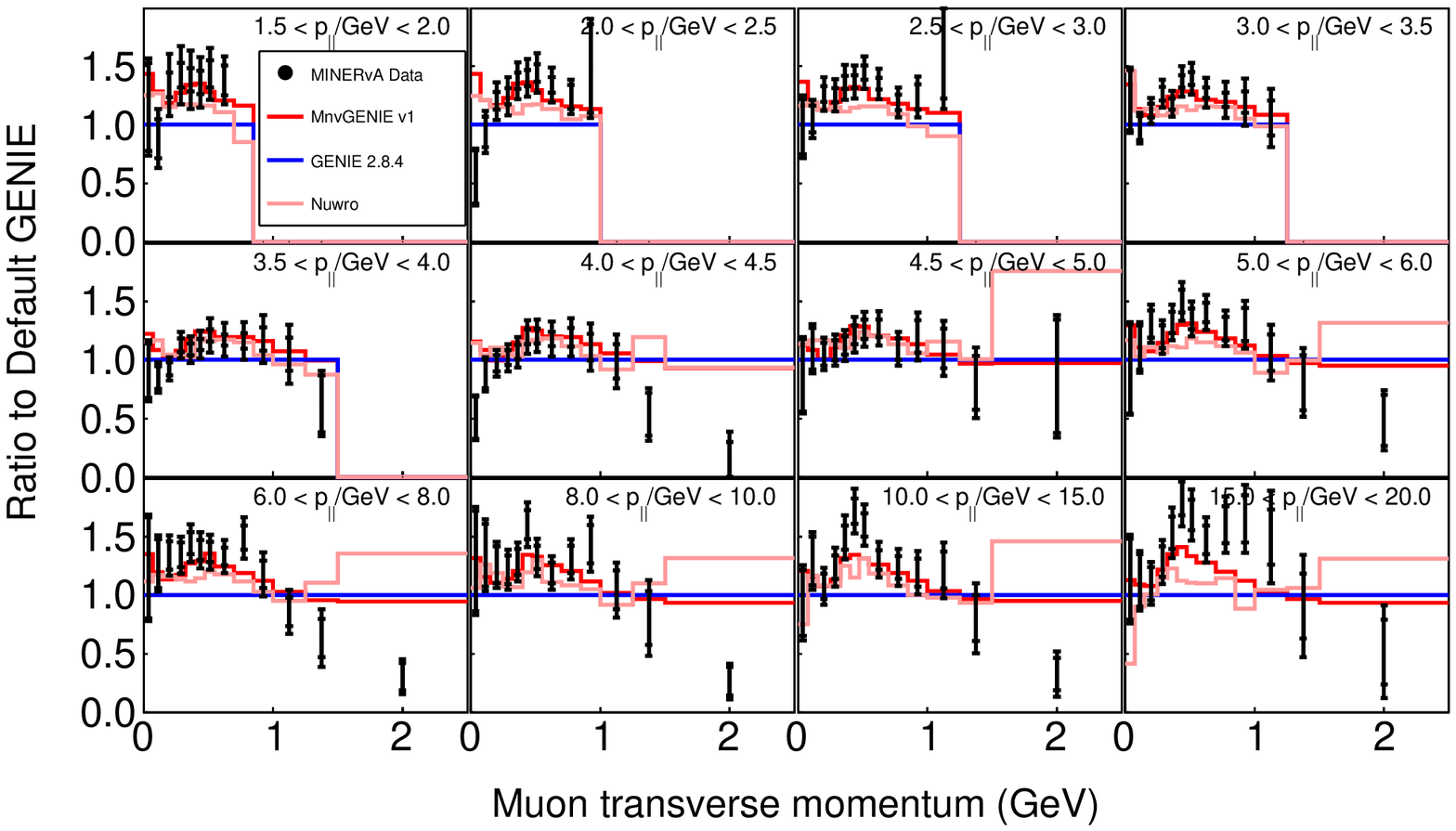}
  \includegraphics[width=0.95\linewidth]{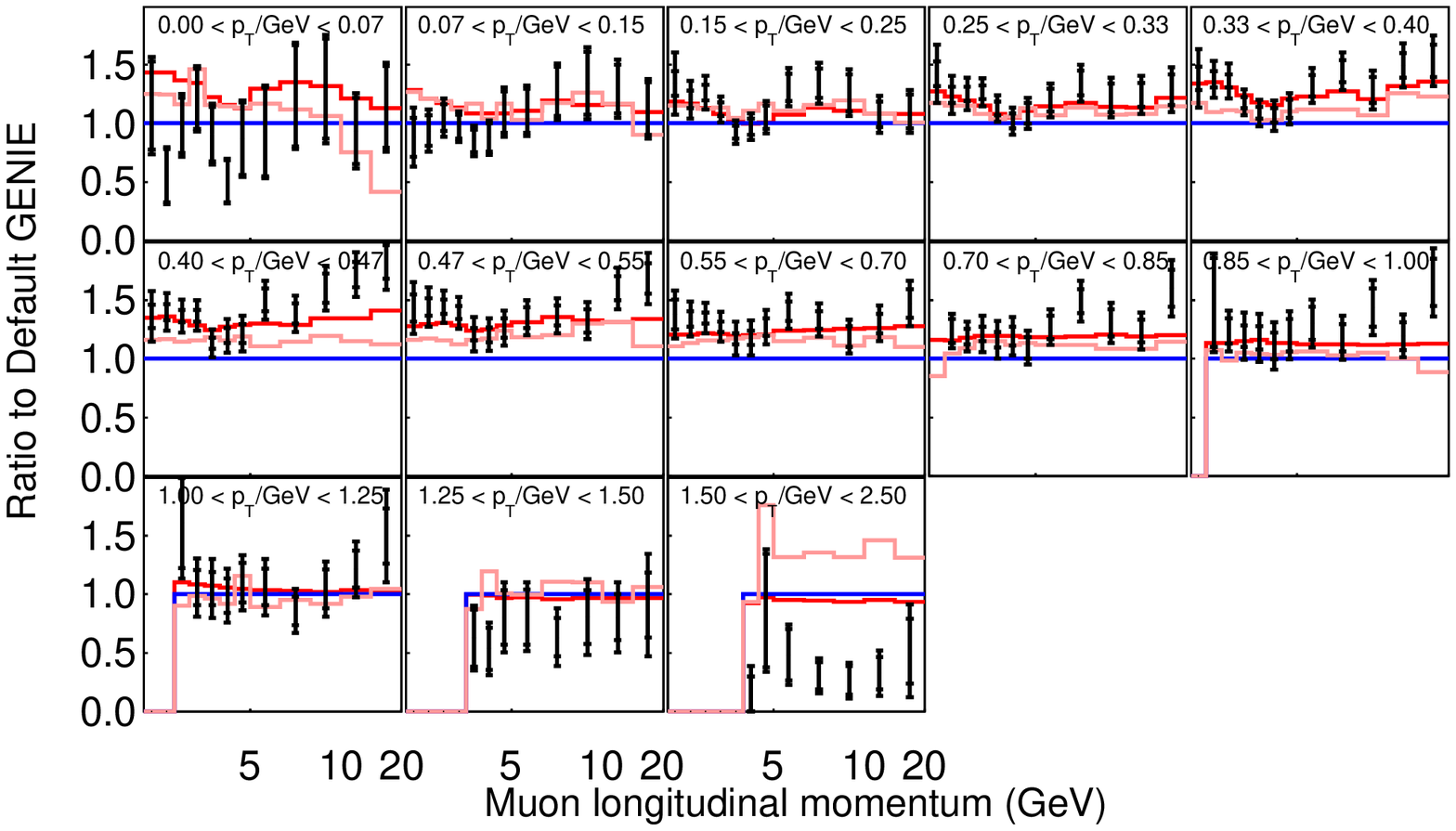}
\caption{Ratio of both the data and NuWro prediction with respect to the \tune}
\label{fig:modcomp_nuwro}
\end{figure*}

A comprehensive set of variations of the \tune~is provided both to demonstrate the effect of each component and to provide guidance on some choices which appear to better simulate the data. Variations applied to the base GENIE model include the application of RPA to CCQE (Fig. \ref{fig:modcomp_rpa}), the addition of 2p2h (Fig. \ref{fig:modcomp_2p2h}), the combination of applying RPA and adding 2p2h (Fig. \ref{fig:modcomp_2p2hrpa}). In addition, a comparison to a different neutrino generator prediction, NuWro\citep{PhysRevC.86.015505} revision 17.09, is shown in Fig. \ref{fig:modcomp_nuwro}. The application of an empirical resonant pion low Q$^{2}$ suppression whose form is based on MINOS data\cite{Adamson:2014pgc} is applied as an addition to the \tune.

Table \ref{tab:DDModelComp} provides a breakdown of $\chi^{2}$ for the various model variations. Comparisons of models to data are presented under two calculations: standard and log-normal $\chi^2$. A pathology referred to as Peelle's-Pertinent-Puzzle(PPP)\citep{Peelle:1987}\citep{BoxCox:1964}\citep{CARLSON20093215} occurs when there are highly correlated systematic uncertainies and a large source of normalization uncertainty. This is observed when the overall scale of the model is well below or above the data. A standard $\chi^2$  calculation assumes the underlying uncertainty is normally distributed, but this isn't the case for all sources of uncertainty in the cross section extraction. Sources of uncertainty which come from multiplicative operations are log-normal. Background subtraction is a source of additive uncertainty while efficiency corrections, normalization factors, such as flux, are multiplicative. Statistical uncertainties are typically additive, but the unfolding procedure makes this less clear. We present both the linear and log versions of the $\chi^2$ as a demonstration of just how different the best fit is under these two fitting prescriptions.


In the case of linear $\chi^2$ a model with the best fit is clearly not modeling the data as seen in Fig. \ref{fig:modcomp_rpa}. While the log-normal case is more consistent with our intuition, it still shows disagreement. In the double differential cross section there are two regions of clear data-Monte Carlo difference, which drive the overall poor $\chi^2$. These are more clearly shown in the single differential cross sections.

\subsection{Single Differential Cross Sections}
\begin{figure}[tp]
\centering
  \includegraphics[width=0.95\linewidth]{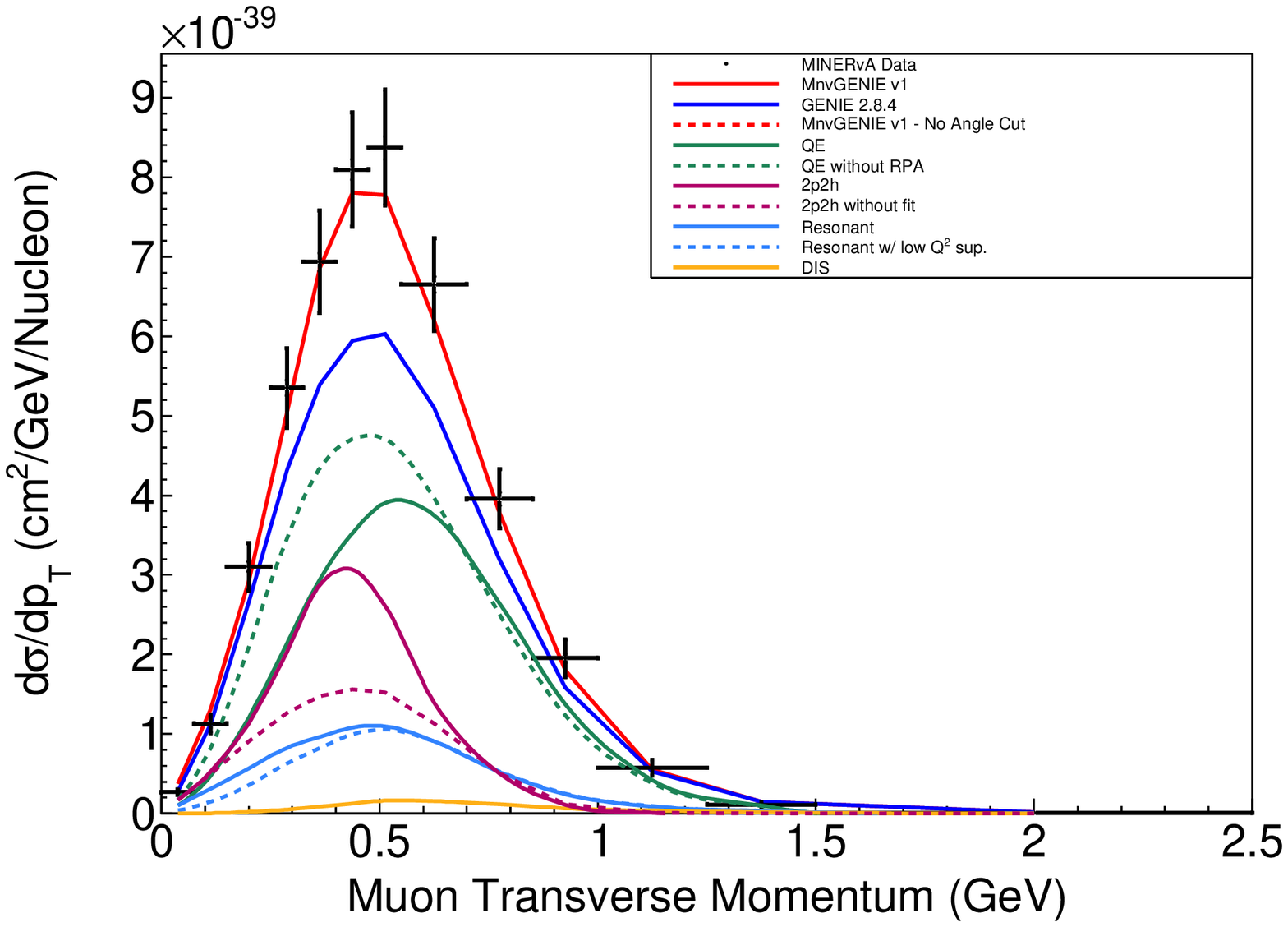} 
  \includegraphics[width=0.95\linewidth]{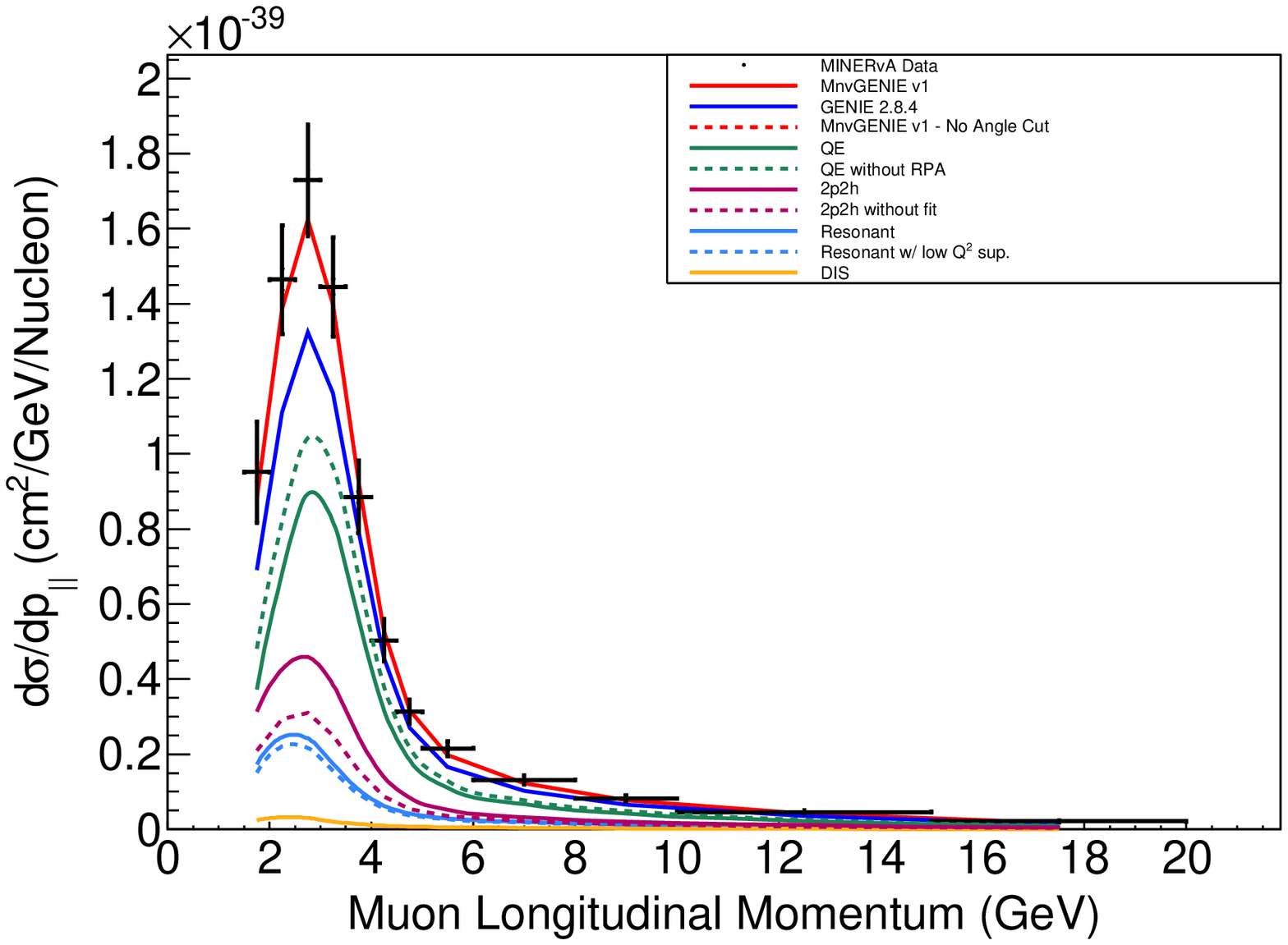} 
\caption{The projections of the double differential result onto the transverse and longitudinal axes. Model components are denoted by the unstacked lines. Solid lines, except the GENIE 2.8.4 line, add up to \tune, while dotted lines are model component modifications. }
\label{fig:ptpzsinglediff}
\end{figure}

The one dimensional projections of the \pt-\pz~result are shown in Fig. \ref{fig:ptpzsinglediff}. 

\begin{figure}[tp]
\centering
  \includegraphics[width=0.95\linewidth]{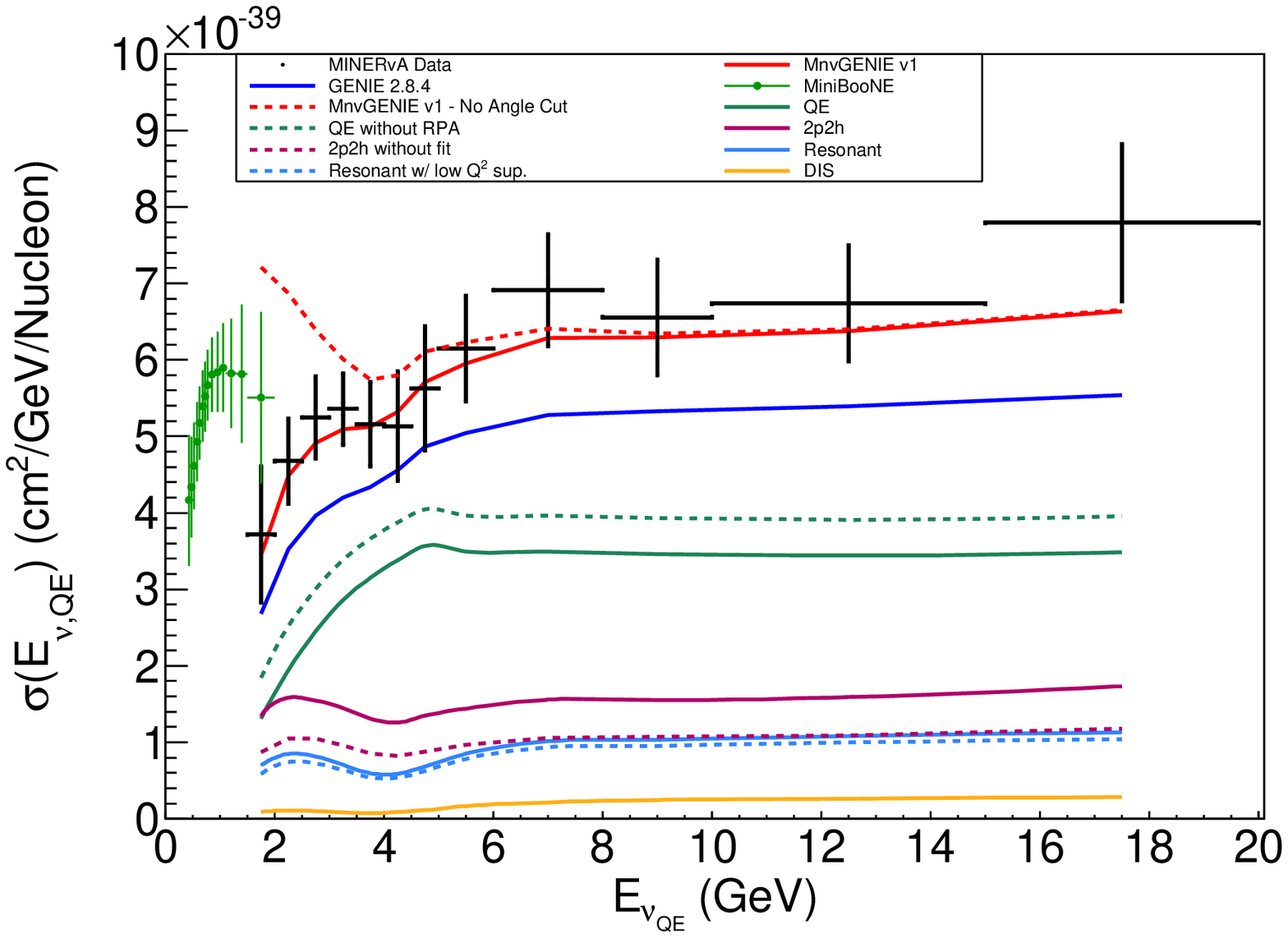} 
  \includegraphics[width=0.95\linewidth]{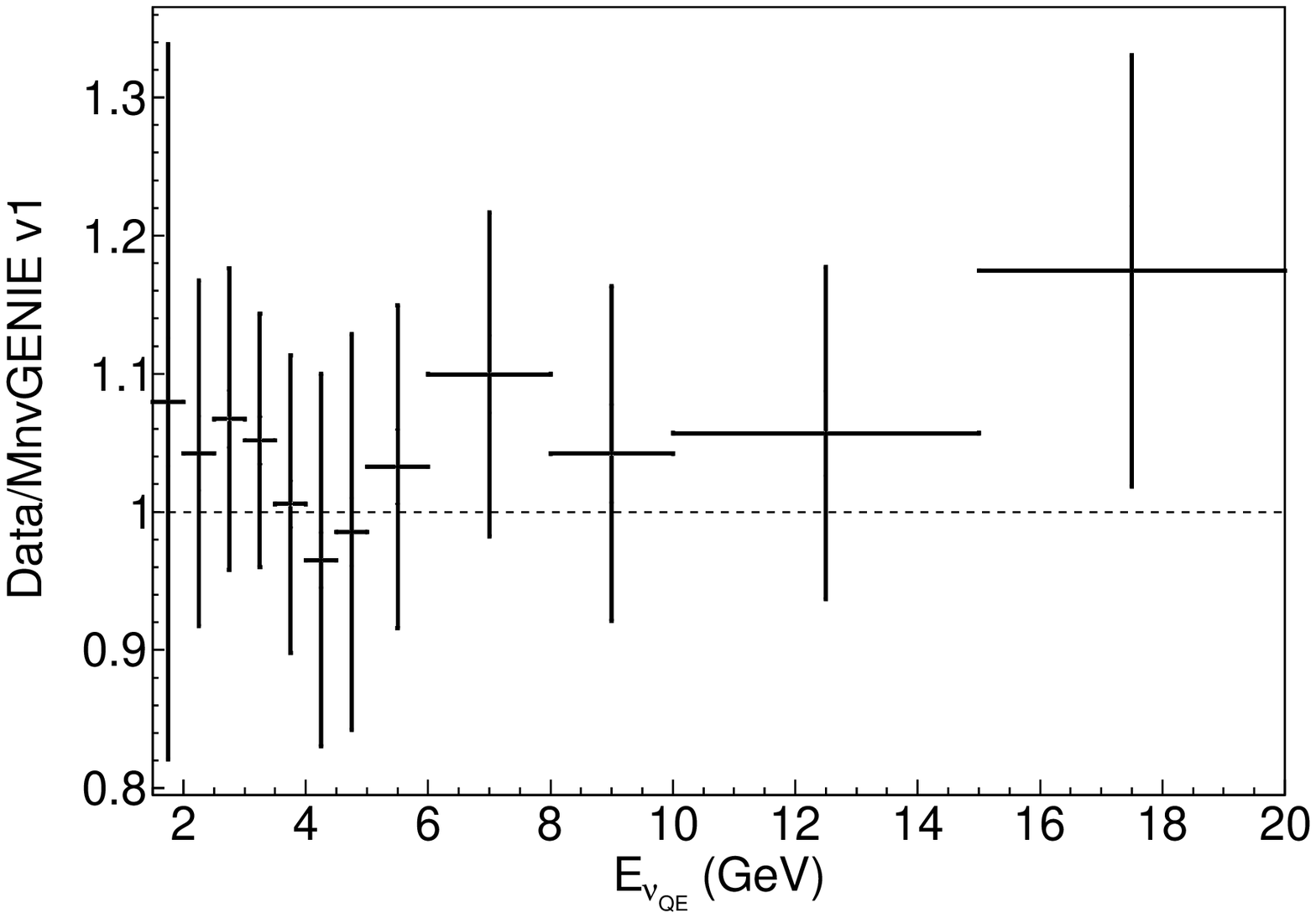} 
\caption{The differential cross section as a function of \enuqe. The curves represent different signal channel contributions to the \tune~prediction. Dotted lines denote the effect of turning on two of the \tune~modifications to the default GENIE prediction. In addition, results from the MiniBooNE measurement are included.}
\label{fig:xsec_enuqe}
\end{figure}

A single differential cross section in \enuqe, which is correlated with \pz, is provided in Fig. \ref{fig:xsec_enuqe}. The \qelike~MiniBooNE result\citep{PhysRevD.81.092005} is shown for comparison at lower energies. The reader should again keep in mind this result is not strictly an absolute cross section, $\sigma(E_\nu)$, but $\frac{d\sigma}{dE_{\nu,QE}}$. Even at a given \enuqe~value, two experiments will measure different cross sections due to different fluxes, detector acceptances, and signal definitions. In particular, the $\theta_\mu$ requirement of the signal definition in this analysis will have a large effect at low \enuqe, demonstrated by the curve labeled ``MnvGENIE v1 - No Angle Cut". 

\begin{figure}[tp]
\centering
  \includegraphics[width=0.95\linewidth]{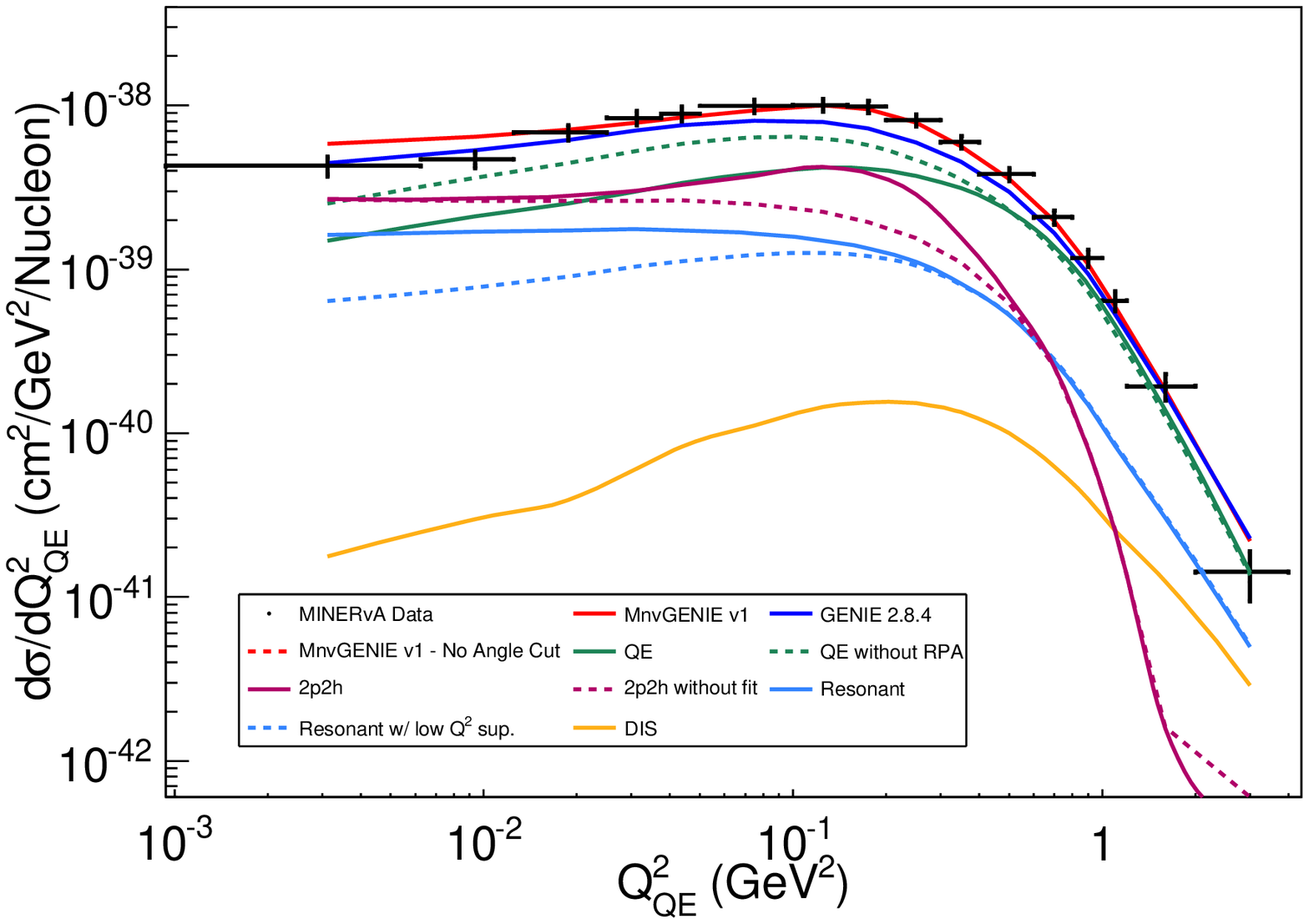} 
  \includegraphics[width=0.95\linewidth]{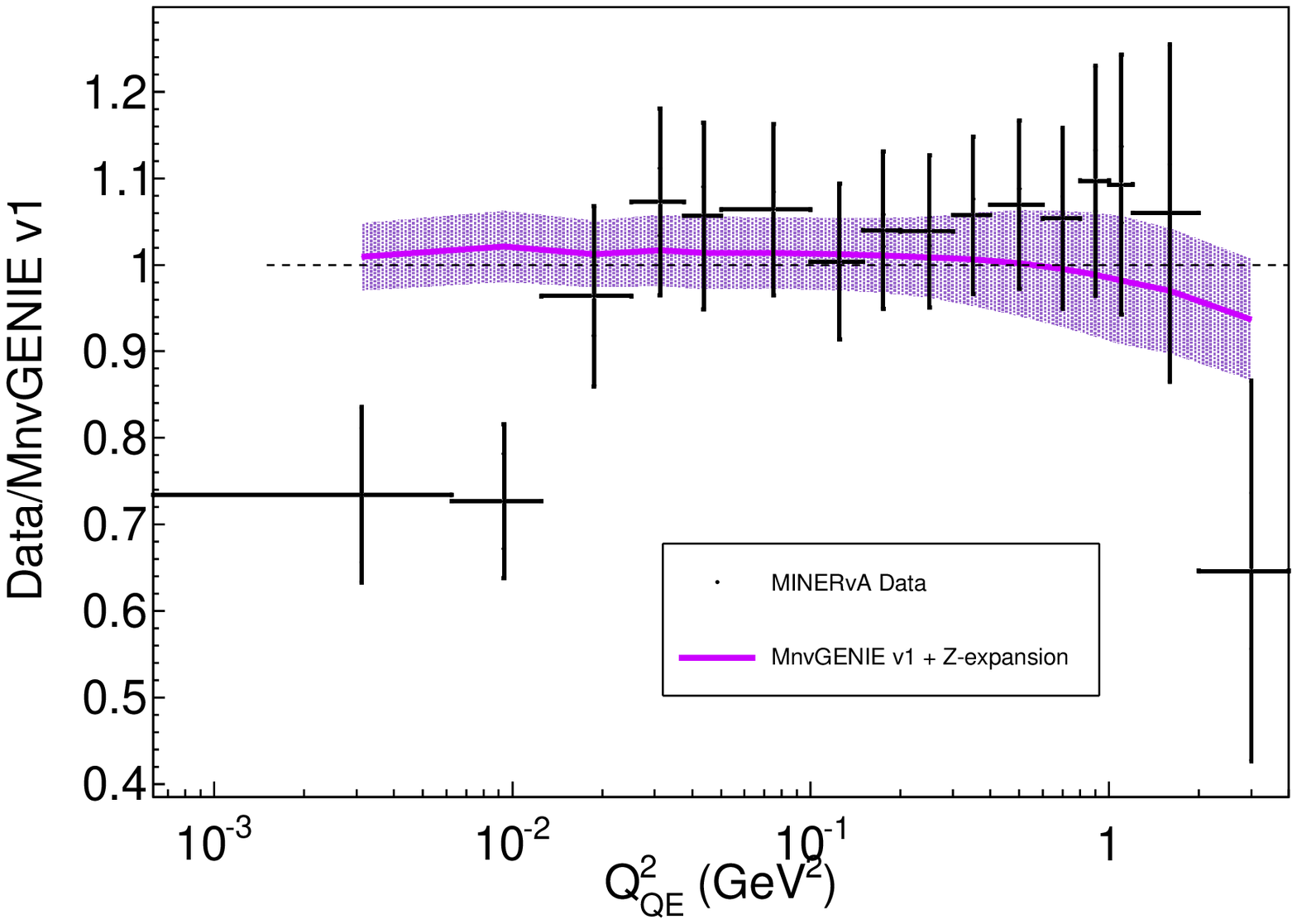} 
\caption{The differential cross section as a function of \qqqe. The curves represent different signal channel contributions to the \tune~prediction. Dotted lines denote the effect of turning on two of the \tune~modifications to the default GENIE prediction.}
\label{fig:qqqe}
\end{figure}

A single differential cross section of \qqqe~is presented in Fig. \ref{fig:qqqe} with the same signal model curves as the double differential result. The addition of the default GENIE 2.8.4 prediction is provided for reference. The effect of the 2p2h enhancement is strongest at moderate \qqqe. The prediction disagrees with the data at very low \qqqe~and high \qqqe. Except at the very edges of the \qqqe~distribution, the rate and shape of the \qqqe~distribution is  well described confirming the anomalous $M_A$ evaluated by \cite{Gran:2006jn,AguilarArevalo:2010zc} for the same range of $Q^2_{QE}$ was due to unmodeled multi-nucleon effects not readily available at the time of those analyses.

For \qqqe$<0.01$ GeV$^2$, one-third of the predicted rate comes from resonant pion production which enter the sample via FSI. The initial state model is missing the effect of a suppression at low $Q^2$, which the background constraint from Sec. \ref{sec:extraction} indicates is necessary to describe the sidebands. Previous \minerva~results have also measured cross sections less than the prediction in comparable phase space\citep{Rodrigues:2015hik}\citep{Eberly:2014mra}\citep{McGivern:2016bwh}\citep{LE2015130}. In addition, MINOS measured a similar effect\citep{Adamson:2014pgc}.

The region where \qqqe$>2$ GeV$^2$ is predicted to be $\sim$70\% true CCQE events. The remaining rate comes from resonant pion events with a small contribution from DIS interactions. The prediction of a larger cross section at high \qqqe~could be due to any combination of the following: the dipole form of the axial form factor or final state interactions or initial state nucleon kinematics.

A variety of data is available to constrain the axial form factor such as data on deuterium\citep{Mann:1973pr}\citep{Barish:1977qk}\citep{Miller:1982qi}\citep{Baker:1981su}\citep{Kitagaki:1983px} and heavy targets\citep{Gran:2006jn}\citep{AguilarArevalo:2010zc}\citep{Brunner:1989kw}\citep{Pohl:1979zm}\citep{Auerbach:2002iy}\citep{Belikov:1983kg}\citep{Bonetti:1977cs}\citep{Lyubushkin:2008pe}\citep{Abe:2014iza}. Only Refs. \citep{Miller:1982qi}\citep{Baker:1981su}\citep{Kitagaki:1983px}\citep{Bonetti:1977cs} have data for Q$^2 >$ 2 GeV$^2$, and they are statistics limited. The z-expansion model\citep{ZExpBhattacharya}\citep{ZExpMeyer} provides a different parameterization of the axial form factor, extracted from the deuterium data listed above, suggests that the current data permits significant variation from the dipole model at large $Q^2$. 

\subsection{Vertex Energy}
\label{subsec:vtxenergy}

This analysis ignores energy within 150 mm of the interaction vertex to ensure it is not strongly sensitive to modeling the particle content in this region. However, the \tune~attempts to provide a more descriptive model of the data and it is important to see how this tune predicts the vertex energy. The prior \minerva~ investigation\citep{Fiorentini:2013ezn} preferred a spectrum with 25\% more protons. At the time a 2p2h model was not part of the reference  model.

\begin{figure}[tp]
\centering
  \includegraphics[width=0.95\linewidth]{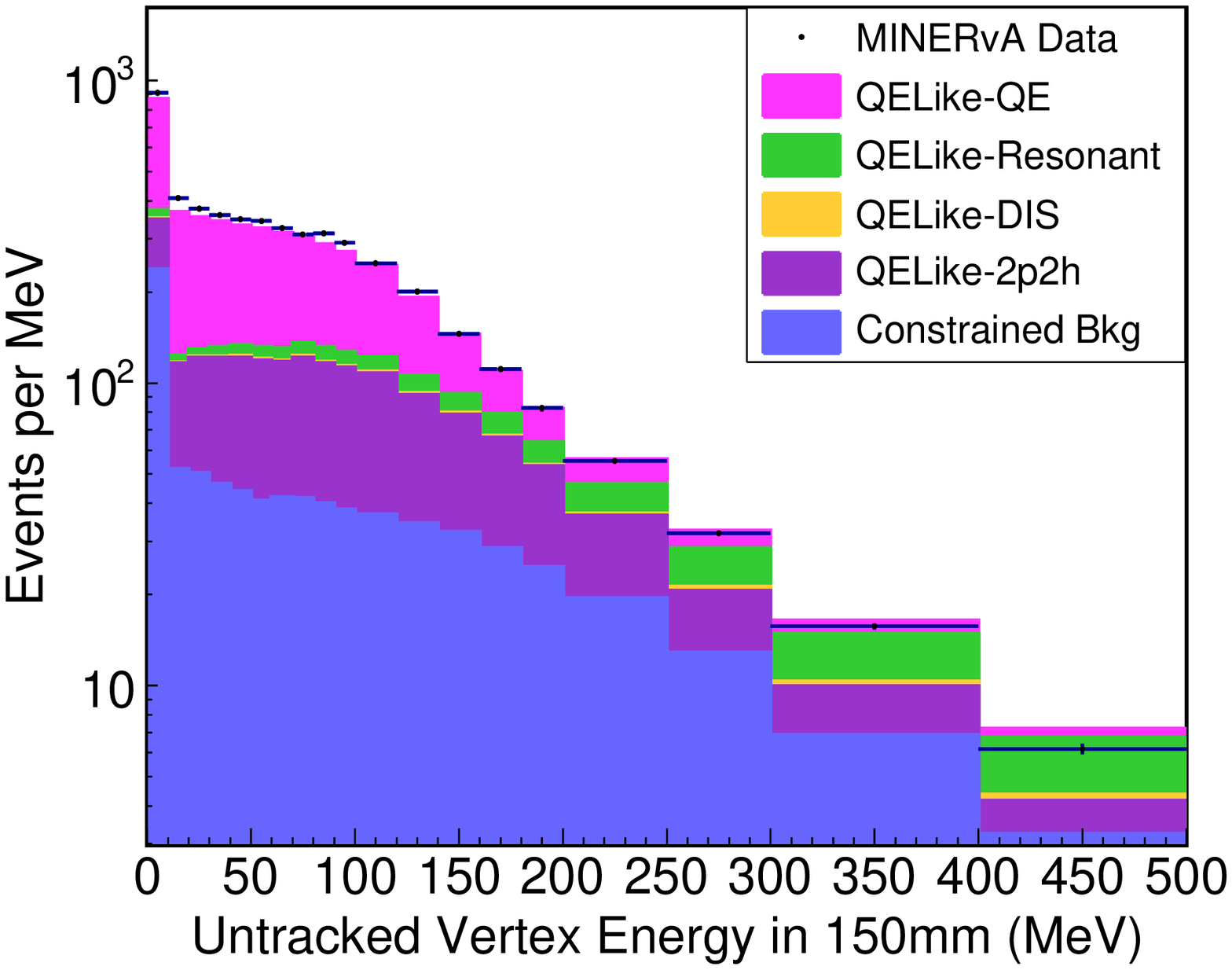}
   \includegraphics[width=0.95\linewidth]{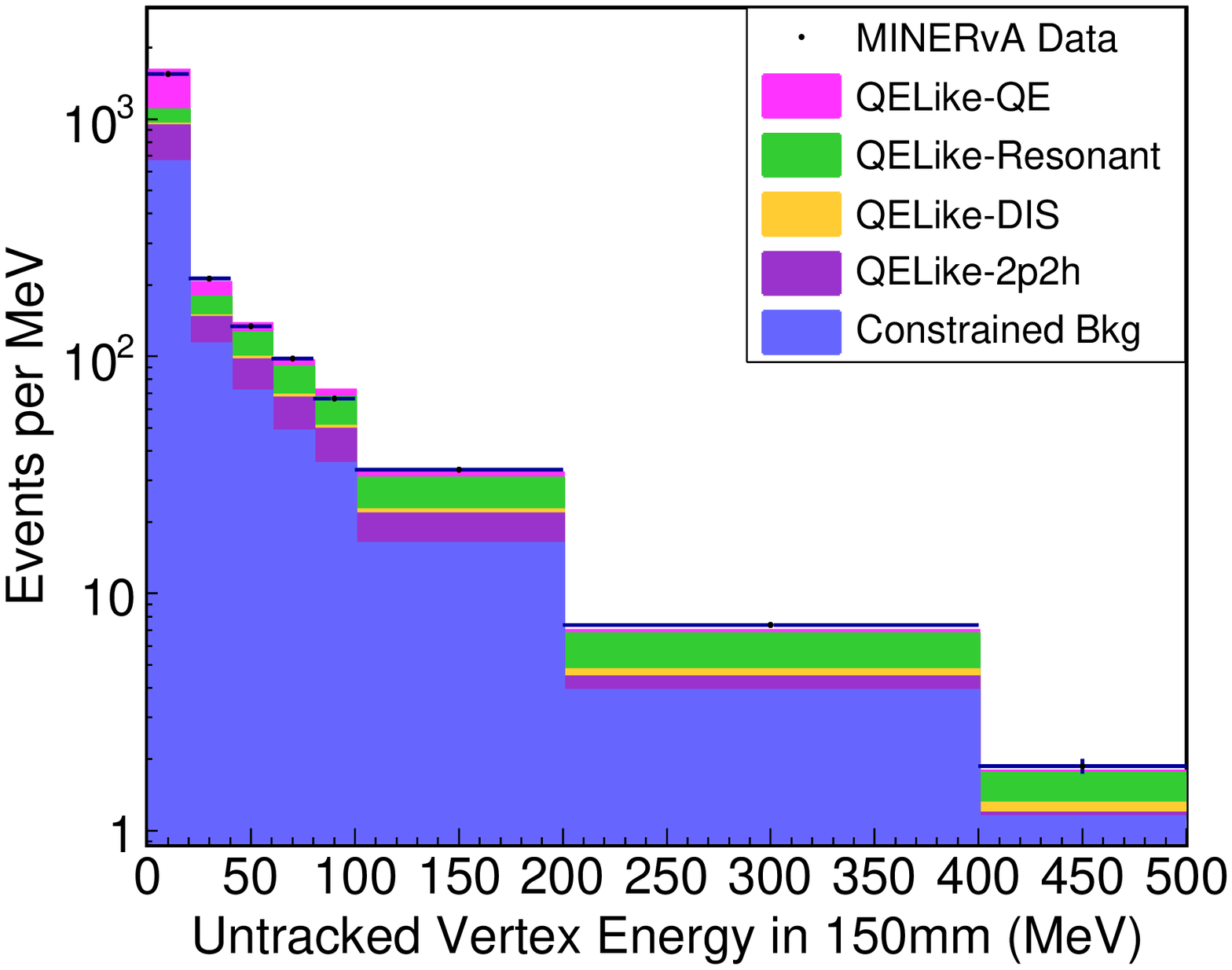}
\caption{Vertex energy within 150 mm of the reconstructed vertex excluding tracked energy for events with just the muon reconstructed (top) and muon plus protons (bottom).}
\label{fig:singlevtxenergy1track}
\end{figure}

\begin{figure}[tp]
\centering
  	\includegraphics[width=0.95\linewidth]{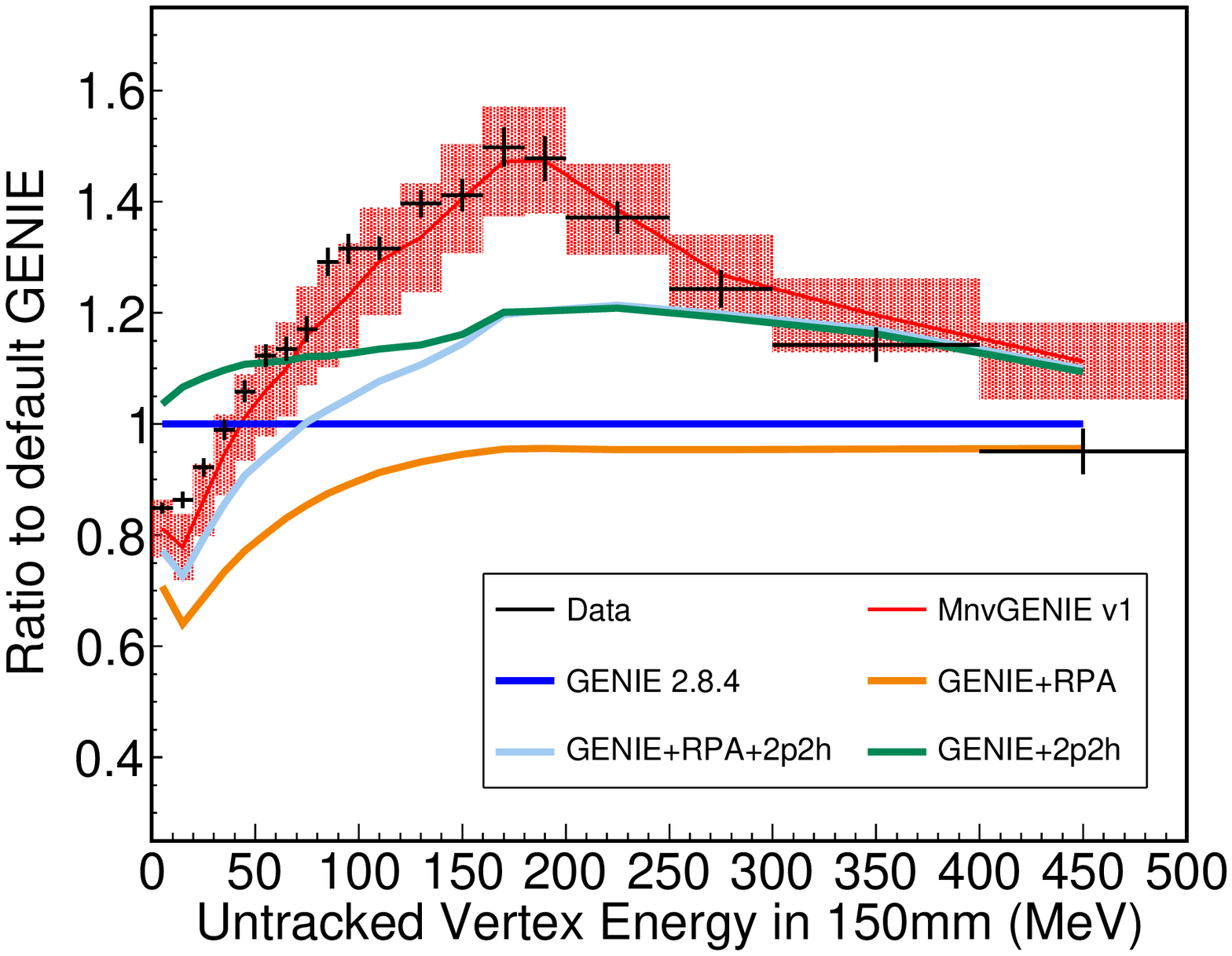}
      \includegraphics[width=0.95\linewidth]{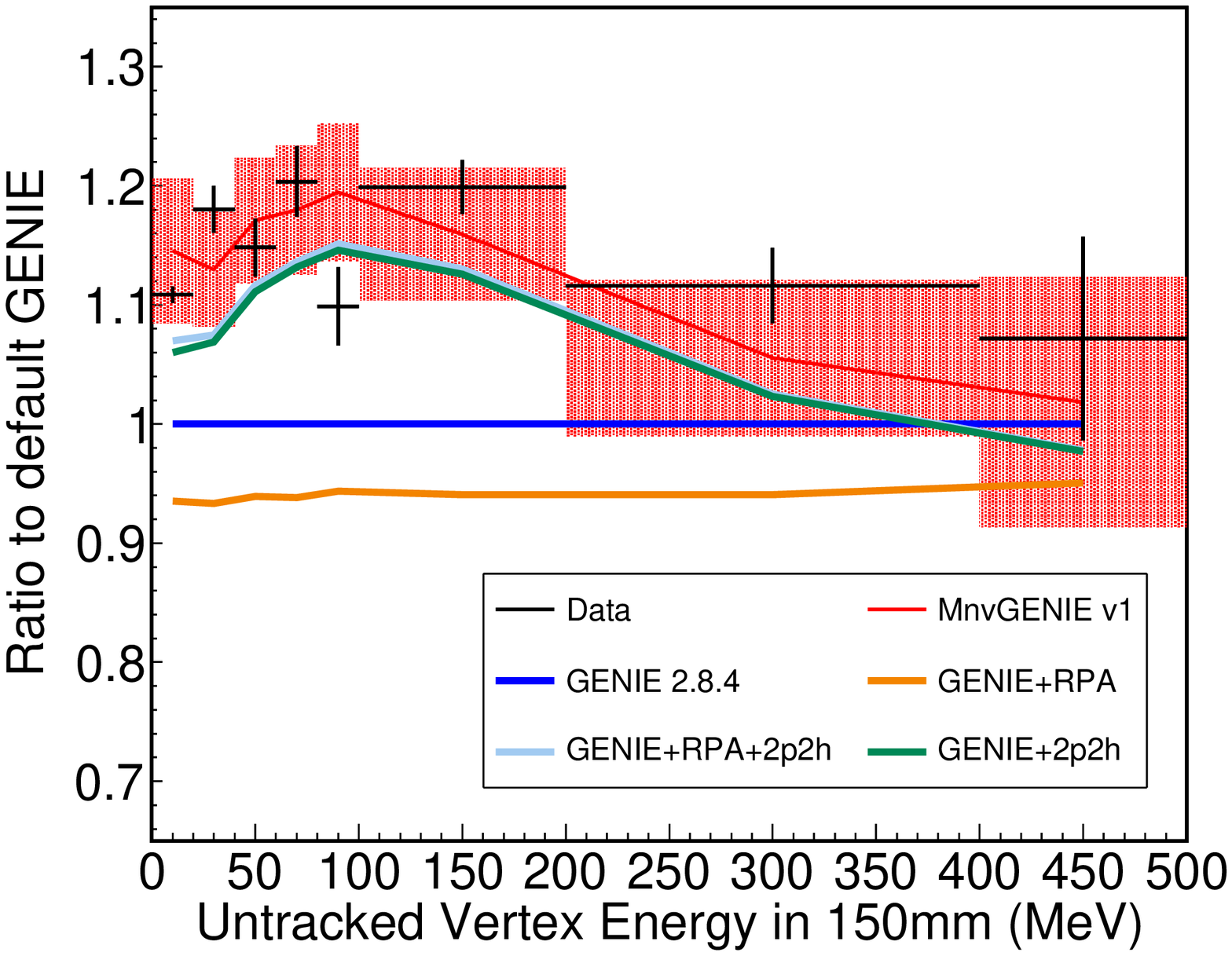}
\caption{Ratio of the data and various GENIE predictions to GENIE 2.8.4 of the vertex energy within 150 mm of the reconstructed vertex excluding tracked energy for events with no proton tracks reconstructed (top) and muon plus proton tracks reconstructed (bottom).}
\label{fig:singleratiovtxenergy1track}
\end{figure}

The vertex energy distribution within a 150 mm region around the vertex with the \tune~is shown for both the 1-track and multi-track samples in Fig. \ref{fig:singlevtxenergy1track}, respectively. Backgrounds have been constrained in these samples. The ratio plot with respect to default GENIE 2.8.4 with full systematic errors are shown in Fig. \ref{fig:singleratiovtxenergy1track}. It is clear without the 2p2h and RPA modeling the data would be poorly described, but the data also favors the empirical enhancement used in the \tune.

\begin{figure}[tp]
\centering
  	\includegraphics[width=0.95\linewidth]{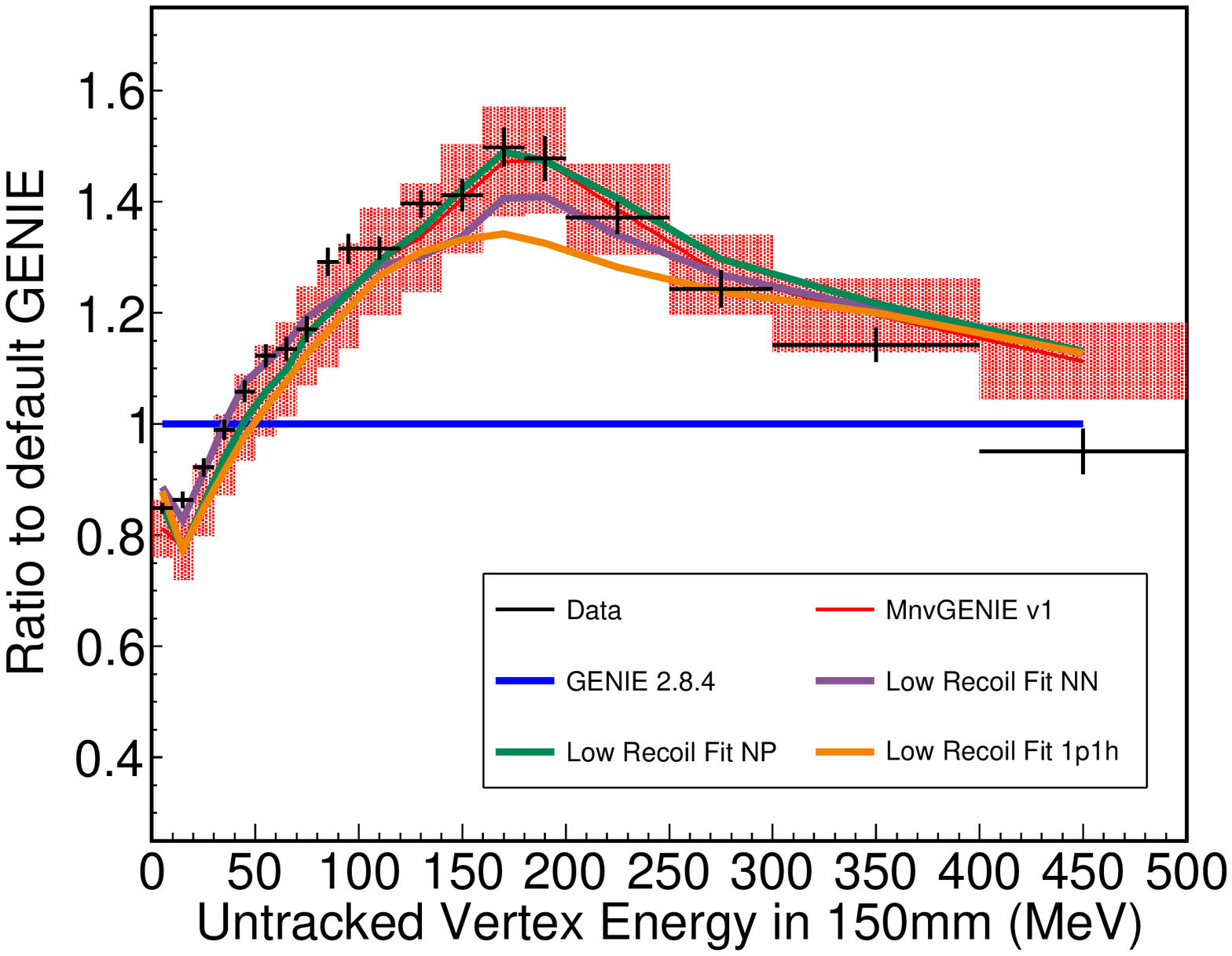}
      \includegraphics[width=0.95\linewidth]{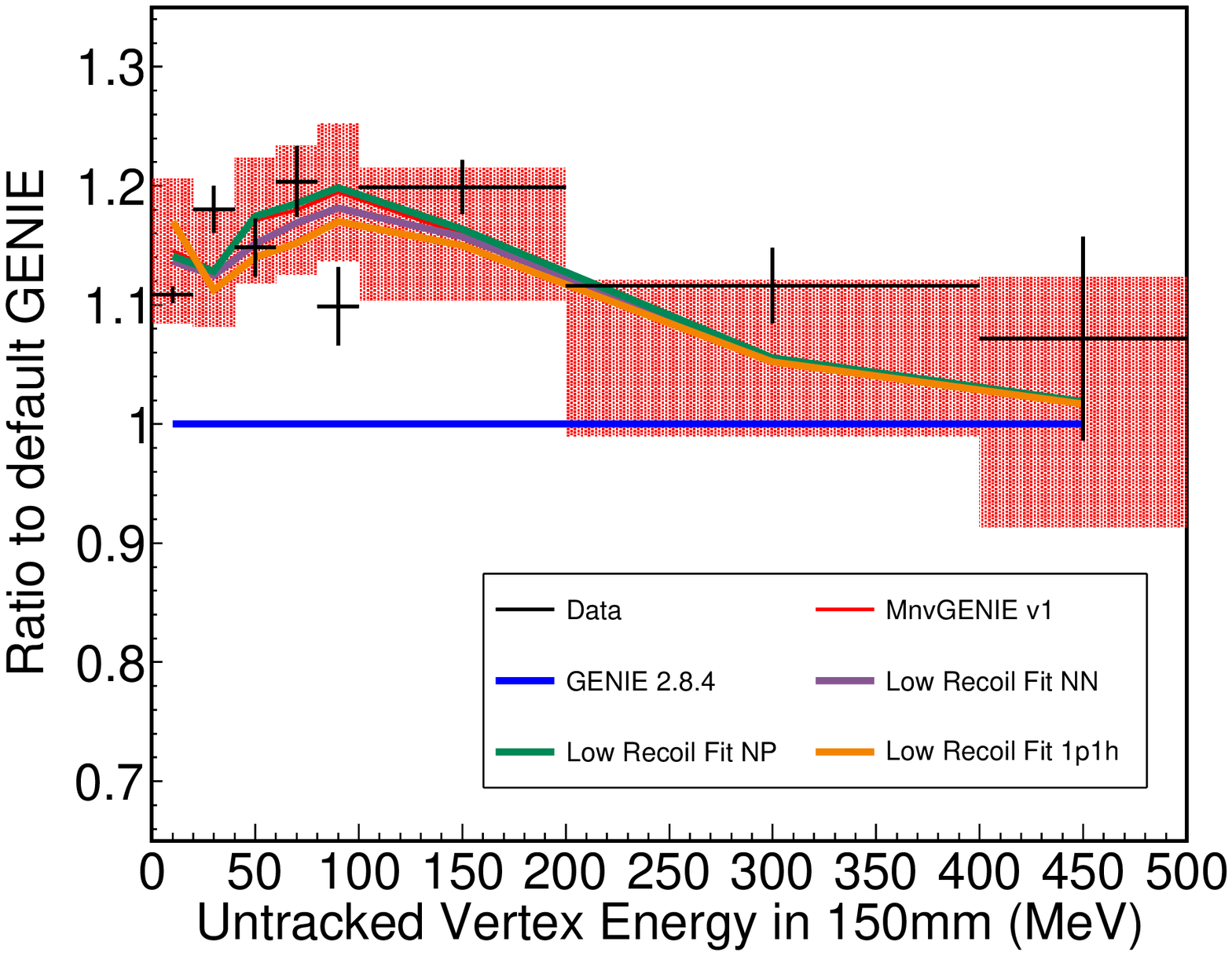}
\caption{Ratio of the data and variants of the enhancement via 2p2h nn, 2p2h np or QE-only events to GENIE 2.8.4 for vertex energy within 150 mm of the reconstructed vertex excluding tracked energy for events with no proton tracks reconstructed (top) and muon plus proton tracks reconstructed (bottom).}
\label{fig:singleratiovar1track}
\end{figure}

As described in Sec. \ref{systematics} the effect of the 2p2h enhancement has a systematic uncertainty derived by three different applications of the fit to various potential contributors, np-pair 2p2h, nn-pair 2p2h, and QE-only. Figure \ref{fig:singleratiovar1track} shows the effect of these variations on the vertex energy distribution.

\begin{figure*}[p]
\centering
  \includegraphics[width=0.95\linewidth]{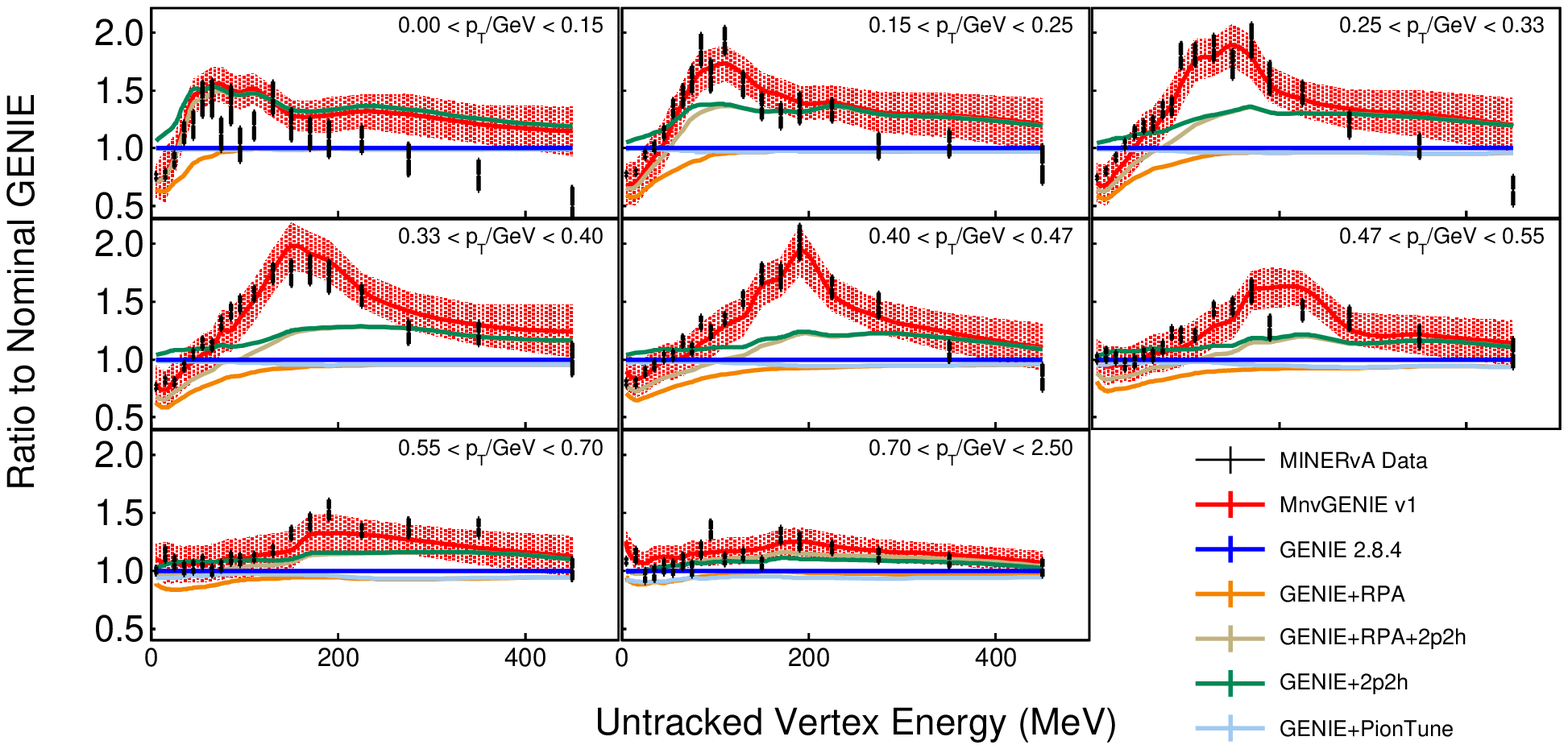}
  \includegraphics[width=0.95\linewidth]{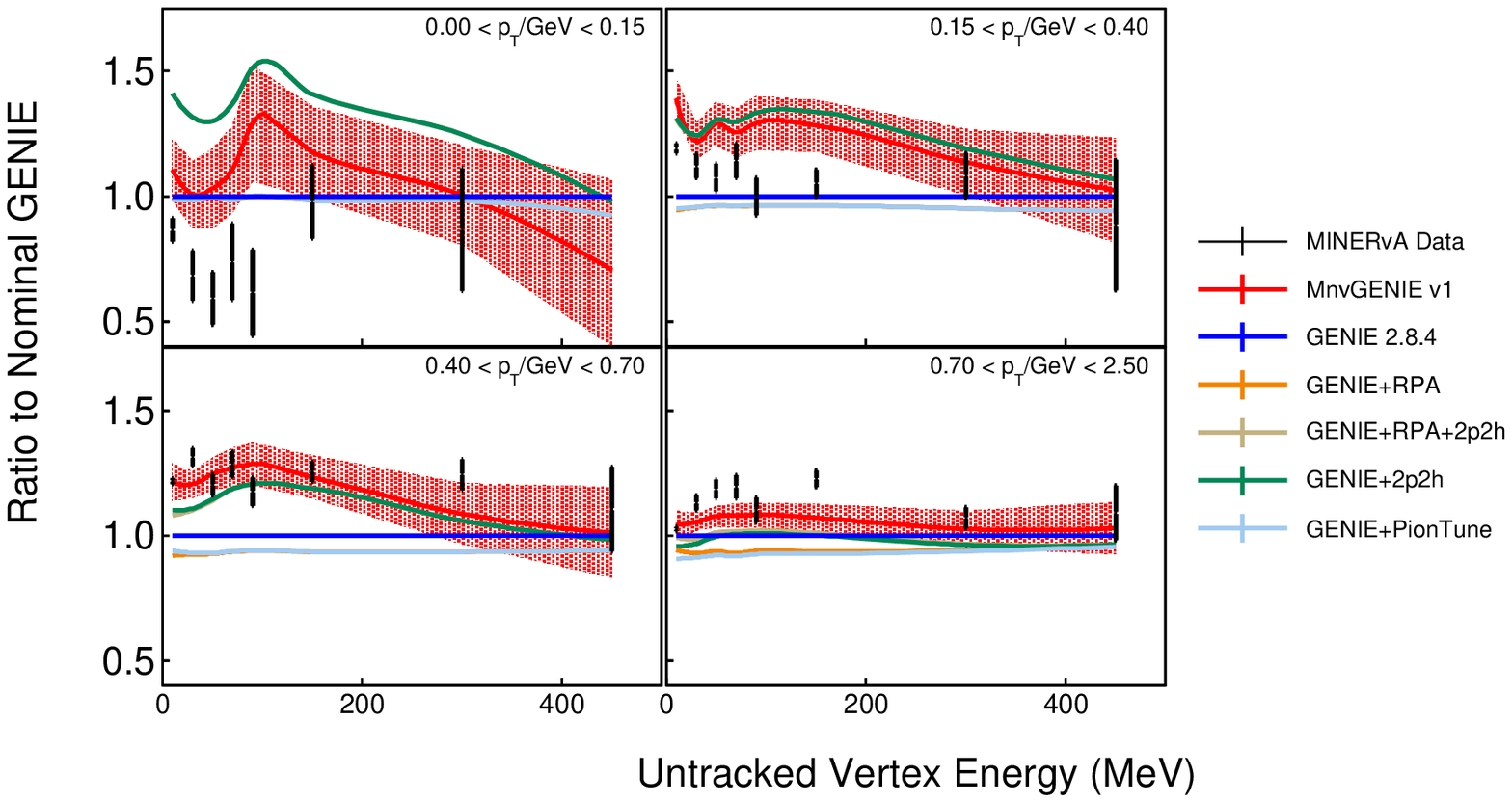}
\caption{Vertex energy in \pt~bins as compared to various GENIE configurations. The single track (top) and multi-track (bottom) samples are shown.}
\label{fig:vtx_bypt_genie}
\end{figure*}

\begin{figure*}[p]
\centering
  \includegraphics[width=0.95\linewidth]{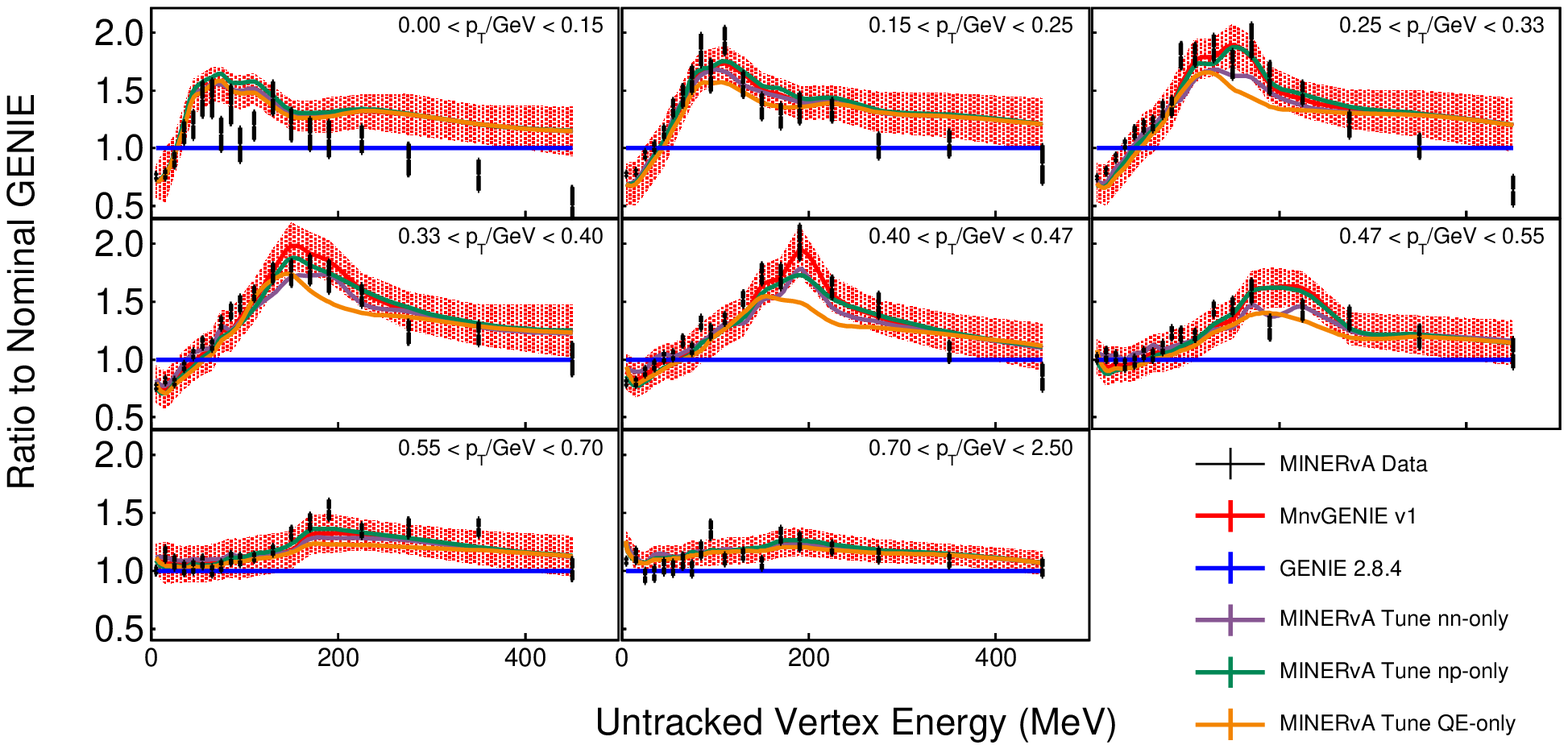}
  \includegraphics[width=0.95\linewidth]{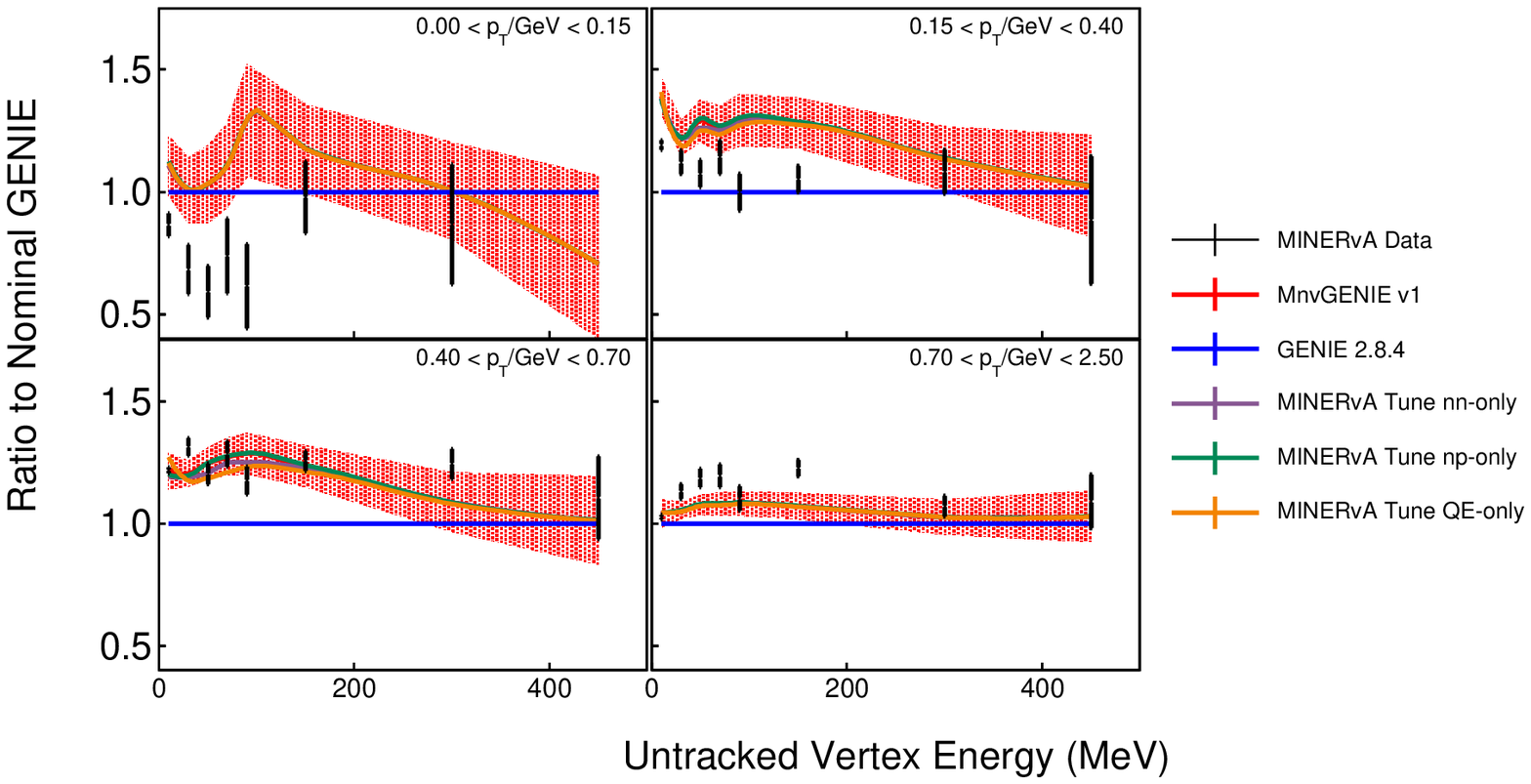}
\caption{Vertex energy in \pt~bins as compared to variants of the enhancement via 2p2h nn, 2p2h np, or QE-only.  The single track (top) and multi-track (bottom) samples are shown. These variations are almost indistinguishable for the multi-track subsample.}
\label{fig:vtx_bypt_enh}
\end{figure*}

The sample has enough events to further break these vertex energy distributions into bins of \pt, shown in Figs. \ref{fig:vtx_bypt_genie} and \ref{fig:vtx_bypt_enh}. Regions with noticeable differences between the simulation and data include low \pt with large vertex energy. Overall, the single track sample has a $\chi^2$ of 355 per 247 degrees of freedom, while the multi-track sample has a $\chi^2$ of 195 per 104 degrees of freedom, so both samples have significant disagreements with the \tune.

\begin{figure*}[p]
\centering
  \includegraphics[width=0.95\linewidth]{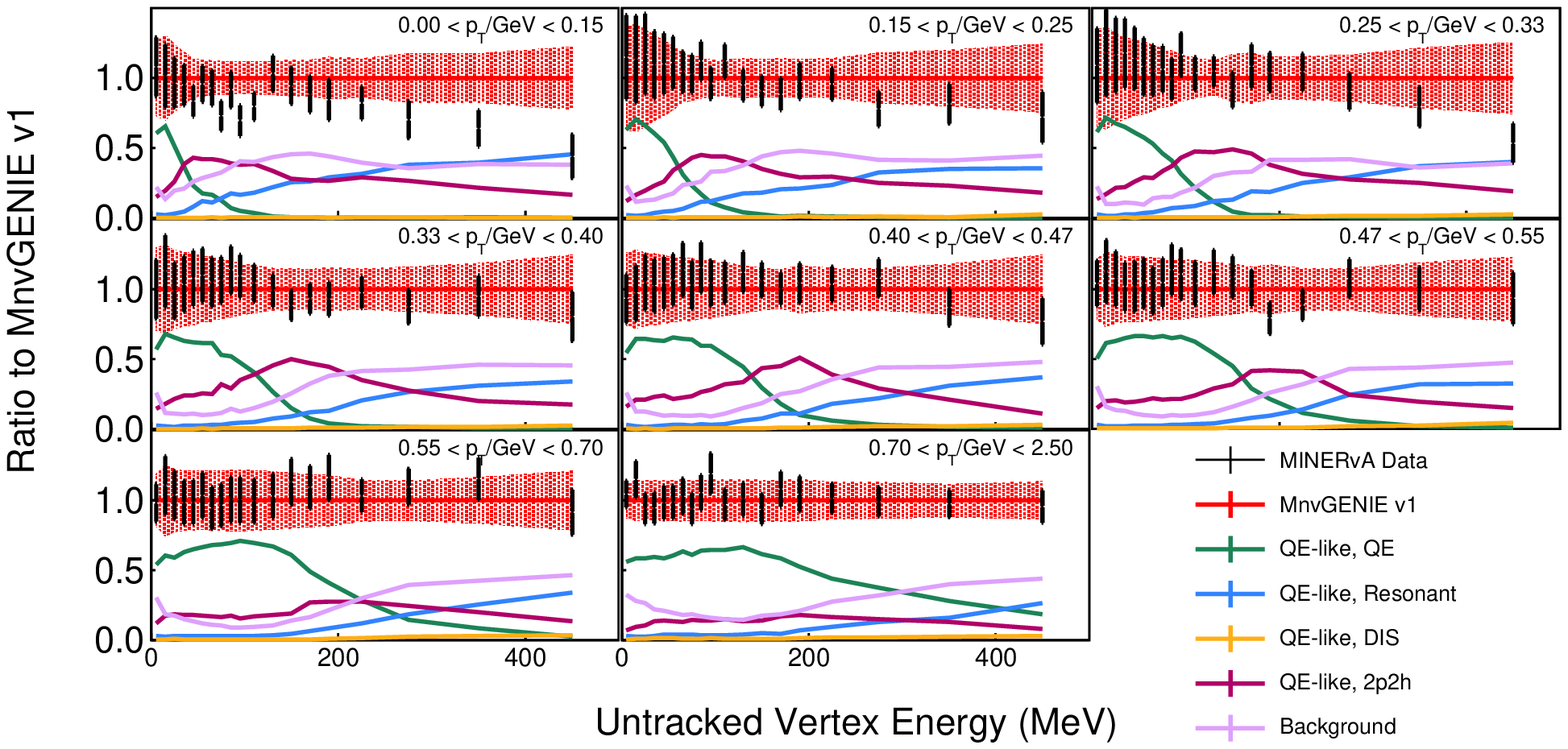}
  \includegraphics[width=0.95\linewidth]{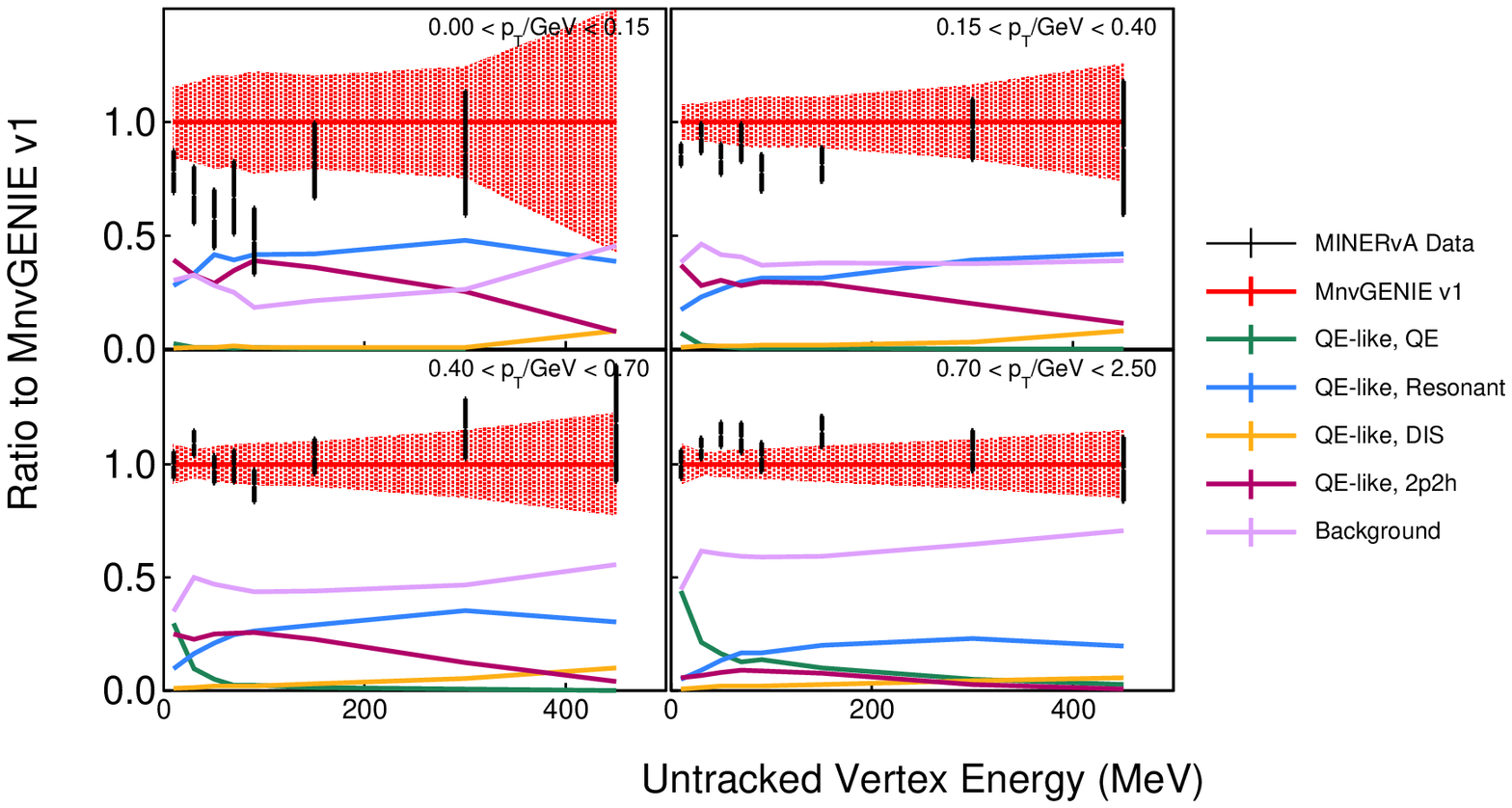}
\caption{The ratio of the data to the \tune~and the fractional event rate by process in the \tune. The single track (top) and multi-track (bottom) samples are shown.}
\label{fig:vtx_rate_ratio}
\end{figure*}

Events with no second track reconstructed and \pt$<$$0.4~GeV^2$ show a prediction of more events with large vertex energy than seen in data. This is also seen in the multi-track events at low vertex energy for \pt$<$$0.4~GeV^2$. The predicted fraction of the event rate by different signal and background processes is shown in Fig. \ref{fig:vtx_rate_ratio}. The regions of Monte Carlo excess correspond to regions of the vertex energy where resonant pion production contributes more to the signal.

%% file: Conclusions.tex
\section{Conclusions}

This analysis measures a double differential cross section with respect to the longitudinal and transverse momentum of the muon for \qelike~events. A suite of various additional processes and models were added to GENIE and are compared to the data.

The \tune~ models the data well except low and high \pt. A low \pt, and in turn low \qqqe, the addition of a low Q$^2$ suppression to resonant events would better replicate the data. At high \pt~a plausible explanation of the Monte Carlo data difference is the current model of the axial form factor does not work in this region.

Finally, a detailed look at the energy deposited near the interaction vertex shows very good agreement with the \tune~for overall vertex energy but deviates when separated into bins of \pt. These results are consistent with the previous \minerva~result\citep{Fiorentini:2013ezn}, and demonstrate that the enhancement of 2p2h processes provides a model for such additional low energy protons.

%% file: Acknowledgments.tex
\begin{acknowledgments}
This document was prepared by members of the MINERvA Collaboration using the resources of the Fermi National Accelerator Laboratory (Fermilab), a U.S. Department of Energy, Office of Science, HEP User Facility. Fermilab is managed by Fermi Research Alliance, LLC (FRA), acting under Contract No. DE-AC02-07CH11359.
These resources included support for the \minerva~construction project, and support
for construction also
was granted by the United States National Science Foundation under
Award No. PHY-0619727 and by the University of Rochester. Support for
participating scientists was provided by NSF and DOE (USA); by CAPES
and CNPq (Brazil); by CoNaCyT (Mexico); by Proyecto Basal FB 0821, CONICYT PIA ACT1413, Fondecyt 3170845 and 11130133 (Chile); by CONCYTEC, DGI-PUCP, and IDI/IGI-UNI (Peru); and by the Latin American Center for Physics (CLAF); NCN Opus Grant No. 2016/21/B/ST2/01092 (Poland).  We thank the MINOS Collaboration for use of its near detector data. Finally, we thank the staff of
Fermilab for support of the beam line, the detector, and computing infrastructure.

%
%
%
%
%
%
%
%
%
%
%

\end{acknowledgments}

%% file: Appendix.tex
\section{Appendix: Quasielastic Result}
\label{sec:appccqe}

A similar analysis was done with a different signal definition to provide a measurement for predictions which cannot produce a post-FSI signal. The model dependence of this result appears in the cross section modeling and FSI systematic uncertainties which are much larger than the \qelike~result. This result applies the same techniques as described in Sec. \ref{sec:extraction}. The signal definition for the result described in this section is: true quasielastic events, including 2p2h events, with a muon angle less than 20 degrees.

\begin{figure}[p]
\centering
  \includegraphics[width=0.95\linewidth]{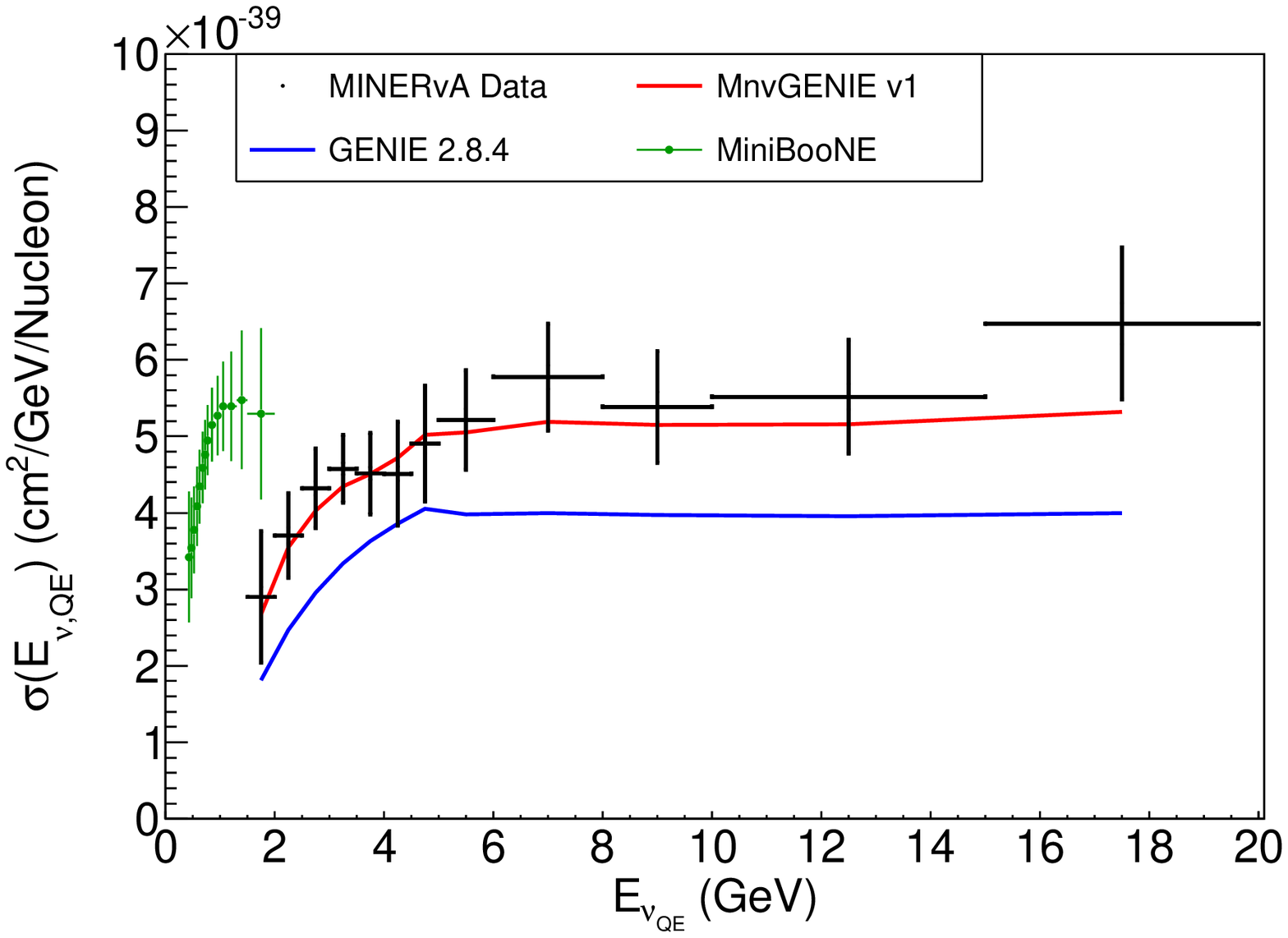}
  \includegraphics[width=0.95\linewidth]{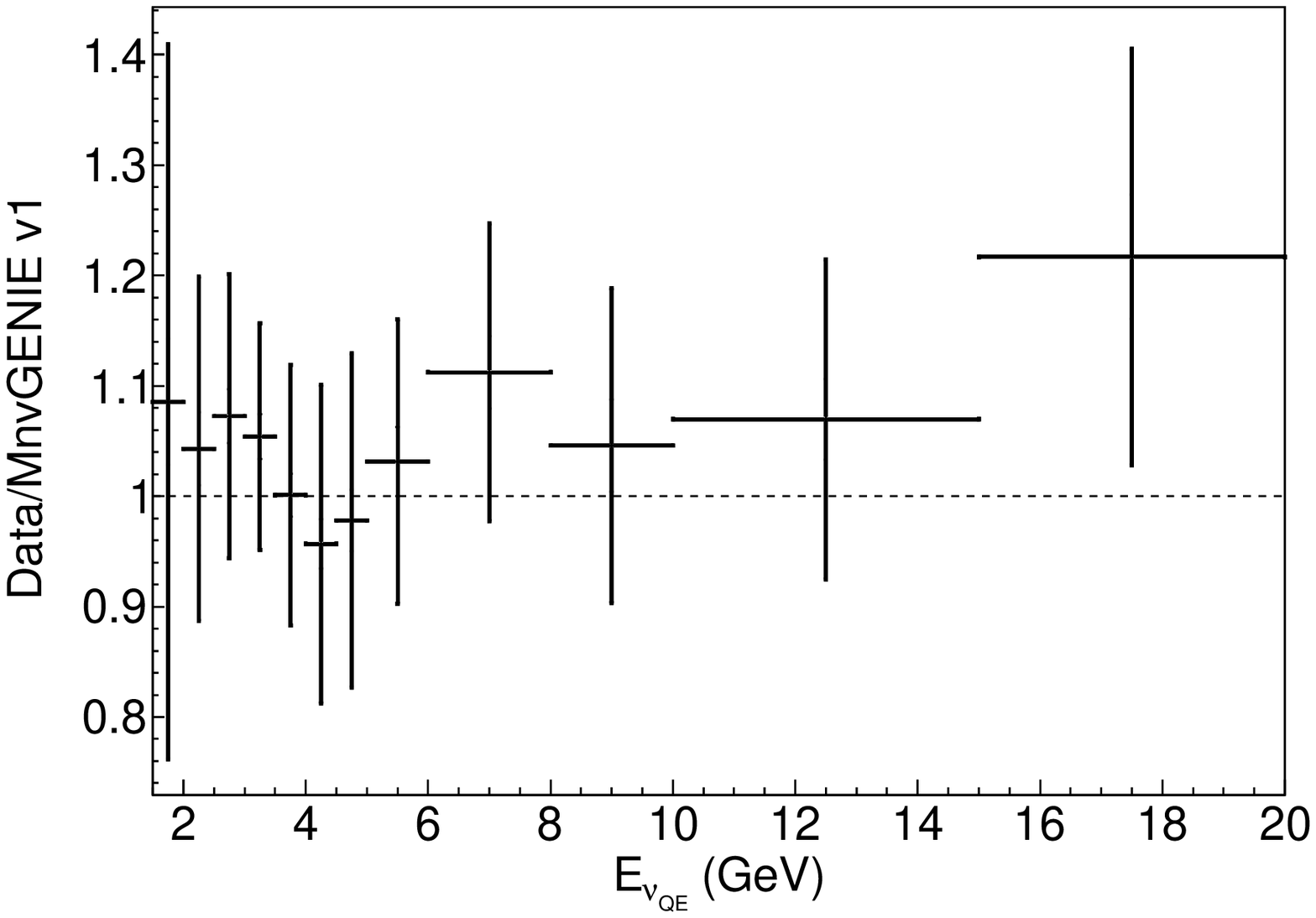}
\caption{The differential cross section as a function of \enuqe~with a quasielastic signal definition. In addition, results from the MiniBooNE measurement are included. }
\label{fig:qeonly_enuqe}
\end{figure}

\begin{figure}[p]
\centering
  \includegraphics[width=0.95\linewidth]{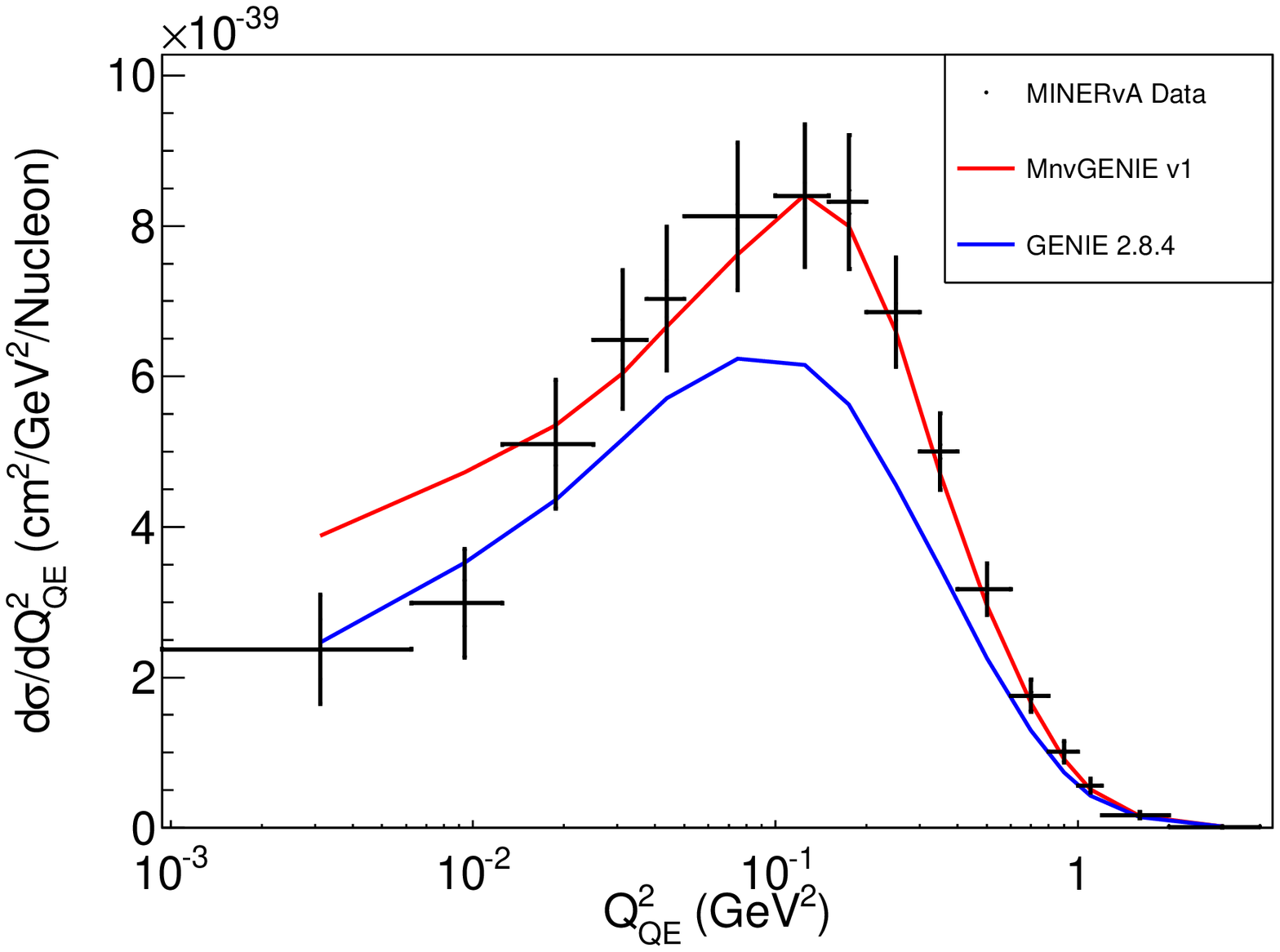}
  \includegraphics[width=0.95\linewidth]{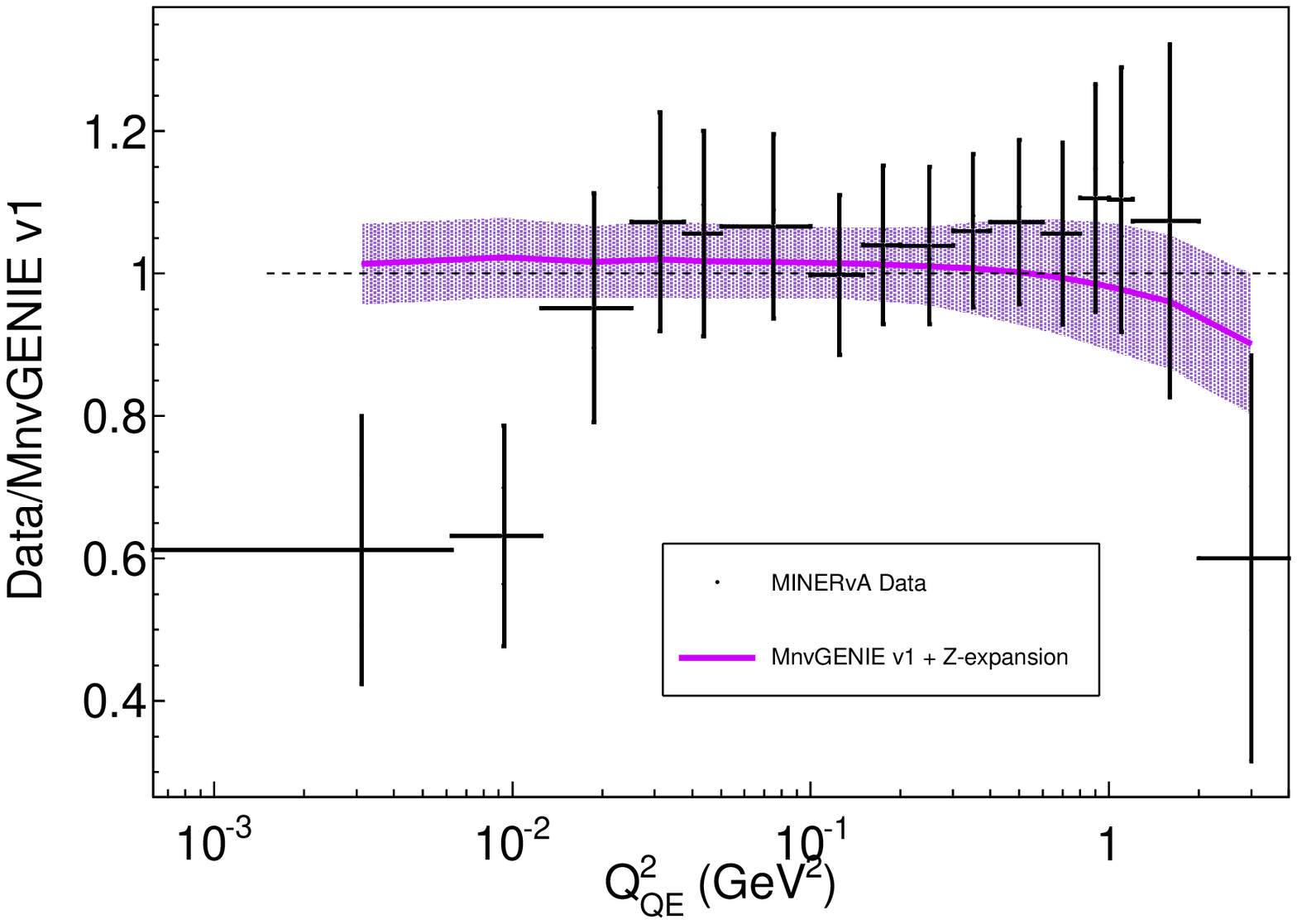}
\caption{The differential cross section as a function of \qqqe~with a quasielastic signal definition. In addition, the prediction for the Z-expansion with the model uncertainty is shown in the ratio. }
\label{fig:qeonly_q2qe}
\end{figure}

\begin{figure}[p]
\centering
  \includegraphics[width=0.95\linewidth]{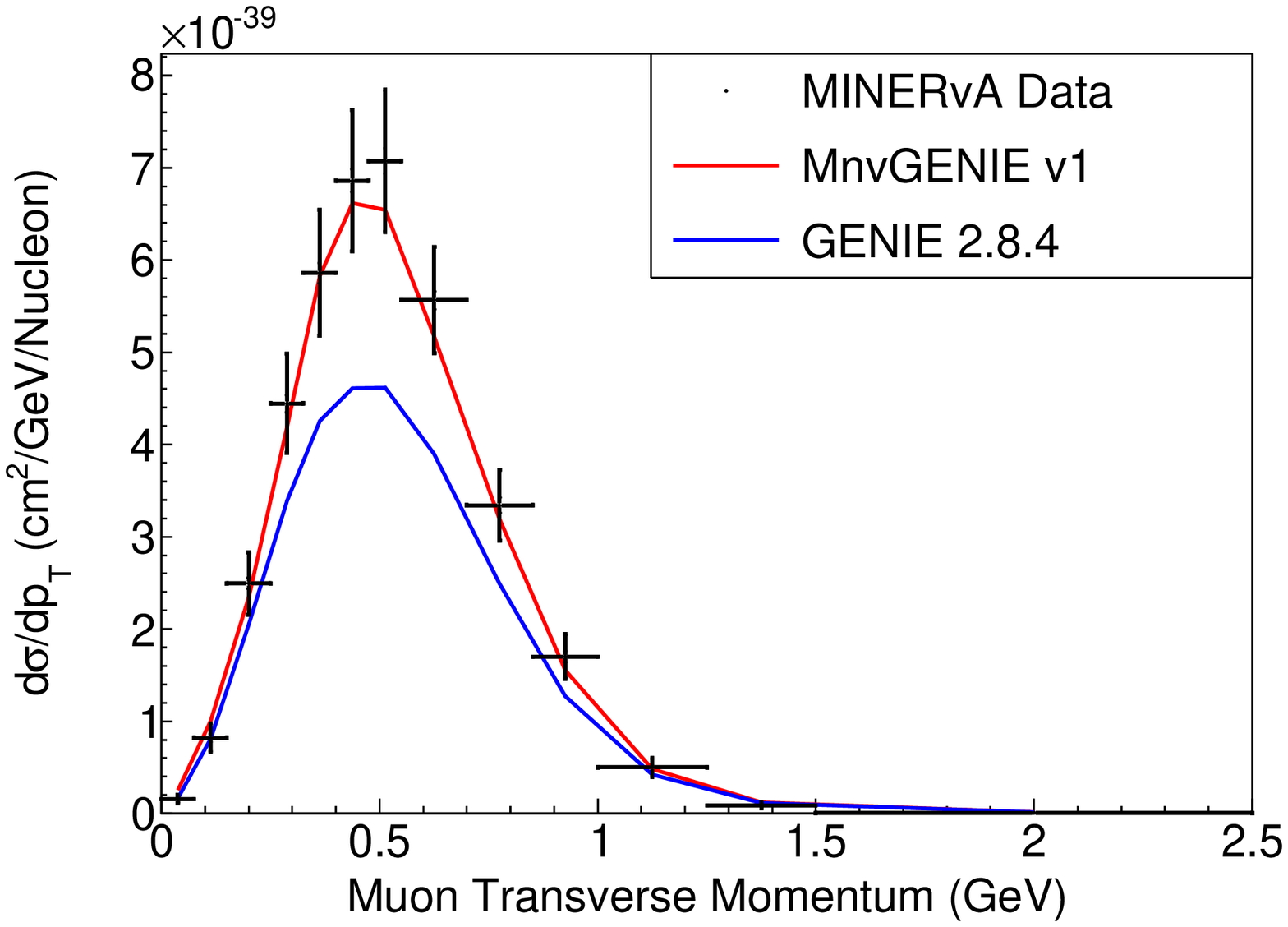}
  \includegraphics[width=0.95\linewidth]{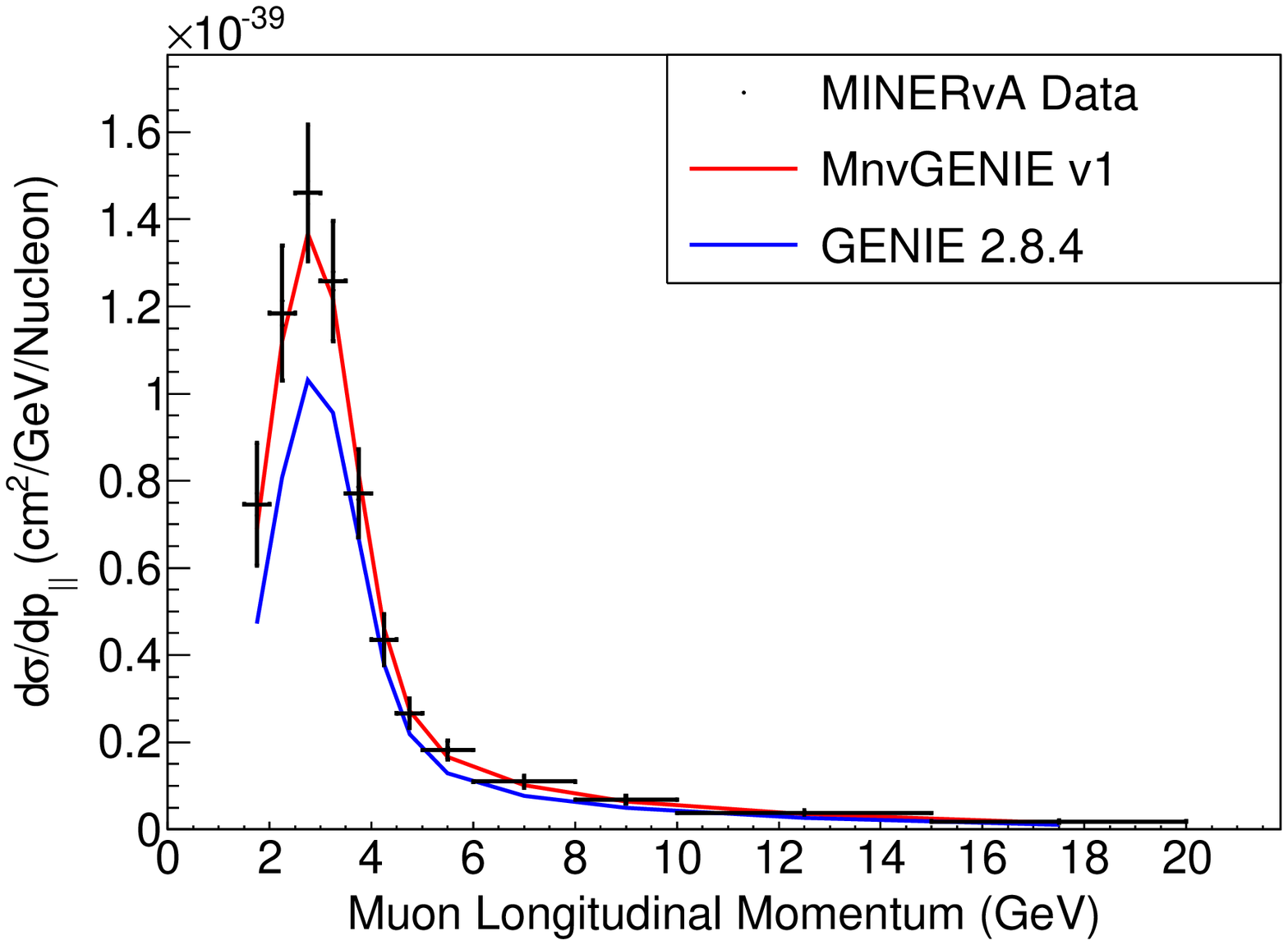}
\caption{The differential cross section as a function of \pt~and\pz~with a quasielastic signal definition.}
\label{fig:qeonly_ptqe}
\end{figure}

\begin{figure*}[p]
\centering
  \includegraphics[width=0.95\linewidth]{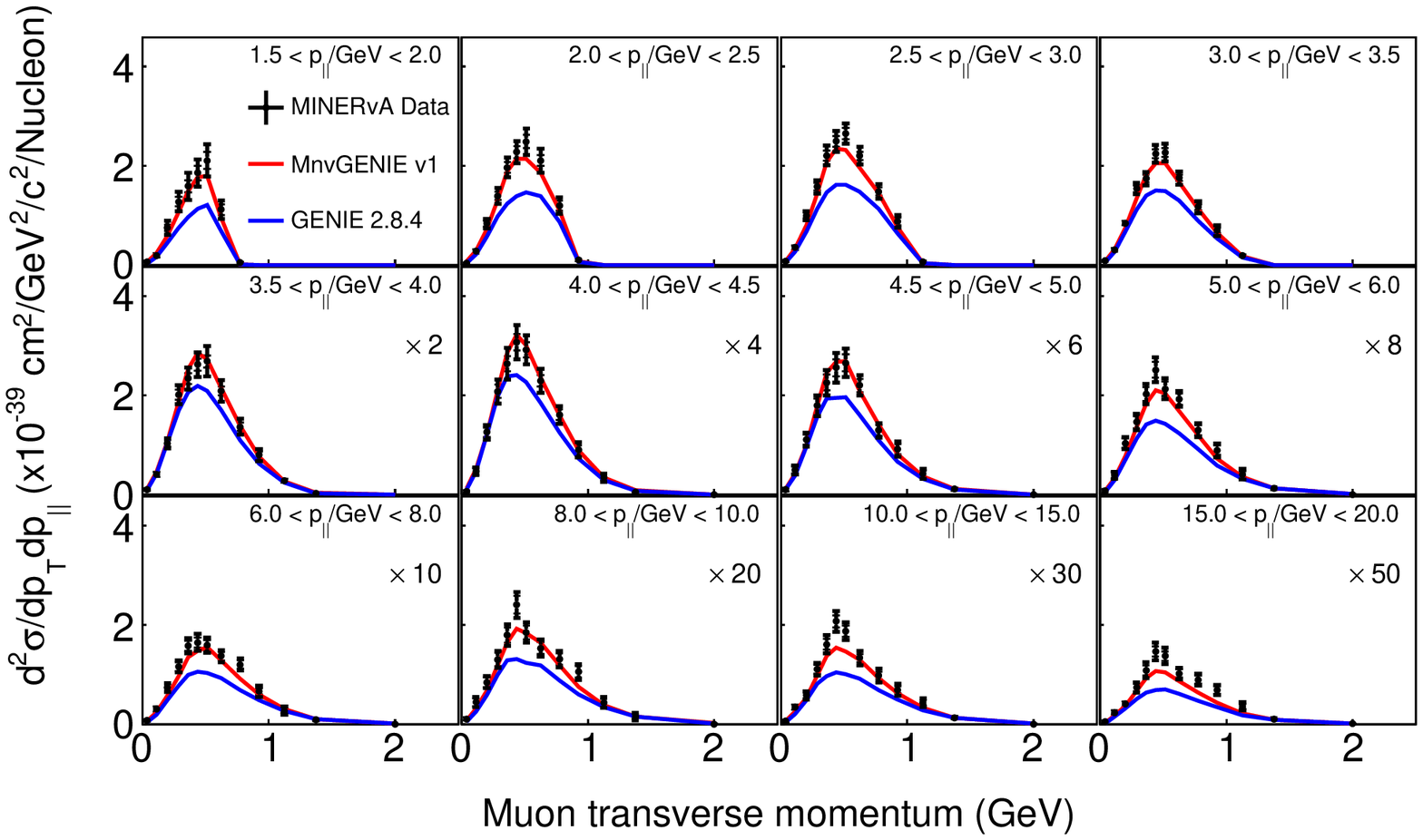}
  \includegraphics[width=0.95\linewidth]{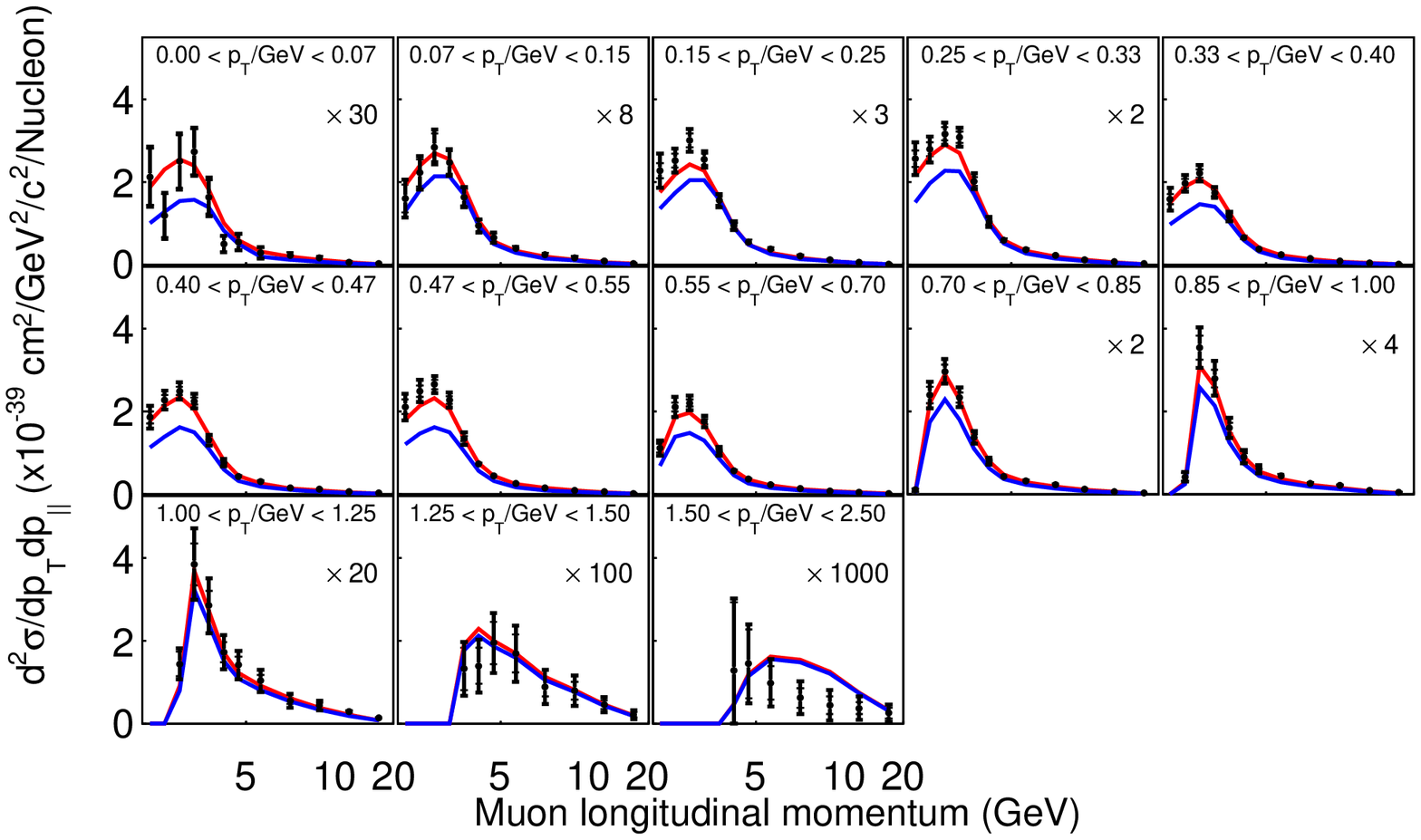}
\caption{Full double differential results, with a quasielastic signal definition, projected as a function of \pt~ and also a function of \pz.}
\label{fig:qeonly_2d}
\end{figure*}

\begin{figure*}[p]
\centering
  \includegraphics[width=0.95\linewidth]{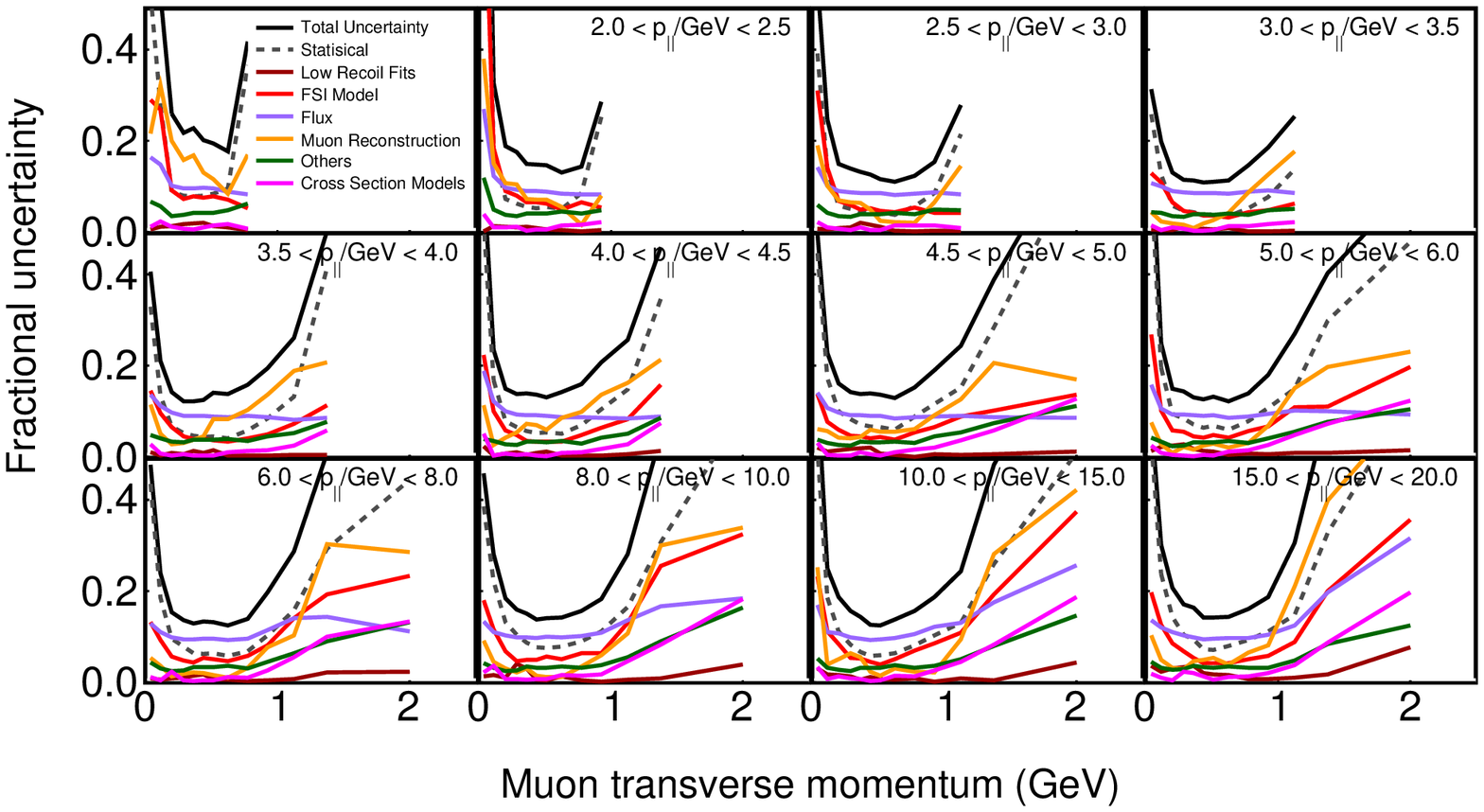}
\caption{Systematic uncertainties for the \pt\pz~result with a quasielastic signal definition.}
\label{fig:qeonly_2dsys}
\end{figure*}

\begin{figure}[p]
\centering
  \includegraphics[width=0.95\linewidth]{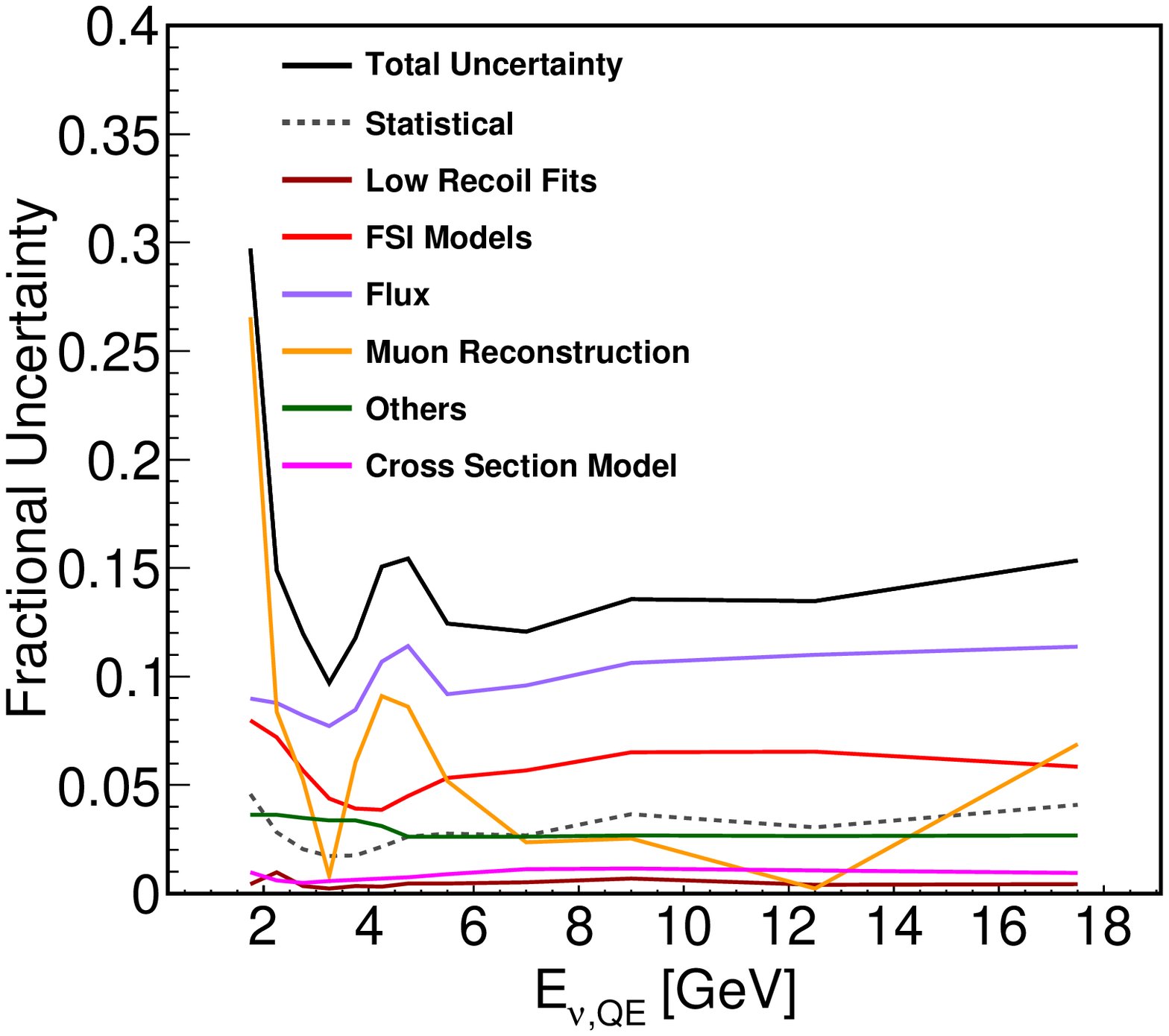}
\caption{Systematic uncertainties for the \enuqe~result with a quasielastic signal definition.}
\label{fig:qeonly_enuqesys}
\end{figure}

\begin{figure}[p]
\centering
  \includegraphics[width=0.95\linewidth]{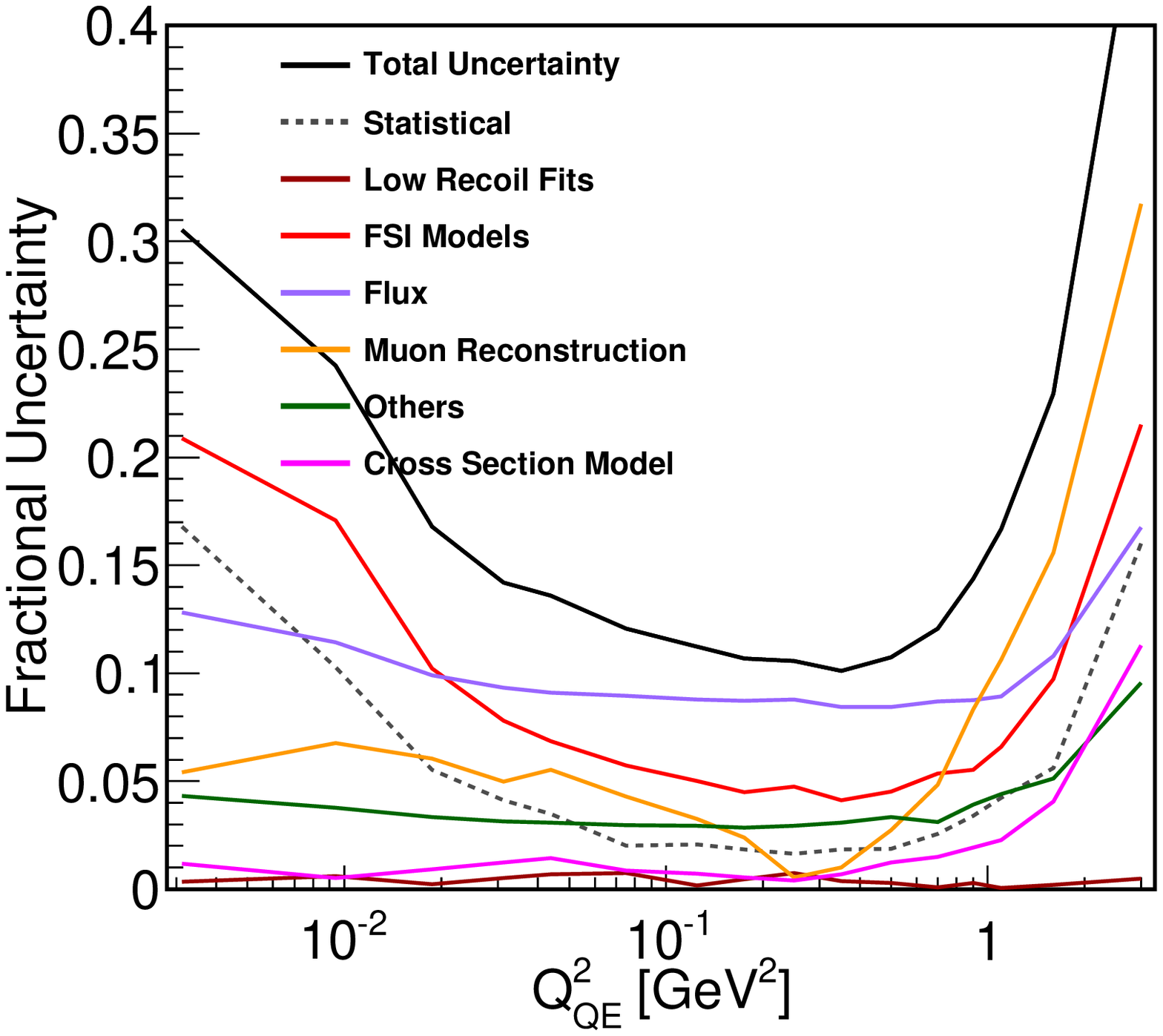}
\caption{Systematic uncertainties for the \qqqe~result with a quasielastic signal definition.}
\label{fig:qeonly_q2qesys}
\end{figure}

The \pt\pz~differential cross section is shown in Fig. \ref{fig:qeonly_2d}. Four single differential cross sections are shown in Figs. \ref{fig:qeonly_enuqe}(\enuqe), \ref{fig:qeonly_q2qe}(\qqqe), and \ref{fig:qeonly_ptqe}(\pt~and\pz). The systematic uncertainties for the \pt\pz, \enuqe, and \qqqe~results are shown in Figs. \ref{fig:qeonly_2dsys},\ref{fig:qeonly_enuqesys}, and \ref{fig:qeonly_q2qesys}.


%% file: Supplemental.tex
\section{Supplemental Material - Analysis Binning,Covariance Elements, and Data Release}
The kinematic bin boundaries are shown in Tab. \ref{tab:analysis_binning}. The bin number is used universally for all results in the the follow tables as well as the data release. These follow the ROOT convention of bin numbering where the first bin in a histogram is labeled as bin 1.

The covariance matrix follows the typical matrix element naming convention which starts counting at 0. For 1D distributions the mapping of bin to matrix element is just a difference of 1 (matrix element = bin number -1). For 2D distributions the mapping is more complex. 

The covariance matrix for the \pt\pl~result is constructed so the large cell is the $n^th$ \pt~bin while the small cells are the \pz~ bins. To determine the corresponding (x,y) bin in the \pt\pl~result the integer division and remainder are used. For instance, if i=5 in the covariance matrix this corresponds to a y bin (\pt) of 5/12 or 0 and a x bin (\pz) of 5 where 12 is the number of \pz~bins in the analysis.

The data release provides both ROOT files and text files. The ROOT files provide at TMatrixD of the covariance and TH1D or TH2D of the data result. The measured data values are also provided in text files labeled $data_result_<var>_<qe/qelike>.txt$ where $<var>$ is the variable and $<qe/qelike>$ is either the quasielastic-like or quasielastic result.The covariance matrix is also provided in text files labeled $cov_<stat+sysetmatics/stat. only>_<var>_<qe/qelike>.txt$ where statistical only covariance is provided as well as statistical plus systematics uncertainties and the other two labels are the same convention as the data result labeling. 

\begin{table}[h]
\centering
\begin{tabular}{| c | c | c | c | c |}
\hline
Bin & $p_{t}$ & $p_{||}$ & \qqqe & \enuqe  \\ \hline\hline
1&[0.000,0.075)&[1.5,2.0)&[0.00000,0.00625)&[1.5,2.0) \\ \hline
2&[0.075,0.150)&[2.0,2.5)&[0.00625,0.01250)&[2.0,2.5) \\ \hline
3&[0.150,0.250)&[2.5,3.0)&[0.01250,0.02500)&[2.5,3.0) \\ \hline
4&[0.250,0.325)&[3.0,3.5)&[0.02500,0.03750)&[3.0,3.5) \\ \hline
5&[0.325,0.400)&[3.5,4.0)&[0.03750,0.05000)&[3.5,4.0) \\ \hline
6&[0.400,0.475)&[4.0,4.5)&[0.05000,0.10000)&[4.0,4.5) \\ \hline
7&[0.475,0.550)&[4.5,5.0)&[0.10000,0.15000)&[4.5,5.0) \\ \hline
8&[0.550,0.700)&[5.0,6.0)&[0.15000,0.20000)&[5.0,6.0) \\ \hline
9&[0.700,0.850)&[6.0,8.0)&[0.20000,0.30000)&[6.0,8.0) \\ \hline
10&[0.850,1.000)&[8.0,10.0)&[0.30000,0.40000)&[8.0,10.0) \\ \hline
11&[1.000,1.250)&[10.0,15.0)&[0.40000,0.60000)&[10.0,15.0) \\ \hline
12&[1.250,1.500)&[15.0,20.0)&[0.60000,0.80000)&[15.0,20.0) \\ \hline
13&[1.500,2.500)& &[0.80000,1.00000)& \\ \hline
14& & &[1.00000,1.20000)& \\ \hline
15& & &[1.20000,2.00000)& \\ \hline
16& & &[2.00000,4.00000)& \\ \hline

\hline
\end{tabular}
\caption{Kinematic binning by bin number for all variables.}
\label{tab:analysis_binning}
\end{table}

\section{Supplemental Material - Quasielastic-like Result}
\subsection{Supplemental: Data Measurement}
This section contains the tabular breakdown of all the various cross section results as well as uncertainty on the measurement. In addition, tables of the model predictions are provided.
\begin{table}[h]
\centering
\begin{tabular}{| c | c | c | c |}
\hline

\hline
Bin & Cross Section & Stat. Unc. & Total. Unc. \\ \hline \hline
1&3.72e-39&1.39e-40&8.88e-40 \\ 
2&4.67e-39&1.08e-40&5.58e-40 \\ 
3&5.24e-39&9.09e-41&5.34e-40 \\ 
4&5.36e-39&7.98e-41&4.66e-40 \\ 
5&5.16e-39&7.94e-41&5.50e-40 \\ 
6&5.13e-39&9.70e-41&7.15e-40 \\ 
7&5.63e-39&1.29e-40&8.15e-40 \\ 
8&6.15e-39&1.46e-40&6.91e-40 \\ 
9&6.91e-39&1.59e-40&7.33e-40 \\ 
10&6.56e-39&2.03e-40&7.56e-40 \\ 
11&6.74e-39&1.74e-40&7.62e-40 \\ 
12&7.79e-39&2.75e-40&1.02e-39 \\ 

\hline
\end{tabular}
\caption{The measured differential cross section as a function of \enuqe. Units are cm$^2$ per GeV$^2$ per nucleon.}
\label{tab:1D_XProjection_CCQELike_CV_2p2h_enu_data}
\end{table}

\begin{table}[h]
\centering
\begin{tabular}{| c | c | c | c |}
\hline

\hline
Bin & Cross Section & Stat. Unc. & Total. Unc. \\ \hline \hline
1&9.51e-40&2.65e-41&1.35e-40 \\ 
2&1.46e-39&2.87e-41&1.45e-40 \\ 
3&1.73e-39&2.64e-41&1.49e-40 \\ 
4&1.45e-39&2.10e-41&1.32e-40 \\ 
5&8.85e-40&1.50e-41&9.63e-41 \\ 
6&5.03e-40&1.06e-41&5.62e-41 \\ 
7&3.14e-40&8.16e-42&3.22e-41 \\ 
8&2.16e-40&5.50e-42&2.01e-41 \\ 
9&1.32e-40&3.16e-42&1.25e-41 \\ 
10&8.29e-41&2.69e-42&8.25e-42 \\ 
11&4.67e-41&1.22e-42&4.59e-42 \\ 
12&2.18e-41&7.77e-43&2.62e-42 \\ 

\hline
\end{tabular}
\caption{The measured differential cross section as a function of $p_{||}$. Units are cm$^2$ per GeV$^2$ per nucleon.}
\label{tab:1D_XProjection_CCQELike_CV_2p2h_pzmu_data}
\end{table}

\begin{table}[h]
\centering
\begin{tabular}{| c | c | c | c |}
\hline

\hline
Bin & Cross Section & Stat. Unc. & Total. Unc. \\ \hline \hline
1&2.69e-40&5.87e-41&6.40e-41 \\ 
2&1.13e-39&1.06e-40&1.51e-40 \\ 
3&3.10e-39&1.30e-40&3.20e-40 \\ 
4&5.35e-39&1.73e-40&5.20e-40 \\ 
5&6.93e-39&2.09e-40&6.67e-40 \\ 
6&8.10e-39&1.75e-40&7.27e-40 \\ 
7&8.37e-39&1.54e-40&7.30e-40 \\ 
8&6.65e-39&1.00e-40&5.81e-40 \\ 
9&3.96e-39&8.30e-41&3.74e-40 \\ 
10&1.95e-39&5.83e-41&2.39e-40 \\ 
11&5.76e-40&2.74e-41&9.77e-41 \\ 
12&1.14e-40&1.25e-41&3.23e-41 \\ 
13&7.49e-42&1.91e-42 &3.25e-42\\

\hline
\end{tabular}
\caption{The measured differential cross section as a function of $p_{t}$. Units are cm$^2$ per GeV$^2$ per nucleon.}
\label{tab:1D_YProjection_CCQELike_CV_2p2h_pzmu_data}
\end{table}

\begin{table}[h]
\centering
\begin{tabular}{| c | c | c | c |}
\hline

\hline
Bin & Cross Section & Stat. Unc. & Total. Unc. \\ \hline \hline
1&4.28e-39&4.37e-40&5.80e-40 \\ 
2&4.69e-39&3.39e-40&5.64e-40 \\ 
3&6.84e-39&3.08e-40&7.33e-40 \\ 
4&8.40e-39&2.79e-40&8.36e-40 \\ 
5&8.91e-39&2.60e-40&9.02e-40 \\ 
6&9.95e-39&1.78e-40&9.25e-40 \\ 
7&1.01e-38&1.82e-40&8.99e-40 \\ 
8&9.84e-39&1.61e-40&8.59e-40 \\ 
9&8.14e-39&1.19e-40&6.84e-40 \\ 
10&5.97e-39&9.65e-41&5.12e-40 \\ 
11&3.81e-39&6.17e-41&3.45e-40 \\ 
12&2.08e-39&4.59e-41&2.05e-40 \\ 
13&1.17e-39&3.47e-41&1.42e-40 \\ 
14&6.42e-40&2.35e-41&8.78e-41 \\ 
15&1.93e-40&9.48e-42&3.53e-41 \\ 
16&1.42e-41&1.87e-42&4.79e-42 \\ 

\hline
\end{tabular}
\caption{The measured differential cross section as a function of \qqqe. Units are cm$^2$ per GeV$^2$ per nucleon.}
\label{tab:1D_XProjection_CCQELike_CV_2p2h_q2_data}
\end{table}

\begin{table*}[h]
\centering
\begin{tabular}{| c | c | c | c | c | c | c | c | c | c | c | c | c |}
\hline
& 1 & 2 & 3 & 4 & 5 & 6 & 7 & 8 & 9 & 10 & 11 & 12 \\ \hline \hline
1 & 7.50e-41 & 4.24e-41 & 8.83e-41 & 9.38e-41 & 5.69e-41 & 1.86e-41 & 1.98e-41 & 1.09e-41 & 8.40e-42 & 6.24e-42 & 3.87e-42 & 2.50e-42 \\ \hline
2 & 2.27e-40 & 3.20e-40 & 4.04e-40 & 3.48e-40 & 2.24e-40 & 1.34e-40 & 9.08e-41 & 5.34e-41 & 3.38e-41 & 2.41e-41 & 1.22e-41 & 5.03e-42 \\ \hline
3 & 8.75e-40 & 9.70e-40 & 1.14e-39 & 9.44e-40 & 5.81e-40 & 3.50e-40 & 2.05e-40 & 1.45e-40 & 8.36e-41 & 4.79e-41 & 2.29e-41 & 9.61e-42 \\ \hline
4 & 1.54e-39 & 1.66e-39 & 1.85e-39 & 1.74e-39 & 1.13e-39 & 5.86e-40 & 3.42e-40 & 2.08e-40 & 1.34e-40 & 7.63e-41 & 4.34e-41 & 1.71e-41 \\ \hline
5 & 1.94e-39 & 2.34e-39 & 2.54e-39 & 1.99e-39 & 1.32e-39 & 7.45e-40 & 4.32e-40 & 2.87e-40 & 1.79e-40 & 1.03e-40 & 6.13e-41 & 2.50e-41 \\ \hline
6 & 2.26e-39 & 2.69e-39 & 2.85e-39 & 2.51e-39 & 1.47e-39 & 8.51e-40 & 4.80e-40 & 3.52e-40 & 1.88e-40 & 1.34e-40 & 7.70e-41 & 3.27e-41 \\ \hline
7 & 2.56e-39 & 2.94e-39 & 3.04e-39 & 2.55e-39 & 1.50e-39 & 8.16e-40 & 4.97e-40 & 3.06e-40 & 1.88e-40 & 1.10e-40 & 7.20e-41 & 3.14e-41 \\
\hline

\hline
\end{tabular}
\caption{The measured double differential cross section as a function of $p_{t}$~(rows) versus $p_{||}$~(columns). Units are cm$^2$ per GeV$^2$ per nucleon.}
\label{tab:2D_CCQELike_CV_2p2h_data}
\end{table*}

\begin{table*}[h]
\centering
\begin{tabular}{| c | c | c | c | c | c | c | c | c | c | c | c | c |}
\hline
& 1 & 2 & 3 & 4 & 5 & 6 & 7 & 8 & 9 & 10 & 11 & 12 \\ \hline \hline
1 & 2.43e-41 & 1.78e-41 & 2.25e-41 & 1.97e-41 & 1.51e-41 & 6.65e-42 & 6.86e-42 & 4.24e-42 & 2.69e-42 & 1.89e-42 & 1.19e-42 & 7.33e-43 \\ \hline
2 & 6.27e-41 & 5.90e-41 & 6.02e-41 & 4.51e-41 & 3.24e-41 & 2.15e-41 & 1.61e-41 & 9.71e-42 & 5.77e-42 & 4.93e-42 & 2.09e-42 & 1.00e-42 \\ \hline
3 & 1.69e-40 & 1.27e-40 & 1.26e-40 & 9.06e-41 & 6.11e-41 & 4.00e-41 & 2.54e-41 & 1.71e-41 & 1.01e-41 & 6.91e-42 & 3.12e-42 & 1.44e-42 \\ \hline
4 & 2.62e-40 & 2.15e-40 & 2.00e-40 & 1.61e-40 & 1.13e-40 & 6.58e-41 & 4.02e-41 & 2.40e-41 & 1.49e-41 & 9.92e-42 & 5.50e-42 & 2.32e-42 \\ \hline
5 & 3.44e-40 & 2.67e-40 & 2.76e-40 & 1.85e-40 & 1.34e-40 & 8.89e-41 & 4.94e-41 & 3.13e-41 & 1.91e-41 & 1.27e-41 & 7.39e-42 & 3.31e-42 \\ \hline
6 & 3.49e-40 & 3.04e-40 & 2.88e-40 & 2.33e-40 & 1.53e-40 & 9.99e-41 & 5.76e-41 & 3.71e-41 & 2.02e-41 & 1.50e-41 & 8.12e-42 & 3.93e-42 \\ \hline
7 & 3.82e-40 & 3.34e-40 & 2.87e-40 & 2.43e-40 & 1.82e-40 & 9.14e-41 & 5.72e-41 & 3.29e-41 & 1.96e-41 & 1.24e-41 & 7.53e-42 & 3.83e-42 \\
\hline

\hline
\end{tabular}
\caption{The measured double differential cross section total uncertainty as a function of $p_{t}$~(rows) versus $p_{||}$~(columns). Units are cm$^2$ per GeV$^2$ per nucleon.}
\label{tab:Unc_2D_CCQELike_CV_2p2h_data}
\end{table*}

\clearpage
\subsection{Supplemental: Monte Carlo Predictions}

\begin{table}[h]
\centering
\begin{tabular}{| c | c |}
\hline

\hline
Bin & Cross Section \\ \hline \hline
1&3.44e-39 \\ 
2&4.48e-39 \\ 
3&4.91e-39 \\ 
4&5.10e-39 \\ 
5&5.13e-39 \\ 
6&5.32e-39 \\ 
7&5.71e-39 \\ 
8&5.95e-39 \\ 
9&6.28e-39 \\ 
10&6.29e-39 \\ 
11&6.37e-39 \\ 
12&6.63e-39 \\ 

\hline
\end{tabular}
\caption{Monte Carlo predicted differential cross section as a function of \enuqe. Units are cm$^2$ per GeV$^2$ per nucleon. This is a prediction for \tune.}
\label{tab:1D_XProjection_CCQELike_CV_2p2h_enu_mc}
\end{table}

\begin{table}[h]
\centering
\begin{tabular}{| c | c |}
\hline

\hline
Bin & Cross Section \\ \hline \hline
1&3.32e-39 \\ 
2&4.36e-39 \\ 
3&4.81e-39 \\ 
4&5.01e-39 \\ 
5&5.06e-39 \\ 
6&5.26e-39 \\ 
7&5.64e-39 \\ 
8&5.87e-39 \\ 
9&6.19e-39 \\ 
10&6.19e-39 \\ 
11&6.27e-39 \\ 
12&6.52e-39 \\ 

\hline
\end{tabular}
\caption{Monte Carlo predicted differential cross section as a function of \enuqe. Units are cm$^2$ per GeV$^2$ per nucleon. This is a prediction for \tune with the addition of the MINOS emperical low $Q^{2}$ suppression.}
\label{tab:1D_XProjection_CCQELike_CV_RPA_Res_MINOS_2p2h_enu_mc}
\end{table}

\begin{table}[h]
\centering
\begin{tabular}{| c | c |}
\hline

\hline
Bin & Cross Section \\ \hline \hline
1&3.48e-39 \\ 
2&4.52e-39 \\ 
3&4.95e-39 \\ 
4&5.12e-39 \\ 
5&5.15e-39 \\ 
6&5.33e-39 \\ 
7&5.72e-39 \\ 
8&5.95e-39 \\ 
9&6.27e-39 \\ 
10&6.28e-39 \\ 
11&6.36e-39 \\ 
12&6.62e-39 \\ 

\hline
\end{tabular}
\caption{Monte Carlo predicted differential cross section as a function of \enuqe. Units are cm$^2$ per GeV$^2$ per nucleon. This is a prediction for \tune reweighted to the Z-expansion prediction.}
\label{tab:1D_XProjection_CCQELike_CV_ZExp_2p2h_enu_mc}
\end{table}

\begin{table}[h]
\centering
\begin{tabular}{| c | c |}
\hline

\hline
Bin & Cross Section \\ \hline \hline
1&2.68e-39 \\ 
2&3.52e-39 \\ 
3&3.96e-39 \\ 
4&4.20e-39 \\ 
5&4.34e-39 \\ 
6&4.56e-39 \\ 
7&4.87e-39 \\ 
8&5.05e-39 \\ 
9&5.28e-39 \\ 
10&5.32e-39 \\ 
11&5.39e-39 \\ 
12&5.54e-39 \\ 

\hline
\end{tabular}
\caption{Monte Carlo predicted differential cross section as a function of \enuqe. Units are cm$^2$ per GeV$^2$ per nucleon. This is a prediction for GENIE 2.8.4.}
\label{tab:1D_XProjection_CCQELike_default_enu_mc}
\end{table}

\begin{table}[h]
\centering
\begin{tabular}{| c | c |}
\hline

\hline
Bin & Cross Section \\ \hline \hline
1&3.94e-39 \\ 
2&5.01e-39 \\ 
3&5.41e-39 \\ 
4&5.55e-39 \\ 
5&5.57e-39 \\ 
6&5.73e-39 \\ 
7&6.11e-39 \\ 
8&6.34e-39 \\ 
9&6.67e-39 \\ 
10&6.69e-39 \\ 
11&6.76e-39 \\ 
12&7.03e-39 \\ 

\hline
\end{tabular}
\caption{Monte Carlo predicted differential cross section as a function of \enuqe. Units are cm$^2$ per GeV$^2$ per nucleon. This is a prediction for GENIE 2.8.4 with the addition of Valencia 2p2h, and the non-resonant pion production reduction.}
\label{tab:1D_XProjection_CCQELike_pion_2p2h_2p2h_enu_mc}
\end{table}

\begin{table}[h]
\centering
\begin{tabular}{| c | c |}
\hline

\hline
Bin & Cross Section \\ \hline \hline
1&2.96e-39 \\ 
2&3.93e-39 \\ 
3&4.40e-39 \\ 
4&4.59e-39 \\ 
5&4.65e-39 \\ 
6&4.87e-39 \\ 
7&5.22e-39 \\ 
8&5.46e-39 \\ 
9&5.76e-39 \\ 
10&5.79e-39 \\ 
11&5.86e-39 \\ 
12&6.06e-39 \\ 

\hline
\end{tabular}
\caption{Monte Carlo predicted differential cross section as a function of \enuqe. Units are cm$^2$ per GeV$^2$ per nucleon. This is a prediction for GENIE 2.8.4 with the addition of RPA, Valencia 2p2h, and the non-resonant pion production reduction.}
\label{tab:1D_XProjection_CCQELike_pion_rpa_2p2h_enu_mc}
\end{table}

\begin{table}[h]
\centering
\begin{tabular}{| c | c |}
\hline

\hline
Bin & Cross Section \\ \hline \hline
1&2.10e-39 \\ 
2&2.90e-39 \\ 
3&3.37e-39 \\ 
4&3.65e-39 \\ 
5&3.82e-39 \\ 
6&4.06e-39 \\ 
7&4.36e-39 \\ 
8&4.51e-39 \\ 
9&4.72e-39 \\ 
10&4.74e-39 \\ 
11&4.80e-39 \\ 
12&4.95e-39 \\ 

\hline
\end{tabular}
\caption{Monte Carlo predicted differential cross section as a function of \enuqe. Units are cm$^2$ per GeV$^2$ per nucleon. This is a prediction for GENIE 2.8.4 with the addition of RPA and the non-resonant pion production reduction.}
\label{tab:1D_XProjection_CCQELike_pion_rpa_enu_mc}
\end{table}

\begin{table}[h]
\centering
\begin{tabular}{| c | c |}
\hline

\hline
Bin & Cross Section \\ \hline \hline
1&3.46e-39 \\ 
2&4.45e-39 \\ 
3&4.89e-39 \\ 
4&5.05e-39 \\ 
5&5.09e-39 \\ 
6&5.28e-39 \\ 
7&5.62e-39 \\ 
8&5.85e-39 \\ 
9&6.15e-39 \\ 
10&6.19e-39 \\ 
11&6.25e-39 \\ 
12&6.45e-39 \\ 

\hline
\end{tabular}
\caption{Monte Carlo predicted differential cross section as a function of \enuqe. Units are cm$^2$ per GeV$^2$ per nucleon. This is a prediction for GENIE 2.8.4 with the addition of Valencia 2p2h.}
\label{tab:1D_XProjection_CCQELike_piontune_2p2h_enu_mc}
\end{table}

\begin{table}[h]
\centering
\begin{tabular}{| c | c |}
\hline

\hline
Bin & Cross Section \\ \hline \hline
1&2.60e-39 \\ 
2&3.42e-39 \\ 
3&3.86e-39 \\ 
4&4.11e-39 \\ 
5&4.26e-39 \\ 
6&4.47e-39 \\ 
7&4.76e-39 \\ 
8&4.90e-39 \\ 
9&5.11e-39 \\ 
10&5.14e-39 \\ 
11&5.19e-39 \\ 
12&5.34e-39 \\ 

\hline
\end{tabular}
\caption{Monte Carlo predicted differential cross section as a function of \enuqe. Units are cm$^2$ per GeV$^2$ per nucleon. This is a prediction for GENIE 2.8.4 with the non-resonant pion production reduction.}
\label{tab:1D_XProjection_CCQELike_piontune_enu_mc}
\end{table}

\begin{table}[h]
\centering
\begin{tabular}{| c | c |}
\hline

\hline
Bin & Cross Section \\ \hline \hline
1&8.81e-40 \\ 
2&1.39e-39 \\ 
3&1.62e-39 \\ 
4&1.40e-39 \\ 
5&9.24e-40 \\ 
6&5.29e-40 \\ 
7&3.19e-40 \\ 
8&2.00e-40 \\ 
9&1.25e-40 \\ 
10&7.90e-41 \\ 
11&4.33e-41 \\ 
12&1.80e-41 \\ 

\hline
\end{tabular}
\caption{Monte Carlo predicted differential cross section as a function of $p_{||}$. Units are cm$^2$ per GeV$^2$ per nucleon. This is a prediction for \tune.}
\label{tab:1D_XProjection_CCQELike_CV_2p2h_pzmu_mc}
\end{table}

\begin{table}[h]
\centering
\begin{tabular}{| c | c |}
\hline

\hline
Bin & Cross Section \\ \hline \hline
1&8.57e-40 \\ 
2&1.36e-39 \\ 
3&1.59e-39 \\ 
4&1.38e-39 \\ 
5&9.10e-40 \\ 
6&5.21e-40 \\ 
7&3.14e-40 \\ 
8&1.97e-40 \\ 
9&1.23e-40 \\ 
10&7.77e-41 \\ 
11&4.25e-41 \\ 
12&1.76e-41 \\ 

\hline
\end{tabular}
\caption{Monte Carlo predicted differential cross section as a function of $p_{||}$. Units are cm$^2$ per GeV$^2$ per nucleon. This is a prediction for \tune with the addition of the MINOS emperical low $Q^{2}$ suppression.}
\label{tab:1D_XProjection_CCQELike_CV_RPA_Res_MINOS_2p2h_pzmu_mc}
\end{table}

\begin{table}[h]
\centering
\begin{tabular}{| c | c |}
\hline

\hline
Bin & Cross Section \\ \hline \hline
1&8.88e-40 \\ 
2&1.39e-39 \\ 
3&1.63e-39 \\ 
4&1.40e-39 \\ 
5&9.29e-40 \\ 
6&5.31e-40 \\ 
7&3.20e-40 \\ 
8&2.00e-40 \\ 
9&1.25e-40 \\ 
10&7.89e-41 \\ 
11&4.32e-41 \\ 
12&1.79e-41 \\ 

\hline
\end{tabular}
\caption{Monte Carlo predicted differential cross section as a function of $p_{||}$. Units are cm$^2$ per GeV$^2$ per nucleon. This is a prediction for \tune reweighted to the Z-expansion prediction.}
\label{tab:1D_XProjection_CCQELike_CV_ZExp_2p2h_pzmu_mc}
\end{table}

\begin{table}[h]
\centering
\begin{tabular}{| c | c |}
\hline

\hline
Bin & Cross Section \\ \hline \hline
1&6.91e-40 \\ 
2&1.11e-39 \\ 
3&1.32e-39 \\ 
4&1.16e-39 \\ 
5&7.92e-40 \\ 
6&4.56e-40 \\ 
7&2.72e-40 \\ 
8&1.69e-40 \\ 
9&1.04e-40 \\ 
10&6.66e-41 \\ 
11&3.64e-41 \\ 
12&1.49e-41 \\ 

\hline
\end{tabular}
\caption{Monte Carlo predicted differential cross section as a function of $p_{||}$. Units are cm$^2$ per GeV$^2$ per nucleon. This is a prediction for GENIE 2.8.4.}
\label{tab:1D_XProjection_CCQELike_default_pzmu_mc}
\end{table}

\begin{table}[h]
\centering
\begin{tabular}{| c | c |}
\hline

\hline
Bin & Cross Section \\ \hline \hline
1&9.82e-40 \\ 
2&1.51e-39 \\ 
3&1.75e-39 \\ 
4&1.52e-39 \\ 
5&1.02e-39 \\ 
6&5.86e-40 \\ 
7&3.49e-40 \\ 
8&2.16e-40 \\ 
9&1.33e-40 \\ 
10&8.45e-41 \\ 
11&4.62e-41 \\ 
12&1.91e-41 \\ 

\hline
\end{tabular}
\caption{Monte Carlo predicted differential cross section as a function of $p_{||}$. Units are cm$^2$ per GeV$^2$ per nucleon. This is a prediction for GENIE 2.8.4 with the addition of Valencia 2p2h, and the non-resonant pion production reduction.}
\label{tab:1D_XProjection_CCQELike_pion_2p2h_2p2h_pzmu_mc}
\end{table}

\begin{table}[h]
\centering
\begin{tabular}{| c | c |}
\hline

\hline
Bin & Cross Section \\ \hline \hline
1&7.74e-40 \\ 
2&1.25e-39 \\ 
3&1.47e-39 \\ 
4&1.26e-39 \\ 
5&8.32e-40 \\ 
6&4.76e-40 \\ 
7&2.88e-40 \\ 
8&1.82e-40 \\ 
9&1.14e-40 \\ 
10&7.25e-41 \\ 
11&3.96e-41 \\ 
12&1.63e-41 \\ 

\hline
\end{tabular}
\caption{Monte Carlo predicted differential cross section as a function of $p_{||}$. Units are cm$^2$ per GeV$^2$ per nucleon. This is a prediction for GENIE 2.8.4 with the addition of RPA, Valencia 2p2h, and the non-resonant pion production reduction.}
\label{tab:1D_XProjection_CCQELike_pion_rpa_2p2h_pzmu_mc}
\end{table}

\begin{table}[h]
\centering
\begin{tabular}{| c | c |}
\hline

\hline
Bin & Cross Section \\ \hline \hline
1&5.68e-40 \\ 
2&9.61e-40 \\ 
3&1.17e-39 \\ 
4&1.01e-39 \\ 
5&6.79e-40 \\ 
6&3.90e-40 \\ 
7&2.35e-40 \\ 
8&1.48e-40 \\ 
9&9.23e-41 \\ 
10&5.89e-41 \\ 
11&3.22e-41 \\ 
12&1.32e-41 \\ 

\hline
\end{tabular}
\caption{Monte Carlo predicted differential cross section as a function of $p_{||}$. Units are cm$^2$ per GeV$^2$ per nucleon. This is a prediction for GENIE 2.8.4 with the addition of RPA and the non-resonant pion production reduction.}
\label{tab:1D_XProjection_CCQELike_pion_rpa_pzmu_mc}
\end{table}

\begin{table}[h]
\centering
\begin{tabular}{| c | c |}
\hline

\hline
Bin & Cross Section \\ \hline \hline
1&8.74e-40 \\ 
2&1.37e-39 \\ 
3&1.60e-39 \\ 
4&1.39e-39 \\ 
5&9.31e-40 \\ 
6&5.33e-40 \\ 
7&3.18e-40 \\ 
8&1.98e-40 \\ 
9&1.22e-40 \\ 
10&7.80e-41 \\ 
11&4.25e-41 \\ 
12&1.75e-41 \\ 

\hline
\end{tabular}
\caption{Monte Carlo predicted differential cross section as a function of $p_{||}$. Units are cm$^2$ per GeV$^2$ per nucleon. This is a prediction for GENIE 2.8.4 with the addition of Valencia 2p2h.}
\label{tab:1D_XProjection_CCQELike_piontune_2p2h_pzmu_mc}
\end{table}

\begin{table}[h]
\centering
\begin{tabular}{| c | c |}
\hline

\hline
Bin & Cross Section \\ \hline \hline
1&6.69e-40 \\ 
2&1.08e-39 \\ 
3&1.29e-39 \\ 
4&1.14e-39 \\ 
5&7.78e-40 \\ 
6&4.47e-40 \\ 
7&2.65e-40 \\ 
8&1.64e-40 \\ 
9&1.01e-40 \\ 
10&6.44e-41 \\ 
11&3.51e-41 \\ 
12&1.44e-41 \\ 

\hline
\end{tabular}
\caption{Monte Carlo predicted differential cross section as a function of $p_{||}$. Units are cm$^2$ per GeV$^2$ per nucleon. This is a prediction for GENIE 2.8.4 with the non-resonant pion production reduction.}
\label{tab:1D_XProjection_CCQELike_piontune_pzmu_mc}
\end{table}

\begin{table}[h]
\centering
\begin{tabular}{| c | c |}
\hline

\hline
Bin & Cross Section \\ \hline \hline
1&3.71e-40 \\ 
2&1.31e-39 \\ 
3&2.93e-39 \\ 
4&5.05e-39 \\ 
5&6.86e-39 \\ 
6&7.80e-39 \\ 
7&7.78e-39 \\ 
8&6.21e-39 \\ 
9&3.78e-39 \\ 
10&1.80e-39 \\ 
11&5.56e-40 \\ 
12&1.48e-40 \\ 

\hline
\end{tabular}
\caption{Monte Carlo predicted differential cross section as a function of $p_{t}$. Units are cm$^2$ per GeV$^2$ per nucleon. This is a prediction for \tune.}
\label{tab:1D_YProjection_CCQELike_CV_2p2h_pzmu_mc}
\end{table}

\begin{table}[h]
\centering
\begin{tabular}{| c | c |}
\hline

\hline
Bin & Cross Section \\ \hline \hline
1&3.08e-40 \\ 
2&1.15e-39 \\ 
3&2.71e-39 \\ 
4&4.84e-39 \\ 
5&6.71e-39 \\ 
6&7.70e-39 \\ 
7&7.72e-39 \\ 
8&6.19e-39 \\ 
9&3.78e-39 \\ 
10&1.80e-39 \\ 
11&5.56e-40 \\ 
12&1.48e-40 \\ 

\hline
\end{tabular}
\caption{Monte Carlo predicted differential cross section as a function of $p_{t}$. Units are cm$^2$ per GeV$^2$ per nucleon. This is a prediction for \tune with the addition of the MINOS emperical low $Q^{2}$ suppression.}
\label{tab:1D_YProjection_CCQELike_CV_RPA_Res_MINOS_2p2h_pzmu_mc}
\end{table}

\begin{table}[h]
\centering
\begin{tabular}{| c | c |}
\hline

\hline
Bin & Cross Section \\ \hline \hline
1&3.74e-40 \\ 
2&1.33e-39 \\ 
3&2.97e-39 \\ 
4&5.12e-39 \\ 
5&6.95e-39 \\ 
6&7.88e-39 \\ 
7&7.84e-39 \\ 
8&6.23e-39 \\ 
9&3.77e-39 \\ 
10&1.77e-39 \\ 
11&5.40e-40 \\ 
12&1.40e-40 \\ 

\hline
\end{tabular}
\caption{Monte Carlo predicted differential cross section as a function of $p_{t}$. Units are cm$^2$ per GeV$^2$ per nucleon. This is a prediction for \tune reweighted to the Z-expansion prediction.}
\label{tab:1D_YProjection_CCQELike_CV_ZExp_2p2h_pzmu_mc}
\end{table}

\begin{table}[h]
\centering
\begin{tabular}{| c | c |}
\hline

\hline
Bin & Cross Section \\ \hline \hline
1&2.82e-40 \\ 
2&1.13e-39 \\ 
3&2.66e-39 \\ 
4&4.32e-39 \\ 
5&5.39e-39 \\ 
6&5.94e-39 \\ 
7&6.03e-39 \\ 
8&5.10e-39 \\ 
9&3.20e-39 \\ 
10&1.58e-39 \\ 
11&5.28e-40 \\ 
12&1.52e-40 \\ 

\hline
\end{tabular}
\caption{Monte Carlo predicted differential cross section as a function of $p_{t}$. Units are cm$^2$ per GeV$^2$ per nucleon. This is a prediction for GENIE 2.8.4.}
\label{tab:1D_YProjection_CCQELike_default_pzmu_mc}
\end{table}

\begin{table}[h]
\centering
\begin{tabular}{| c | c |}
\hline

\hline
Bin & Cross Section \\ \hline \hline
1&4.36e-40 \\ 
2&1.65e-39 \\ 
3&3.77e-39 \\ 
4&6.31e-39 \\ 
5&8.18e-39 \\ 
6&8.89e-39 \\ 
7&8.48e-39 \\ 
8&6.35e-39 \\ 
9&3.59e-39 \\ 
10&1.63e-39 \\ 
11&5.08e-40 \\ 
12&1.41e-40 \\ 

\hline
\end{tabular}
\caption{Monte Carlo predicted differential cross section as a function of $p_{t}$. Units are cm$^2$ per GeV$^2$ per nucleon. This is a prediction for GENIE 2.8.4 with the addition of Valencia 2p2h, and the non-resonant pion production reduction.}
\label{tab:1D_YProjection_CCQELike_pion_2p2h_2p2h_pzmu_mc}
\end{table}

\begin{table}[h]
\centering
\begin{tabular}{| c | c |}
\hline

\hline
Bin & Cross Section \\ \hline \hline
1&3.66e-40 \\ 
2&1.27e-39 \\ 
3&2.68e-39 \\ 
4&4.26e-39 \\ 
5&5.44e-39 \\ 
6&6.26e-39 \\ 
7&6.66e-39 \\ 
8&5.93e-39 \\ 
9&3.77e-39 \\ 
10&1.80e-39 \\ 
11&5.56e-40 \\ 
12&1.48e-40 \\ 

\hline
\end{tabular}
\caption{Monte Carlo predicted differential cross section as a function of $p_{t}$. Units are cm$^2$ per GeV$^2$ per nucleon. This is a prediction for GENIE 2.8.4 with the addition of RPA, Valencia 2p2h, and the non-resonant pion production reduction.}
\label{tab:1D_YProjection_CCQELike_pion_rpa_2p2h_pzmu_mc}
\end{table}

\begin{table}[h]
\centering
\begin{tabular}{| c | c |}
\hline

\hline
Bin & Cross Section \\ \hline \hline
1&2.17e-40 \\ 
2&7.86e-40 \\ 
3&1.79e-39 \\ 
4&3.02e-39 \\ 
5&3.99e-39 \\ 
6&4.74e-39 \\ 
7&5.17e-39 \\ 
8&4.81e-39 \\ 
9&3.28e-39 \\ 
10&1.69e-39 \\ 
11&5.49e-40 \\ 
12&1.48e-40 \\ 

\hline
\end{tabular}
\caption{Monte Carlo predicted differential cross section as a function of $p_{t}$. Units are cm$^2$ per GeV$^2$ per nucleon. This is a prediction for GENIE 2.8.4 with the addition of RPA and the non-resonant pion production reduction.}
\label{tab:1D_YProjection_CCQELike_pion_rpa_pzmu_mc}
\end{table}

\begin{table}[h]
\centering
\begin{tabular}{| c | c |}
\hline

\hline
Bin & Cross Section \\ \hline \hline
1&4.31e-40 \\ 
2&1.61e-39 \\ 
3&3.53e-39 \\ 
4&5.51e-39 \\ 
5&6.75e-39 \\ 
6&7.35e-39 \\ 
7&7.36e-39 \\ 
8&6.07e-39 \\ 
9&3.58e-39 \\ 
10&1.63e-39 \\ 
11&5.08e-40 \\ 
12&1.41e-40 \\ 

\hline
\end{tabular}
\caption{Monte Carlo predicted differential cross section as a function of $p_{t}$. Units are cm$^2$ per GeV$^2$ per nucleon. This is a prediction for GENIE 2.8.4 with the addition of Valencia 2p2h.}
\label{tab:1D_YProjection_CCQELike_piontune_2p2h_pzmu_mc}
\end{table}

\begin{table}[h]
\centering
\begin{tabular}{| c | c |}
\hline

\hline
Bin & Cross Section \\ \hline \hline
1&2.82e-40 \\ 
2&1.12e-39 \\ 
3&2.64e-39 \\ 
4&4.27e-39 \\ 
5&5.31e-39 \\ 
6&5.82e-39 \\ 
7&5.87e-39 \\ 
8&4.95e-39 \\ 
9&3.09e-39 \\ 
10&1.52e-39 \\ 
11&5.02e-40 \\ 
12&1.41e-40 \\ 

\hline
\end{tabular}
\caption{Monte Carlo predicted differential cross section as a function of $p_{t}$. Units are cm$^2$ per GeV$^2$ per nucleon. This is a prediction for GENIE 2.8.4 with the non-resonant pion production reduction.}
\label{tab:1D_YProjection_CCQELike_piontune_pzmu_mc}
\end{table}

\begin{table}[h]
\centering
\begin{tabular}{| c | c |}
\hline

\hline
Bin & Cross Section \\ \hline \hline
1&5.83e-39 \\ 
2&6.46e-39 \\ 
3&7.10e-39 \\ 
4&7.83e-39 \\ 
5&8.43e-39 \\ 
6&9.35e-39 \\ 
7&1.00e-38 \\ 
8&9.47e-39 \\ 
9&7.83e-39 \\ 
10&5.65e-39 \\ 
11&3.56e-39 \\ 
12&1.97e-39 \\ 
13&1.07e-39 \\ 
14&5.88e-40 \\ 
15&1.83e-40 \\ 
16&2.20e-41 \\ 

\hline
\end{tabular}
\caption{Monte Carlo predicted differential cross section as a function of \qqqe. Units are cm$^2$ per GeV$^2$ per nucleon. This is a prediction for \tune.}
\label{tab:1D_XProjection_CCQELike_CV_2p2h_q2_mc}
\end{table}

\begin{table}[h]
\centering
\begin{tabular}{| c | c |}
\hline

\hline
Bin & Cross Section \\ \hline \hline
1&4.83e-39 \\ 
2&5.53e-39 \\ 
3&6.26e-39 \\ 
4&7.09e-39 \\ 
5&7.80e-39 \\ 
6&8.90e-39 \\ 
7&9.76e-39 \\ 
8&9.31e-39 \\ 
9&7.75e-39 \\ 
10&5.62e-39 \\ 
11&3.55e-39 \\ 
12&1.97e-39 \\ 
13&1.07e-39 \\ 
14&5.89e-40 \\ 
15&1.83e-40 \\ 
16&2.20e-41 \\ 

\hline
\end{tabular}
\caption{Monte Carlo predicted differential cross section as a function of \qqqe. Units are cm$^2$ per GeV$^2$ per nucleon. This is a prediction for \tune with the addition of the MINOS emperical low $Q^{2}$ suppression.}
\label{tab:1D_XProjection_CCQELike_CV_RPA_Res_MINOS_2p2h_q2_mc}
\end{table}

\begin{table}[h]
\centering
\begin{tabular}{| c | c |}
\hline

\hline
Bin & Cross Section \\ \hline \hline
1&5.88e-39 \\ 
2&6.53e-39 \\ 
3&7.19e-39 \\ 
4&7.93e-39 \\ 
5&8.55e-39 \\ 
6&9.48e-39 \\ 
7&1.02e-38 \\ 
8&9.57e-39 \\ 
9&7.90e-39 \\ 
10&5.68e-39 \\ 
11&3.57e-39 \\ 
12&1.96e-39 \\ 
13&1.06e-39 \\ 
14&5.77e-40 \\ 
15&1.77e-40 \\ 
16&2.06e-41 \\ 

\hline
\end{tabular}
\caption{Monte Carlo predicted differential cross section as a function of \qqqe. Units are cm$^2$ per GeV$^2$ per nucleon. This is a prediction for \tune reweighted to the Z-expansion prediction.}
\label{tab:1D_XProjection_CCQELike_CV_ZExp_2p2h_q2_mc}
\end{table}

\begin{table}[h]
\centering
\begin{tabular}{| c | c |}
\hline

\hline
Bin & Cross Section \\ \hline \hline
1&4.45e-39 \\ 
2&5.31e-39 \\ 
3&6.17e-39 \\ 
4&7.03e-39 \\ 
5&7.60e-39 \\ 
6&8.11e-39 \\ 
7&7.93e-39 \\ 
8&7.26e-39 \\ 
9&5.96e-39 \\ 
10&4.54e-39 \\ 
11&2.96e-39 \\ 
12&1.67e-39 \\ 
13&9.35e-40 \\ 
14&5.28e-40 \\ 
15&1.75e-40 \\ 
16&2.28e-41 \\ 

\hline
\end{tabular}
\caption{Monte Carlo predicted differential cross section as a function of \qqqe. Units are cm$^2$ per GeV$^2$ per nucleon. This is a prediction for GENIE 2.8.4.}
\label{tab:1D_XProjection_CCQELike_default_q2_mc}
\end{table}

\begin{table}[h]
\centering
\begin{tabular}{| c | c |}
\hline

\hline
Bin & Cross Section \\ \hline \hline
1&6.79e-39 \\ 
2&7.90e-39 \\ 
3&8.94e-39 \\ 
4&9.99e-39 \\ 
5&1.08e-38 \\ 
6&1.17e-38 \\ 
7&1.20e-38 \\ 
8&1.10e-38 \\ 
9&8.69e-39 \\ 
10&5.98e-39 \\ 
11&3.55e-39 \\ 
12&1.86e-39 \\ 
13&9.79e-40 \\ 
14&5.32e-40 \\ 
15&1.67e-40 \\ 
16&2.12e-41 \\ 

\hline
\end{tabular}
\caption{Monte Carlo predicted differential cross section as a function of \qqqe. Units are cm$^2$ per GeV$^2$ per nucleon. This is a prediction for GENIE 2.8.4 with the addition of Valencia 2p2h, and the non-resonant pion production reduction.}
\label{tab:1D_XProjection_CCQELike_pion_2p2h_2p2h_q2_mc}
\end{table}

\begin{table}[h]
\centering
\begin{tabular}{| c | c |}
\hline

\hline
Bin & Cross Section \\ \hline \hline
1&5.75e-39 \\ 
2&6.32e-39 \\ 
3&6.84e-39 \\ 
4&7.37e-39 \\ 
5&7.76e-39 \\ 
6&8.07e-39 \\ 
7&8.01e-39 \\ 
8&7.52e-39 \\ 
9&6.48e-39 \\ 
10&5.15e-39 \\ 
11&3.49e-39 \\ 
12&1.97e-39 \\ 
13&1.07e-39 \\ 
14&5.88e-40 \\ 
15&1.83e-40 \\ 
16&2.20e-41 \\ 

\hline
\end{tabular}
\caption{Monte Carlo predicted differential cross section as a function of \qqqe. Units are cm$^2$ per GeV$^2$ per nucleon. This is a prediction for GENIE 2.8.4 with the addition of RPA, Valencia 2p2h, and the non-resonant pion production reduction.}
\label{tab:1D_XProjection_CCQELike_pion_rpa_2p2h_q2_mc}
\end{table}

\begin{table}[h]
\centering
\begin{tabular}{| c | c |}
\hline

\hline
Bin & Cross Section \\ \hline \hline
1&3.48e-39 \\ 
2&3.85e-39 \\ 
3&4.29e-39 \\ 
4&4.81e-39 \\ 
5&5.17e-39 \\ 
6&5.63e-39 \\ 
7&5.81e-39 \\ 
8&5.61e-39 \\ 
9&4.96e-39 \\ 
10&4.08e-39 \\ 
11&2.88e-39 \\ 
12&1.72e-39 \\ 
13&9.91e-40 \\ 
14&5.63e-40 \\ 
15&1.81e-40 \\ 
16&2.20e-41 \\ 

\hline
\end{tabular}
\caption{Monte Carlo predicted differential cross section as a function of \qqqe. Units are cm$^2$ per GeV$^2$ per nucleon. This is a prediction for GENIE 2.8.4 with the addition of RPA and the non-resonant pion production reduction.}
\label{tab:1D_XProjection_CCQELike_pion_rpa_q2_mc}
\end{table}

\begin{table}[h]
\centering
\begin{tabular}{| c | c |}
\hline

\hline
Bin & Cross Section \\ \hline \hline
1&6.71e-39 \\ 
2&7.76e-39 \\ 
3&8.68e-39 \\ 
4&9.54e-39 \\ 
5&1.01e-38 \\ 
6&1.05e-38 \\ 
7&1.00e-38 \\ 
8&9.02e-39 \\ 
9&7.34e-39 \\ 
10&5.49e-39 \\ 
11&3.47e-39 \\ 
12&1.86e-39 \\ 
13&9.79e-40 \\ 
14&5.32e-40 \\ 
15&1.67e-40 \\ 
16&2.12e-41 \\ 

\hline
\end{tabular}
\caption{Monte Carlo predicted differential cross section as a function of \qqqe. Units are cm$^2$ per GeV$^2$ per nucleon. This is a prediction for GENIE 2.8.4 with the addition of Valencia 2p2h.}
\label{tab:1D_XProjection_CCQELike_piontune_2p2h_q2_mc}
\end{table}

\begin{table}[h]
\centering
\begin{tabular}{| c | c |}
\hline

\hline
Bin & Cross Section \\ \hline \hline
1&4.44e-39 \\ 
2&5.29e-39 \\ 
3&6.13e-39 \\ 
4&6.98e-39 \\ 
5&7.53e-39 \\ 
6&8.01e-39 \\ 
7&7.80e-39 \\ 
8&7.12e-39 \\ 
9&5.82e-39 \\ 
10&4.41e-39 \\ 
11&2.87e-39 \\ 
12&1.61e-39 \\ 
13&9.00e-40 \\ 
14&5.07e-40 \\ 
15&1.66e-40 \\ 
16&2.12e-41 \\ 

\hline
\end{tabular}
\caption{Monte Carlo predicted differential cross section as a function of \qqqe. Units are cm$^2$ per GeV$^2$ per nucleon. This is a prediction for GENIE 2.8.4 with the non-resonant pion production reduction.}
\label{tab:1D_XProjection_CCQELike_piontune_q2_mc}
\end{table}

\begin{table*}[h]
\centering
\begin{tabular}{| c | c | c | c | c | c | c | c | c | c | c | c | c |}
\hline
& 1 & 2 & 3 & 4 & 5 & 6 & 7 & 8 & 9 & 10 & 11 & 12 \\ \hline \hline
1 & 9.33e-41 & 1.10e-40 & 1.23e-40 & 1.04e-40 & 7.64e-41 & 4.23e-41 & 2.58e-41 & 1.52e-41 & 9.17e-42 & 6.35e-42 & 5.01e-42 & 2.47e-42 \\ \hline
2 & 3.31e-40 & 4.13e-40 & 4.52e-40 & 4.08e-40 & 2.86e-40 & 1.66e-40 & 8.98e-41 & 5.35e-41 & 3.18e-41 & 2.08e-41 & 1.09e-41 & 4.88e-42 \\ \hline
3 & 7.74e-40 & 9.44e-40 & 1.02e-39 & 9.10e-40 & 6.39e-40 & 3.60e-40 & 1.96e-40 & 1.19e-40 & 7.07e-41 & 4.22e-41 & 2.30e-41 & 9.42e-42 \\ \hline
4 & 1.37e-39 & 1.64e-39 & 1.76e-39 & 1.58e-39 & 1.08e-39 & 6.03e-40 & 3.48e-40 & 1.99e-40 & 1.17e-40 & 7.14e-41 & 3.97e-41 & 1.62e-41 \\ \hline
5 & 1.87e-39 & 2.30e-39 & 2.44e-39 & 2.10e-39 & 1.43e-39 & 8.06e-40 & 4.67e-40 & 2.66e-40 & 1.57e-40 & 9.72e-41 & 5.40e-41 & 2.21e-41 \\ \hline
6 & 2.24e-39 & 2.62e-39 & 2.76e-39 & 2.35e-39 & 1.61e-39 & 9.05e-40 & 5.09e-40 & 3.05e-40 & 1.75e-40 & 1.12e-40 & 6.02e-41 & 2.52e-41 \\ \hline
7 & 2.29e-39 & 2.64e-39 & 2.75e-39 & 2.35e-39 & 1.54e-39 & 8.48e-40 & 5.07e-40 & 2.98e-40 & 1.85e-40 & 1.11e-40 & 5.89e-41 & 2.50e-41 \\
\hline

\hline
\end{tabular}
\caption{Monte Carlo predicted double differential cross section as a function of $p_{t}$~(rows) versus $p_{||}$~(columns). Units are cm$^2$ per GeV$^2$ per nucleon. This is the prediction for \tune.}
\label{tab:2D_CCQELike_CV_2p2h_mc}
\end{table*}

\begin{table*}[h]
\centering
\begin{tabular}{| c | c | c | c | c | c | c | c | c | c | c | c | c |}
\hline
& 1 & 2 & 3 & 4 & 5 & 6 & 7 & 8 & 9 & 10 & 11 & 12 \\ \hline \hline
1 & 7.43e-41 & 8.89e-41 & 9.86e-41 & 8.87e-41 & 6.61e-41 & 3.67e-41 & 2.21e-41 & 1.22e-41 & 7.52e-42 & 4.94e-42 & 4.44e-42 & 2.23e-42 \\ \hline
2 & 2.84e-40 & 3.54e-40 & 3.91e-40 & 3.60e-40 & 2.58e-40 & 1.48e-40 & 7.98e-41 & 4.73e-41 & 2.79e-41 & 1.82e-41 & 9.62e-42 & 4.30e-42 \\ \hline
3 & 7.10e-40 & 8.66e-40 & 9.39e-40 & 8.51e-40 & 6.00e-40 & 3.39e-40 & 1.84e-40 & 1.11e-40 & 6.53e-41 & 3.89e-41 & 2.10e-41 & 8.59e-42 \\ \hline
4 & 1.31e-39 & 1.56e-39 & 1.69e-39 & 1.52e-39 & 1.05e-39 & 5.84e-40 & 3.35e-40 & 1.92e-40 & 1.12e-40 & 6.81e-41 & 3.79e-41 & 1.55e-41 \\ \hline
5 & 1.83e-39 & 2.24e-39 & 2.38e-39 & 2.06e-39 & 1.40e-39 & 7.90e-40 & 4.57e-40 & 2.60e-40 & 1.53e-40 & 9.46e-41 & 5.25e-41 & 2.14e-41 \\ \hline
6 & 2.21e-39 & 2.58e-39 & 2.72e-39 & 2.32e-39 & 1.59e-39 & 8.95e-40 & 5.03e-40 & 3.01e-40 & 1.72e-40 & 1.10e-40 & 5.92e-41 & 2.48e-41 \\ \hline
7 & 2.27e-39 & 2.62e-39 & 2.73e-39 & 2.33e-39 & 1.53e-39 & 8.42e-40 & 5.03e-40 & 2.95e-40 & 1.84e-40 & 1.10e-40 & 5.83e-41 & 2.47e-41 \\
\hline

\hline
\end{tabular}
\caption{Monte Carlo predicted double differential cross section as a function of $p_{t}$~(rows) versus $p_{||}$~(columns). Units are cm$^2$ per GeV$^2$ per nucleon. This is the prediction for \tune with the addition of the MINOS emperical low $Q^{2}$ suppression.}
\label{tab:2D_CCQELike_CV_RPA_Res_MINOS_2p2h_mc}
\end{table*}

\begin{table*}[h]
\centering
\begin{tabular}{| c | c | c | c | c | c | c | c | c | c | c | c | c |}
\hline
& 1 & 2 & 3 & 4 & 5 & 6 & 7 & 8 & 9 & 10 & 11 & 12 \\ \hline \hline
1 & 9.40e-41 & 1.10e-40 & 1.24e-40 & 1.06e-40 & 7.74e-41 & 4.30e-41 & 2.62e-41 & 1.54e-41 & 9.26e-42 & 6.40e-42 & 5.04e-42 & 2.48e-42 \\ \hline
2 & 3.34e-40 & 4.18e-40 & 4.58e-40 & 4.13e-40 & 2.90e-40 & 1.68e-40 & 9.12e-41 & 5.43e-41 & 3.22e-41 & 2.10e-41 & 1.10e-41 & 4.95e-42 \\ \hline
3 & 7.84e-40 & 9.56e-40 & 1.03e-39 & 9.24e-40 & 6.49e-40 & 3.66e-40 & 1.99e-40 & 1.21e-40 & 7.17e-41 & 4.28e-41 & 2.34e-41 & 9.55e-42 \\ \hline
4 & 1.39e-39 & 1.66e-39 & 1.79e-39 & 1.60e-39 & 1.10e-39 & 6.13e-40 & 3.53e-40 & 2.02e-40 & 1.19e-40 & 7.24e-41 & 4.03e-41 & 1.64e-41 \\ \hline
5 & 1.89e-39 & 2.32e-39 & 2.47e-39 & 2.13e-39 & 1.45e-39 & 8.17e-40 & 4.73e-40 & 2.70e-40 & 1.59e-40 & 9.85e-41 & 5.46e-41 & 2.23e-41 \\ \hline
6 & 2.26e-39 & 2.64e-39 & 2.78e-39 & 2.38e-39 & 1.63e-39 & 9.15e-40 & 5.15e-40 & 3.08e-40 & 1.77e-40 & 1.13e-40 & 6.08e-41 & 2.55e-41 \\ \hline
7 & 2.30e-39 & 2.66e-39 & 2.78e-39 & 2.37e-39 & 1.56e-39 & 8.55e-40 & 5.11e-40 & 3.01e-40 & 1.87e-40 & 1.12e-40 & 5.94e-41 & 2.52e-41 \\
\hline

\hline
\end{tabular}
\caption{Monte Carlo predicted double differential cross section as a function of $p_{t}$~(rows) versus $p_{||}$~(columns). Units are cm$^2$ per GeV$^2$ per nucleon. This is the prediction for \tune reweighted to the Z-expansion prediction.}
\label{tab:2D_CCQELike_CV_ZExp_2p2h_mc}
\end{table*}

\begin{table*}[h]
\centering
\begin{tabular}{| c | c | c | c | c | c | c | c | c | c | c | c | c |}
\hline
& 1 & 2 & 3 & 4 & 5 & 6 & 7 & 8 & 9 & 10 & 11 & 12 \\ \hline \hline
1 & 6.52e-41 & 7.66e-41 & 8.99e-41 & 7.76e-41 & 6.26e-41 & 3.66e-41 & 2.28e-41 & 1.18e-41 & 6.81e-42 & 4.83e-42 & 4.15e-42 & 2.20e-42 \\ \hline
2 & 2.58e-40 & 3.41e-40 & 3.85e-40 & 3.59e-40 & 2.64e-40 & 1.53e-40 & 8.29e-41 & 4.83e-41 & 2.66e-41 & 1.80e-41 & 9.40e-42 & 4.47e-42 \\ \hline
3 & 6.54e-40 & 8.08e-40 & 9.04e-40 & 8.43e-40 & 6.28e-40 & 3.61e-40 & 1.95e-40 & 1.11e-40 & 6.24e-41 & 3.81e-41 & 2.13e-41 & 8.73e-42 \\ \hline
4 & 1.08e-39 & 1.34e-39 & 1.48e-39 & 1.37e-39 & 1.00e-39 & 5.75e-40 & 3.15e-40 & 1.74e-40 & 1.00e-40 & 6.27e-41 & 3.50e-41 & 1.33e-41 \\ \hline
5 & 1.40e-39 & 1.70e-39 & 1.87e-39 & 1.69e-39 & 1.21e-39 & 6.99e-40 & 3.85e-40 & 2.17e-40 & 1.24e-40 & 8.05e-41 & 4.11e-41 & 1.63e-41 \\ \hline
6 & 1.66e-39 & 1.92e-39 & 2.09e-39 & 1.84e-39 & 1.30e-39 & 7.11e-40 & 3.96e-40 & 2.35e-40 & 1.36e-40 & 8.34e-41 & 4.48e-41 & 1.79e-41 \\ \hline
7 & 1.79e-39 & 2.04e-39 & 2.11e-39 & 1.83e-39 & 1.25e-39 & 6.78e-40 & 3.95e-40 & 2.27e-40 & 1.37e-40 & 8.34e-41 & 4.50e-41 & 1.87e-41 \\
\hline

\hline
\end{tabular}
\caption{Monte Carlo predicted double differential cross section as a function of $p_{t}$~(rows) versus $p_{||}$~(columns). Units are cm$^2$ per GeV$^2$ per nucleon. This is the prediction for GENIE 2.8.4.}
\label{tab:2D_CCQELike_default_mc}
\end{table*}

\begin{table*}[h]
\centering
\begin{tabular}{| c | c | c | c | c | c | c | c | c | c | c | c | c |}
\hline
& 1 & 2 & 3 & 4 & 5 & 6 & 7 & 8 & 9 & 10 & 11 & 12 \\ \hline \hline
1 & 1.06e-40 & 1.26e-40 & 1.43e-40 & 1.26e-40 & 9.49e-41 & 5.29e-41 & 3.24e-41 & 1.80e-41 & 1.09e-41 & 7.45e-42 & 5.52e-42 & 2.62e-42 \\ \hline
2 & 3.98e-40 & 5.05e-40 & 5.65e-40 & 5.20e-40 & 3.75e-40 & 2.16e-40 & 1.16e-40 & 6.92e-41 & 3.98e-41 & 2.62e-41 & 1.38e-41 & 6.21e-42 \\ \hline
3 & 9.58e-40 & 1.18e-39 & 1.30e-39 & 1.19e-39 & 8.55e-40 & 4.86e-40 & 2.63e-40 & 1.56e-40 & 9.05e-41 & 5.44e-41 & 2.97e-41 & 1.21e-41 \\ \hline
4 & 1.64e-39 & 1.99e-39 & 2.18e-39 & 2.00e-39 & 1.40e-39 & 7.92e-40 & 4.47e-40 & 2.53e-40 & 1.47e-40 & 8.99e-41 & 5.00e-41 & 2.01e-41 \\ \hline
5 & 2.16e-39 & 2.67e-39 & 2.89e-39 & 2.54e-39 & 1.75e-39 & 9.96e-40 & 5.70e-40 & 3.22e-40 & 1.88e-40 & 1.17e-40 & 6.42e-41 & 2.61e-41 \\ \hline
6 & 2.49e-39 & 2.94e-39 & 3.13e-39 & 2.71e-39 & 1.88e-39 & 1.05e-39 & 5.88e-40 & 3.50e-40 & 2.01e-40 & 1.28e-40 & 6.87e-41 & 2.84e-41 \\ \hline
7 & 2.45e-39 & 2.86e-39 & 3.00e-39 & 2.58e-39 & 1.71e-39 & 9.41e-40 & 5.60e-40 & 3.27e-40 & 2.02e-40 & 1.21e-40 & 6.45e-41 & 2.73e-41 \\
\hline

\hline
\end{tabular}
\caption{Monte Carlo predicted double differential cross section as a function of $p_{t}$~(rows) versus $p_{||}$~(columns). Units are cm$^2$ per GeV$^2$ per nucleon. This is the prediction for GENIE 2.8.4 with the addition of Valencia 2p2h, and the non-resonant pion production reduction.}
\label{tab:2D_CCQELike_pion_2p2h_2p2h_mc}
\end{table*}

\begin{table*}[h]
\centering
\begin{tabular}{| c | c | c | c | c | c | c | c | c | c | c | c | c |}
\hline
& 1 & 2 & 3 & 4 & 5 & 6 & 7 & 8 & 9 & 10 & 11 & 12 \\ \hline \hline
1 & 9.21e-41 & 1.08e-40 & 1.21e-40 & 1.03e-40 & 7.52e-41 & 4.18e-41 & 2.54e-41 & 1.50e-41 & 9.07e-42 & 6.24e-42 & 4.96e-42 & 2.46e-42 \\ \hline
2 & 3.21e-40 & 4.02e-40 & 4.39e-40 & 3.95e-40 & 2.76e-40 & 1.60e-40 & 8.72e-41 & 5.19e-41 & 3.07e-41 & 2.01e-41 & 1.06e-41 & 4.74e-42 \\ \hline
3 & 7.10e-40 & 8.67e-40 & 9.33e-40 & 8.32e-40 & 5.81e-40 & 3.27e-40 & 1.81e-40 & 1.09e-40 & 6.43e-41 & 3.89e-41 & 2.13e-41 & 8.74e-42 \\ \hline
4 & 1.16e-39 & 1.39e-39 & 1.49e-39 & 1.31e-39 & 9.11e-40 & 5.12e-40 & 2.90e-40 & 1.67e-40 & 9.79e-41 & 6.17e-41 & 3.40e-41 & 1.35e-41 \\ \hline
5 & 1.51e-39 & 1.81e-39 & 1.93e-39 & 1.65e-39 & 1.14e-39 & 6.44e-40 & 3.65e-40 & 2.13e-40 & 1.24e-40 & 8.04e-41 & 4.21e-41 & 1.70e-41 \\ \hline
6 & 1.84e-39 & 2.12e-39 & 2.21e-39 & 1.88e-39 & 1.28e-39 & 7.07e-40 & 4.04e-40 & 2.44e-40 & 1.44e-40 & 8.77e-41 & 4.80e-41 & 1.95e-41 \\ \hline
7 & 2.02e-39 & 2.31e-39 & 2.35e-39 & 1.99e-39 & 1.32e-39 & 7.16e-40 & 4.24e-40 & 2.50e-40 & 1.53e-40 & 9.32e-41 & 4.95e-41 & 2.10e-41 \\
\hline

\hline
\end{tabular}
\caption{Monte Carlo predicted double differential cross section as a function of $p_{t}$~(rows) versus $p_{||}$~(columns). Units are cm$^2$ per GeV$^2$ per nucleon. This is the prediction for GENIE 2.8.4 with the addition of RPA, Valencia 2p2h, and the non-resonant pion production reduction.}
\label{tab:2D_CCQELike_pion_rpa_2p2h_mc}
\end{table*}

\begin{table*}[h]
\centering
\begin{tabular}{| c | c | c | c | c | c | c | c | c | c | c | c | c |}
\hline
& 1 & 2 & 3 & 4 & 5 & 6 & 7 & 8 & 9 & 10 & 11 & 12 \\ \hline \hline
1 & 5.24e-41 & 6.03e-41 & 6.95e-41 & 5.64e-41 & 4.41e-41 & 2.60e-41 & 1.61e-41 & 8.86e-42 & 5.07e-42 & 3.73e-42 & 3.64e-42 & 2.05e-42 \\ \hline
2 & 1.89e-40 & 2.47e-40 & 2.72e-40 & 2.46e-40 & 1.74e-40 & 1.02e-40 & 5.62e-41 & 3.25e-41 & 1.85e-41 & 1.24e-41 & 6.47e-42 & 3.15e-42 \\ \hline
3 & 4.61e-40 & 5.62e-40 & 6.16e-40 & 5.58e-40 & 4.09e-40 & 2.32e-40 & 1.26e-40 & 7.44e-41 & 4.23e-41 & 2.56e-41 & 1.45e-41 & 5.94e-42 \\ \hline
4 & 7.90e-40 & 9.58e-40 & 1.04e-39 & 9.43e-40 & 6.76e-40 & 3.84e-40 & 2.13e-40 & 1.19e-40 & 6.94e-41 & 4.36e-41 & 2.42e-41 & 9.21e-42 \\ \hline
5 & 1.07e-39 & 1.29e-39 & 1.39e-39 & 1.24e-39 & 8.71e-40 & 5.00e-40 & 2.79e-40 & 1.58e-40 & 9.10e-41 & 5.91e-41 & 3.04e-41 & 1.21e-41 \\ \hline
6 & 1.35e-39 & 1.56e-39 & 1.67e-39 & 1.45e-39 & 1.01e-39 & 5.55e-40 & 3.10e-40 & 1.85e-40 & 1.08e-40 & 6.58e-41 & 3.55e-41 & 1.45e-41 \\ \hline
7 & 1.55e-39 & 1.76e-39 & 1.81e-39 & 1.57e-39 & 1.06e-39 & 5.73e-40 & 3.35e-40 & 1.93e-40 & 1.17e-40 & 7.12e-41 & 3.81e-41 & 1.59e-41 \\
\hline

\hline
\end{tabular}
\caption{Monte Carlo predicted double differential cross section as a function of $p_{t}$~(rows) versus $p_{||}$~(columns). Units are cm$^2$ per GeV$^2$ per nucleon. This is the prediction for GENIE 2.8.4 with the addition of RPA and the non-resonant pion production reduction.}
\label{tab:2D_CCQELike_pion_rpa_mc}
\end{table*}

\begin{table*}[h]
\centering
\begin{tabular}{| c | c | c | c | c | c | c | c | c | c | c | c | c |}
\hline
& 1 & 2 & 3 & 4 & 5 & 6 & 7 & 8 & 9 & 10 & 11 & 12 \\ \hline \hline
1 & 1.05e-40 & 1.24e-40 & 1.41e-40 & 1.24e-40 & 9.37e-41 & 5.24e-41 & 3.20e-41 & 1.79e-41 & 1.08e-41 & 7.34e-42 & 5.46e-42 & 2.61e-42 \\ \hline
2 & 3.88e-40 & 4.94e-40 & 5.52e-40 & 5.06e-40 & 3.65e-40 & 2.11e-40 & 1.14e-40 & 6.75e-41 & 3.87e-41 & 2.56e-41 & 1.34e-41 & 6.06e-42 \\ \hline
3 & 8.94e-40 & 1.11e-39 & 1.21e-39 & 1.11e-39 & 7.97e-40 & 4.53e-40 & 2.49e-40 & 1.46e-40 & 8.41e-41 & 5.12e-41 & 2.80e-41 & 1.14e-41 \\ \hline
4 & 1.43e-39 & 1.75e-39 & 1.91e-39 & 1.73e-39 & 1.23e-39 & 7.00e-40 & 3.89e-40 & 2.21e-40 & 1.27e-40 & 8.02e-41 & 4.42e-41 & 1.75e-41 \\ \hline
5 & 1.80e-39 & 2.18e-39 & 2.37e-39 & 2.09e-39 & 1.46e-39 & 8.34e-40 & 4.68e-40 & 2.69e-40 & 1.55e-40 & 1.01e-40 & 5.23e-41 & 2.11e-41 \\ \hline
6 & 2.09e-39 & 2.43e-39 & 2.59e-39 & 2.24e-39 & 1.55e-39 & 8.55e-40 & 4.83e-40 & 2.89e-40 & 1.70e-40 & 1.04e-40 & 5.66e-41 & 2.27e-41 \\ \hline
7 & 2.19e-39 & 2.52e-39 & 2.59e-39 & 2.22e-39 & 1.49e-39 & 8.10e-40 & 4.77e-40 & 2.79e-40 & 1.70e-40 & 1.03e-40 & 5.51e-41 & 2.34e-41 \\
\hline

\hline
\end{tabular}
\caption{Monte Carlo predicted double differential cross section as a function of $p_{t}$~(rows) versus $p_{||}$~(columns). Units are cm$^2$ per GeV$^2$ per nucleon. This is the prediction for GENIE 2.8.4 with the addition of Valencia 2p2h.}
\label{tab:2D_CCQELike_piontune_2p2h_mc}
\end{table*}

\begin{table*}[h]
\centering
\begin{tabular}{| c | c | c | c | c | c | c | c | c | c | c | c | c |}
\hline
& 1 & 2 & 3 & 4 & 5 & 6 & 7 & 8 & 9 & 10 & 11 & 12 \\ \hline \hline
1 & 6.51e-41 & 7.64e-41 & 8.98e-41 & 7.76e-41 & 6.26e-41 & 3.66e-41 & 2.27e-41 & 1.17e-41 & 6.79e-42 & 4.83e-42 & 4.15e-42 & 2.20e-42 \\ \hline
2 & 2.57e-40 & 3.39e-40 & 3.84e-40 & 3.58e-40 & 2.63e-40 & 1.53e-40 & 8.26e-41 & 4.82e-41 & 2.65e-41 & 1.79e-41 & 9.35e-42 & 4.47e-42 \\ \hline
3 & 6.45e-40 & 8.01e-40 & 8.97e-40 & 8.39e-40 & 6.25e-40 & 3.59e-40 & 1.94e-40 & 1.11e-40 & 6.21e-41 & 3.78e-41 & 2.12e-41 & 8.64e-42 \\ \hline
4 & 1.06e-39 & 1.32e-39 & 1.46e-39 & 1.36e-39 & 9.98e-40 & 5.72e-40 & 3.13e-40 & 1.72e-40 & 9.89e-41 & 6.21e-41 & 3.45e-41 & 1.32e-41 \\ \hline
5 & 1.36e-39 & 1.67e-39 & 1.84e-39 & 1.67e-39 & 1.20e-39 & 6.90e-40 & 3.81e-40 & 2.14e-40 & 1.22e-40 & 7.94e-41 & 4.05e-41 & 1.62e-41 \\ \hline
6 & 1.60e-39 & 1.87e-39 & 2.05e-39 & 1.81e-39 & 1.28e-39 & 7.02e-40 & 3.89e-40 & 2.30e-40 & 1.34e-40 & 8.19e-41 & 4.40e-41 & 1.76e-41 \\ \hline
7 & 1.71e-39 & 1.98e-39 & 2.06e-39 & 1.80e-39 & 1.22e-39 & 6.66e-40 & 3.88e-40 & 2.23e-40 & 1.34e-40 & 8.13e-41 & 4.37e-41 & 1.83e-41 \\
\hline

\hline
\end{tabular}
\caption{Monte Carlo predicted double differential cross section as a function of $p_{t}$~(rows) versus $p_{||}$~(columns). Units are cm$^2$ per GeV$^2$ per nucleon. This is the prediction for GENIE 2.8.4 with the non-resonant pion production reduction.}
\label{tab:2D_CCQELike_piontune_mc}
\end{table*}

%% file: SupplementalQE.tex
\section{Supplemental Material - Quasielastic Result}
\subsection{Supplemental: Data Measurement}
This section contains the tabular breakdown of all the various cross section results as well as uncertainty on the measurement. In addition, tables of the model predictions are provided.
\begin{table}[h]
\centering
\begin{tabular}{| c | c | c | c |}
\hline

\hline
Bin & Cross Section & Stat. Unc. & Total. Unc. \\ \hline \hline
1&2.90e-39&1.33e-40&8.62e-40 \\ 
2&3.70e-39&1.04e-40&5.51e-40 \\ 
3&4.32e-39&8.86e-41&5.17e-40 \\ 
4&4.58e-39&7.87e-41&4.44e-40 \\ 
5&4.51e-39&7.92e-41&5.32e-40 \\ 
6&4.51e-39&9.69e-41&6.80e-40 \\ 
7&4.91e-39&1.28e-40&7.58e-40 \\ 
8&5.21e-39&1.43e-40&6.49e-40 \\ 
9&5.78e-39&1.53e-40&6.97e-40 \\ 
10&5.39e-39&1.96e-40&7.31e-40 \\ 
11&5.52e-39&1.68e-40&7.43e-40 \\ 
12&6.47e-39&2.65e-40&9.95e-40 \\ 

\hline
\end{tabular}
\caption{The measured differential cross section as a function of \enuqe. Units are cm$^2$ per GeV$^2$ per nucleon.}
\label{tab:1D_XProjection_CCQE_CV_2p2h_enu_data}
\end{table}

\begin{table}[h]
\centering
\begin{tabular}{| c | c | c | c |}
\hline

\hline
Bin & Cross Section & Stat. Unc. & Total. Unc. \\ \hline \hline
1&7.46e-40&2.53e-41&1.40e-40 \\ 
2&1.18e-39&2.79e-41&1.56e-40 \\ 
3&1.46e-39&2.60e-41&1.57e-40 \\ 
4&1.26e-39&2.09e-41&1.38e-40 \\ 
5&7.71e-40&1.49e-41&1.01e-40 \\ 
6&4.36e-40&1.05e-41&5.87e-41 \\ 
7&2.68e-40&7.93e-42&3.37e-41 \\ 
8&1.83e-40&5.35e-42&2.17e-41 \\ 
9&1.10e-40&3.06e-42&1.37e-41 \\ 
10&6.83e-41&2.60e-42&9.12e-42 \\ 
11&3.83e-41&1.18e-42&4.98e-42 \\ 
12&1.81e-41&7.48e-43&2.69e-42 \\ 

\hline
\end{tabular}
\caption{The measured differential cross section as a function of $p_{||}$. Units are cm$^2$ per GeV$^2$ per nucleon.}
\label{tab:1D_XProjection_CCQE_CV_2p2h_pzmu_data}
\end{table}

\begin{table}[h]
\centering
\begin{tabular}{| c | c | c | c |}
\hline

\hline
Bin & Cross Section & Stat. Unc. & Total. Unc. \\ \hline \hline
1&1.53e-40&5.57e-41&6.89e-41 \\ 
2&8.20e-40&1.06e-40&1.82e-40 \\ 
3&2.49e-39&1.24e-40&3.54e-40 \\ 
4&4.44e-39&1.64e-40&5.55e-40 \\ 
5&5.86e-39&2.08e-40&7.04e-40 \\ 
6&6.86e-39&1.74e-40&7.75e-40 \\ 
7&7.07e-39&1.50e-40&7.75e-40 \\ 
8&5.57e-39&9.70e-41&5.76e-40 \\ 
9&3.34e-39&8.20e-41&3.86e-40 \\ 
10&1.70e-39&5.85e-41&2.44e-40 \\ 
11&5.00e-40&2.74e-41&1.05e-40 \\ 
12&8.71e-41&1.15e-41&3.33e-41 \\ 
13& 4.75e-42&1.67e-42 &2.78e-42 \\
\hline
\end{tabular}
\caption{The measured differential cross section as a function of $p_{t}$. Units are cm$^2$ per GeV$^2$ per nucleon.}
\label{tab:1D_YProjection_CCQE_CV_2p2h_pzmu_data}
\end{table}

\begin{table}[h]
\centering
\begin{tabular}{| c | c | c | c |}
\hline

\hline
Bin & Cross Section & Stat. Unc. & Total. Unc. \\ \hline \hline
1&2.37e-39&3.99e-40&7.25e-40 \\ 
2&2.99e-39&3.07e-40&7.24e-40 \\ 
3&5.10e-39&2.82e-40&8.56e-40 \\ 
4&6.49e-39&2.68e-40&9.21e-40 \\ 
5&7.03e-39&2.45e-40&9.55e-40 \\ 
6&8.13e-39&1.64e-40&9.81e-40 \\ 
7&8.40e-39&1.74e-40&9.44e-40 \\ 
8&8.32e-39&1.53e-40&8.88e-40 \\ 
9&6.86e-39&1.12e-40&7.25e-40 \\ 
10&5.00e-39&9.21e-41&5.06e-40 \\ 
11&3.17e-39&5.90e-41&3.41e-40 \\ 
12&1.76e-39&4.49e-41&2.12e-40 \\ 
13&1.01e-39&3.43e-41&1.45e-40 \\ 
14&5.61e-40&2.37e-41&9.36e-41 \\ 
15&1.69e-40&9.46e-42&3.88e-41 \\ 
16&1.03e-41&1.65e-42&4.87e-42 \\ 

\hline
\end{tabular}
\caption{The measured differential cross section as a function of \qqqe. Units are cm$^2$ per GeV$^2$ per nucleon.}
\label{tab:1D_XProjection_CCQE_CV_2p2h_q2_data}
\end{table}

\begin{table*}[h]
\centering
\begin{tabular}{| c | c | c | c | c | c | c | c | c | c | c | c | c |}
\hline
& 1 & 2 & 3 & 4 & 5 & 6 & 7 & 8 & 9 & 10 & 11 & 12 \\ \hline \hline
1 & 4.01e-41 & 1.16e-41 & 4.79e-41 & 6.79e-41 & 3.99e-41 & 1.08e-41 & 1.35e-41 & 6.17e-42 & 5.77e-42 & 4.05e-42 & 1.50e-42 & 8.51e-43 \\ \hline
2 & 1.34e-40 & 2.06e-40 & 2.87e-40 & 2.60e-40 & 1.73e-40 & 1.01e-40 & 7.32e-41 & 4.23e-41 & 2.68e-41 & 1.93e-41 & 9.82e-42 & 3.88e-42 \\ \hline
3 & 6.70e-40 & 7.47e-40 & 9.16e-40 & 7.83e-40 & 4.77e-40 & 2.93e-40 & 1.72e-40 & 1.22e-40 & 6.90e-41 & 3.92e-41 & 1.78e-41 & 7.44e-42 \\ \hline
4 & 1.21e-39 & 1.32e-39 & 1.51e-39 & 1.50e-39 & 9.82e-40 & 5.07e-40 & 2.91e-40 & 1.78e-40 & 1.14e-40 & 6.33e-41 & 3.59e-41 & 1.42e-41 \\ \hline
5 & 1.54e-39 & 1.91e-39 & 2.17e-39 & 1.71e-39 & 1.15e-39 & 6.51e-40 & 3.71e-40 & 2.49e-40 & 1.55e-40 & 8.79e-41 & 5.28e-41 & 2.13e-41 \\ \hline
6 & 1.77e-39 & 2.19e-39 & 2.42e-39 & 2.20e-39 & 1.28e-39 & 7.51e-40 & 4.17e-40 & 3.07e-40 & 1.61e-40 & 1.17e-40 & 6.77e-41 & 2.89e-41 \\ \hline
7 & 2.03e-39 & 2.41e-39 & 2.59e-39 & 2.24e-39 & 1.32e-39 & 7.17e-40 & 4.36e-40 & 2.62e-40 & 1.57e-40 & 9.08e-41 & 6.16e-41 & 2.72e-41 \\ \hline
8 & 1.11e-39 & 2.10e-39 & 2.20e-39 & 1.75e-39 & 1.05e-39 & 5.72e-40 & 3.68e-40 & 2.41e-40 & 1.38e-40 & 7.67e-41 & 4.46e-41 & 2.05e-41 \\ \hline
9 & 5.27e-41 & 1.17e-39 & 1.46e-39 & 1.16e-39 & 6.75e-40 & 3.97e-40 & 2.12e-40 & 1.61e-40 & 1.19e-40 & 6.46e-41 & 3.24e-41 & 1.79e-41 \\ \hline
10 & 0.00e+00 & 1.01e-40 & 8.48e-40 & 6.75e-40 & 3.86e-40 & 2.18e-40 & 1.46e-40 & 1.07e-40 & 6.33e-41 & 5.10e-41 & 2.22e-41 & 1.35e-41 \\ \hline
11 & 0.00e+00 & 0.00e+00 & 6.95e-41 & 1.79e-40 & 1.33e-40 & 7.97e-41 & 6.46e-41 & 4.68e-41 & 2.47e-41 & 2.01e-41 & 1.34e-41 & 6.77e-42 \\ \hline
12 & 0.00e+00 & 0.00e+00 & 0.00e+00 & 0.00e+00 & 1.15e-41 & 1.10e-41 & 1.63e-41 & 1.38e-41 & 6.87e-42 & 6.23e-42 & 3.70e-42 & 1.82e-42 \\ \hline
13 & 0.00e+00 & 0.00e+00 & 0.00e+00 & 0.00e+00 & 0.00e+00 & 1.01e-42 & 1.14e-42 & 6.56e-43 & 3.87e-43 & 2.33e-43 & 2.01e-43 & 1.56e-43 \\ \hline

\hline
\end{tabular}
\caption{The measured double differential cross section as a function of $p_{t}$~versus $p_{||}$. Units are cm$^2$ per GeV$^2$ per nucleon.}
\label{tab:2D_CCQE_CV_2p2h_data}
\end{table*}

\begin{table*}[h]
\centering
\begin{tabular}{| c | c | c | c | c | c | c | c | c | c | c | c | c |}
\hline
& 1 & 2 & 3 & 4 & 5 & 6 & 7 & 8 & 9 & 10 & 11 & 12 \\ \hline \hline
1 & 2.65e-41 & 1.60e-41 & 2.66e-41 & 2.13e-41 & 1.62e-41 & 6.48e-42 & 6.68e-42 & 4.38e-42 & 2.76e-42 & 1.85e-42 & 9.65e-43 & 5.13e-43 \\ \hline
2 & 7.05e-41 & 6.74e-41 & 7.04e-41 & 5.15e-41 & 3.64e-41 & 2.37e-41 & 1.67e-41 & 1.06e-41 & 6.43e-42 & 5.44e-42 & 2.44e-42 & 1.17e-42 \\ \hline
3 & 1.74e-40 & 1.40e-40 & 1.36e-40 & 1.02e-40 & 6.81e-41 & 4.39e-41 & 2.77e-41 & 1.82e-41 & 1.06e-41 & 7.24e-42 & 3.32e-42 & 1.54e-42 \\ \hline
4 & 2.63e-40 & 2.32e-40 & 2.09e-40 & 1.71e-40 & 1.21e-40 & 7.00e-41 & 4.15e-41 & 2.55e-41 & 1.55e-41 & 1.01e-41 & 5.61e-42 & 2.43e-42 \\ \hline
5 & 3.49e-40 & 2.84e-40 & 2.86e-40 & 1.92e-40 & 1.40e-40 & 9.10e-41 & 5.15e-41 & 3.26e-41 & 2.00e-41 & 1.33e-41 & 7.55e-42 & 3.48e-42 \\ \hline
6 & 3.55e-40 & 3.23e-40 & 3.03e-40 & 2.38e-40 & 1.60e-40 & 1.02e-40 & 5.85e-41 & 3.89e-41 & 2.15e-41 & 1.62e-41 & 8.50e-42 & 4.09e-42 \\ \hline
7 & 3.92e-40 & 3.52e-40 & 3.00e-40 & 2.46e-40 & 1.87e-40 & 9.38e-41 & 5.94e-41 & 3.47e-41 & 2.06e-41 & 1.28e-41 & 7.64e-42 & 3.85e-42 \\ \hline
8 & 1.96e-40 & 2.74e-40 & 2.42e-40 & 1.97e-40 & 1.45e-40 & 8.14e-41 & 4.77e-41 & 2.94e-41 & 1.72e-41 & 1.09e-41 & 6.11e-42 & 2.94e-42 \\ \hline
9 & 2.20e-41 & 1.67e-40 & 1.79e-40 & 1.69e-40 & 1.07e-40 & 6.34e-41 & 3.24e-41 & 2.26e-41 & 1.65e-41 & 1.02e-41 & 5.07e-42 & 2.89e-42 \\ \hline
10 & 0.00e+00 & 2.88e-41 & 1.29e-40 & 1.25e-40 & 7.53e-41 & 4.51e-41 & 2.79e-41 & 1.93e-41 & 1.25e-41 & 9.35e-42 & 4.15e-42 & 2.52e-42 \\ \hline
11 & 0.00e+00 & 0.00e+00 & 1.94e-41 & 4.53e-41 & 3.48e-41 & 2.05e-41 & 1.58e-41 & 1.26e-41 & 7.10e-42 & 5.66e-42 & 3.25e-42 & 2.07e-42 \\ \hline
12 & 0.00e+00 & 0.00e+00 & 0.00e+00 & 0.00e+00 & 5.67e-42 & 5.07e-42 & 6.35e-42 & 5.59e-42 & 3.47e-42 & 3.37e-42 & 1.77e-42 & 1.09e-42 \\ \hline
13 & 0.00e+00 & 0.00e+00 & 0.00e+00 & 0.00e+00 & 0.00e+00 & 1.49e-42 & 7.96e-43 & 3.88e-43 & 2.39e-43 & 1.91e-43 & 1.67e-43 & 1.59e-43 \\ \hline

\hline
\end{tabular}
\caption{The measured double differential cross section total uncertainty as a function of $p_{t}$~versus $p_{||}$. Units are cm$^2$ per GeV$^2$ per nucleon.}
\label{tab:Unc_2D_CCQE_CV_2p2h_data}
\end{table*}

\clearpage
\subsection{Supplemental: Monte Carlo Predictions}

\begin{table}[h]
\centering
\begin{tabular}{| c | c |}
\hline

\hline
Bin & Cross Section \\ \hline \hline
1&2.67e-39 \\ 
2&3.55e-39 \\ 
3&4.03e-39 \\ 
4&4.34e-39 \\ 
5&4.51e-39 \\ 
6&4.72e-39 \\ 
7&5.02e-39 \\ 
8&5.06e-39 \\ 
9&5.19e-39 \\ 
10&5.15e-39 \\ 
11&5.16e-39 \\ 
12&5.32e-39 \\ 

\hline
\end{tabular}
\caption{Monte Carlo predicted differential cross section as a function of \enuqe. Units are cm$^2$ per GeV$^2$ per nucleon. This is a prediction for \tune.}
\label{tab:1D_XProjection_CCQE_CV_2p2h_enu_mc}
\end{table}

\begin{table}[h]
\centering
\begin{tabular}{| c | c |}
\hline

\hline
Bin & Cross Section \\ \hline \hline
1&1.81e-39 \\ 
2&2.47e-39 \\ 
3&2.96e-39 \\ 
4&3.34e-39 \\ 
5&3.63e-39 \\ 
6&3.85e-39 \\ 
7&4.05e-39 \\ 
8&3.98e-39 \\ 
9&4.00e-39 \\ 
10&3.97e-39 \\ 
11&3.95e-39 \\ 
12&4.00e-39 \\ 

\hline
\end{tabular}
\caption{Monte Carlo predicted differential cross section as a function of \enuqe. Units are cm$^2$ per GeV$^2$ per nucleon. This is a prediction for GENIE 2.8.4.}
\label{tab:1D_XProjection_CCQE_default_enu_mc}
\end{table}

\begin{table}[h]
\centering
\begin{tabular}{| c | c |}
\hline

\hline
Bin & Cross Section \\ \hline \hline
1&6.89e-40 \\ 
2&1.12e-39 \\ 
3&1.37e-39 \\ 
4&1.22e-39 \\ 
5&8.14e-40 \\ 
6&4.63e-40 \\ 
7&2.73e-40 \\ 
8&1.68e-40 \\ 
9&1.02e-40 \\ 
10&6.48e-41 \\ 
11&3.50e-41 \\ 
12&1.44e-41 \\ 

\hline
\end{tabular}
\caption{Monte Carlo predicted differential cross section as a function of $p_{||}$. Units are cm$^2$ per GeV$^2$ per nucleon. This is a prediction for \tune.}
\label{tab:1D_XProjection_CCQE_CV_2p2h_pzmu_mc}
\end{table}

\begin{table}[h]
\centering
\begin{tabular}{| c | c |}
\hline

\hline
Bin & Cross Section \\ \hline \hline
1&4.73e-40 \\ 
2&8.07e-40 \\ 
3&1.03e-39 \\ 
4&9.57e-40 \\ 
5&6.66e-40 \\ 
6&3.80e-40 \\ 
7&2.19e-40 \\ 
8&1.31e-40 \\ 
9&7.84e-41 \\ 
10&4.98e-41 \\ 
11&2.67e-41 \\ 
12&1.08e-41 \\ 

\hline
\end{tabular}
\caption{Monte Carlo predicted differential cross section as a function of $p_{||}$. Units are cm$^2$ per GeV$^2$ per nucleon. This is a prediction for GENIE 2.8.4.}
\label{tab:1D_XProjection_CCQE_default_pzmu_mc}
\end{table}

\begin{table}[h]
\centering
\begin{tabular}{| c | c |}
\hline

\hline
Bin & Cross Section \\ \hline \hline
1&2.56e-40 \\ 
2&1.00e-39 \\ 
3&2.35e-39 \\ 
4&4.19e-39 \\ 
5&5.83e-39 \\ 
6&6.62e-39 \\ 
7&6.54e-39 \\ 
8&5.18e-39 \\ 
9&3.20e-39 \\ 
10&1.55e-39 \\ 
11&4.80e-40 \\ 
12&1.20e-40 \\ 

\hline
\end{tabular}
\caption{Monte Carlo predicted differential cross section as a function of $p_{t}$. Units are cm$^2$ per GeV$^2$ per nucleon. This is a prediction for \tune.}
\label{tab:1D_YProjection_CCQE_CV_2p2h_pzmu_mc}
\end{table}

\begin{table}[h]
\centering
\begin{tabular}{| c | c |}
\hline

\hline
Bin & Cross Section \\ \hline \hline
1&1.65e-40 \\ 
2&8.11e-40 \\ 
3&2.05e-39 \\ 
4&3.40e-39 \\ 
5&4.25e-39 \\ 
6&4.61e-39 \\ 
7&4.62e-39 \\ 
8&3.90e-39 \\ 
9&2.49e-39 \\ 
10&1.27e-39 \\ 
11&4.22e-40 \\ 
12&1.12e-40 \\ 

\hline
\end{tabular}
\caption{Monte Carlo predicted differential cross section as a function of $p_{t}$. Units are cm$^2$ per GeV$^2$ per nucleon. This is a prediction for GENIE 2.8.4.}
\label{tab:1D_YProjection_CCQE_default_pzmu_mc}
\end{table}

\begin{table}[h]
\centering
\begin{tabular}{| c | c |}
\hline

\hline
Bin & Cross Section \\ \hline \hline
1&3.88e-39 \\ 
2&4.73e-39 \\ 
3&5.35e-39 \\ 
4&6.05e-39 \\ 
5&6.66e-39 \\ 
6&7.62e-39 \\ 
7&8.42e-39 \\ 
8&8.00e-39 \\ 
9&6.60e-39 \\ 
10&4.72e-39 \\ 
11&2.96e-39 \\ 
12&1.66e-39 \\ 
13&9.14e-40 \\ 
14&5.08e-40 \\ 
15&1.57e-40 \\ 
16&1.72e-41 \\ 

\hline
\end{tabular}
\caption{Monte Carlo predicted differential cross section as a function of \qqqe. Units are cm$^2$ per GeV$^2$ per nucleon. This is a prediction for \tune.}
\label{tab:1D_XProjection_CCQE_CV_2p2h_q2_mc}
\end{table}

\begin{table}[h]
\centering
\begin{tabular}{| c | c |}
\hline

\hline
Bin & Cross Section \\ \hline \hline
1&2.46e-39 \\ 
2&3.53e-39 \\ 
3&4.36e-39 \\ 
4&5.17e-39 \\ 
5&5.71e-39 \\ 
6&6.24e-39 \\ 
7&6.15e-39 \\ 
8&5.63e-39 \\ 
9&4.56e-39 \\ 
10&3.46e-39 \\ 
11&2.26e-39 \\ 
12&1.29e-39 \\ 
13&7.39e-40 \\ 
14&4.25e-40 \\ 
15&1.39e-40 \\ 
16&1.63e-41 \\ 

\hline
\end{tabular}
\caption{Monte Carlo predicted differential cross section as a function of \qqqe. Units are cm$^2$ per GeV$^2$ per nucleon. This is a prediction for GENIE 2.8.4.}
\label{tab:1D_XProjection_CCQE_default_q2_mc}
\end{table}

\begin{table*}[h]
\centering
\begin{tabular}{| c | c | c | c | c | c | c | c | c | c | c | c | c |}
\hline
& 1 & 2 & 3 & 4 & 5 & 6 & 7 & 8 & 9 & 10 & 11 & 12 \\ \hline \hline
1 & 6.27e-41 & 7.66e-41 & 8.49e-41 & 8.01e-41 & 6.06e-41 & 3.37e-41 & 2.00e-41 & 1.06e-41 & 6.65e-42 & 4.18e-42 & 2.25e-42 & 6.77e-43 \\ \hline
2 & 2.40e-40 & 3.00e-40 & 3.37e-40 & 3.18e-40 & 2.34e-40 & 1.33e-40 & 7.15e-41 & 4.21e-41 & 2.47e-41 & 1.62e-41 & 8.53e-42 & 3.58e-42 \\ \hline
3 & 5.88e-40 & 7.29e-40 & 8.06e-40 & 7.56e-40 & 5.40e-40 & 3.05e-40 & 1.66e-40 & 9.72e-41 & 5.72e-41 & 3.39e-41 & 1.81e-41 & 7.36e-42 \\ \hline
4 & 1.08e-39 & 1.31e-39 & 1.45e-39 & 1.35e-39 & 9.41e-40 & 5.28e-40 & 2.98e-40 & 1.70e-40 & 9.77e-41 & 5.94e-41 & 3.28e-41 & 1.35e-41 \\ \hline
5 & 1.50e-39 & 1.89e-39 & 2.08e-39 & 1.83e-39 & 1.27e-39 & 7.14e-40 & 4.10e-40 & 2.30e-40 & 1.35e-40 & 8.26e-41 & 4.60e-41 & 1.88e-41 \\ \hline
6 & 1.78e-39 & 2.14e-39 & 2.34e-39 & 2.05e-39 & 1.43e-39 & 8.08e-40 & 4.47e-40 & 2.63e-40 & 1.49e-40 & 9.63e-41 & 5.14e-41 & 2.14e-41 \\ \hline
7 & 1.80e-39 & 2.14e-39 & 2.33e-39 & 2.05e-39 & 1.37e-39 & 7.51e-40 & 4.48e-40 & 2.56e-40 & 1.56e-40 & 9.19e-41 & 4.90e-41 & 2.09e-41 \\ \hline
8 & 9.54e-40 & 1.87e-39 & 1.97e-39 & 1.68e-39 & 1.08e-39 & 5.89e-40 & 3.49e-40 & 2.11e-40 & 1.29e-40 & 8.22e-41 & 4.34e-41 & 1.75e-41 \\ \hline
9 & 2.26e-41 & 1.10e-39 & 1.45e-39 & 1.14e-39 & 6.98e-40 & 4.00e-40 & 2.36e-40 & 1.50e-40 & 9.21e-41 & 5.95e-41 & 3.11e-41 & 1.28e-41 \\ \hline
10 & 0.00e+00 & 7.45e-41 & 7.74e-40 & 6.49e-40 & 3.83e-40 & 2.24e-40 & 1.37e-40 & 9.08e-41 & 5.98e-41 & 3.77e-41 & 2.08e-41 & 8.71e-42 \\ \hline
11 & 0.00e+00 & 0.00e+00 & 4.59e-41 & 1.85e-40 & 1.37e-40 & 8.58e-41 & 6.15e-41 & 4.59e-41 & 3.05e-41 & 1.95e-41 & 1.10e-41 & 4.32e-42 \\ \hline
12 & 0.00e+00 & 0.00e+00 & 0.00e+00 & 0.00e+00 & 1.91e-41 & 2.29e-41 & 1.99e-41 & 1.70e-41 & 1.13e-41 & 8.12e-42 & 4.60e-42 & 1.97e-42 \\ \hline
13 & 0.00e+00 & 0.00e+00 & 0.00e+00 & 0.00e+00 & 0.00e+00 & 4.85e-43 & 1.19e-42 & 1.62e-42 & 1.55e-42 & 1.25e-42 & 7.59e-43 & 3.24e-43 \\ \hline

\hline
\end{tabular}
\caption{Monte Carlo predicted double differential cross section as a function of $p_{t}$~versus $p_{||}$. Units are cm$^2$ per GeV$^2$ per nucleon. This is the prediction for \tune.}
\label{tab:2D_CCQE_CV_2p2h_mc}
\end{table*}

\begin{table*}[h]
\centering
\begin{tabular}{| c | c | c | c | c | c | c | c | c | c | c | c | c |}
\hline
& 1 & 2 & 3 & 4 & 5 & 6 & 7 & 8 & 9 & 10 & 11 & 12 \\ \hline \hline
1 & 3.36e-41 & 4.29e-41 & 5.15e-41 & 5.27e-41 & 4.65e-41 & 2.77e-41 & 1.70e-41 & 6.96e-42 & 4.20e-42 & 2.66e-42 & 1.35e-42 & 4.00e-43 \\ \hline
2 & 1.63e-40 & 2.25e-40 & 2.67e-40 & 2.67e-40 & 2.11e-40 & 1.19e-40 & 6.41e-41 & 3.65e-41 & 1.95e-41 & 1.33e-41 & 6.96e-42 & 3.12e-42 \\ \hline
3 & 4.53e-40 & 5.81e-40 & 6.81e-40 & 6.82e-40 & 5.25e-40 & 3.03e-40 & 1.63e-40 & 8.79e-41 & 4.83e-41 & 2.94e-41 & 1.62e-41 & 6.57e-42 \\ \hline
4 & 7.57e-40 & 9.84e-40 & 1.14e-39 & 1.13e-39 & 8.52e-40 & 4.95e-40 & 2.62e-40 & 1.43e-40 & 7.87e-41 & 4.99e-41 & 2.75e-41 & 1.04e-41 \\ \hline
5 & 9.75e-40 & 1.24e-39 & 1.47e-39 & 1.40e-39 & 1.04e-39 & 5.98e-40 & 3.23e-40 & 1.78e-40 & 9.93e-41 & 6.46e-41 & 3.24e-41 & 1.28e-41 \\ \hline
6 & 1.13e-39 & 1.39e-39 & 1.62e-39 & 1.50e-39 & 1.10e-39 & 6.04e-40 & 3.27e-40 & 1.87e-40 & 1.06e-40 & 6.58e-41 & 3.51e-41 & 1.38e-41 \\ \hline
7 & 1.21e-39 & 1.47e-39 & 1.62e-39 & 1.49e-39 & 1.05e-39 & 5.68e-40 & 3.28e-40 & 1.79e-40 & 1.04e-40 & 6.22e-41 & 3.37e-41 & 1.41e-41 \\ \hline
8 & 6.97e-40 & 1.39e-39 & 1.48e-39 & 1.30e-39 & 8.53e-40 & 4.62e-40 & 2.69e-40 & 1.55e-40 & 9.36e-41 & 5.92e-41 & 3.09e-41 & 1.22e-41 \\ \hline
9 & 1.76e-41 & 8.68e-40 & 1.14e-39 & 8.99e-40 & 5.51e-40 & 3.14e-40 & 1.83e-40 & 1.14e-40 & 6.99e-41 & 4.42e-41 & 2.31e-41 & 9.53e-42 \\ \hline
10 & 0.00e+00 & 6.12e-41 & 6.44e-40 & 5.35e-40 & 3.12e-40 & 1.84e-40 & 1.11e-40 & 7.36e-41 & 4.84e-41 & 3.03e-41 & 1.65e-41 & 6.96e-42 \\ \hline
11 & 0.00e+00 & 0.00e+00 & 3.99e-41 & 1.62e-40 & 1.20e-40 & 7.55e-41 & 5.42e-41 & 4.05e-41 & 2.70e-41 & 1.73e-41 & 9.63e-42 & 3.78e-42 \\ \hline
12 & 0.00e+00 & 0.00e+00 & 0.00e+00 & 0.00e+00 & 1.78e-41 & 2.12e-41 & 1.86e-41 & 1.59e-41 & 1.06e-41 & 7.66e-42 & 4.29e-42 & 1.82e-42 \\ \hline
13 & 0.00e+00 & 0.00e+00 & 0.00e+00 & 0.00e+00 & 0.00e+00 & 4.69e-43 & 1.15e-42 & 1.57e-42 & 1.49e-42 & 1.20e-42 & 7.33e-43 & 3.11e-43 \\ \hline

\hline
\end{tabular}
\caption{Monte Carlo predicted double differential cross section as a function of $p_{t}$~versus $p_{||}$. Units are cm$^2$ per GeV$^2$ per nucleon. This is the prediction for GENIE 2.8.4.}
\label{tab:2D_CCQE_default_mc}
\end{table*}